\documentclass[amsmath,amssymb,twocolumn]{revtex4}
\usepackage{amsmath}
\usepackage[latin1]{inputenc}
\usepackage[english]{babel}
\usepackage{latexsym}
\usepackage{graphicx}

\def\be{\begin{equation}}
\def\ee{\end{equation}}
\def\bc{\begin{center}}
\def\ec{\end{center}}
\def\bea{\begin{eqnarray}}
\def\eea{\end{eqnarray}}
\def\bean{\begin{eqnarray*}}
\def\eean{\end{eqnarray*}}

\def\log{{\rm \; log}}

\begin{document}

\title[]{Supercooled Liquids for Pedestrians}

\author{Andrea Cavagna}

\affiliation{Centre for Statistical Mechanics and Complexity, INFM-CNR \\ 
Istituto Sistemi Complessi, CNR \\
via dei Taurini 19, 00185 Rome, Italy
\\ \hfill }

\begin{abstract}

When we lower the temperature of a liquid, at some point we meet a first order phase
transition to the crystal. Yet, under certain conditions it is possible to keep the system in its
metastable phase and to avoid crystallization. In this way the
liquid enters in the supercooled phase.
Supercooled liquids have a very rich phenomenology, which is
still far from being completely understood. To begin with, there is
the problem of how to prevent crystallization and how deeply the liquid
can be supercooled before a metastability limit is hit. But by far the
most interesting feature of supercooled liquids is the dynamic glass
transition: when the temperature is decreased below a certain point,
the relaxation time increases so much that a dramatic dynamical arrest
intervenes and we are unable to equilibrate the system within
reasonable experimental times. The glass transition is a phenomenon
whose physical origin has stirred an enormous interest in the last
hundred years. Why does it occur? Is it just a conventional reference
point, or does it have a more profound physical meaning? Is it a
purely dynamical event, or the manifestation of a true thermodynamic
transition? What is the correlation length associated to the
sharp increase of the relaxation time? Can we define a new kind of
amorphous order?  A shared theory of supercooled liquids and the glass
transition does not yet exist and  these questions are still
largely open.  Here, I will illustrate in the most elementary fashion the
main phenomenological traits of supercooled liquids and discuss in a
very partial way a few theoretical ideas on the subject.

\end{abstract}

%\pacs{00.00, 00.00, 00.00}

\maketitle

\tableofcontents

\newpage

\section{Introduction}

In March 2007 I was invited at the Weizmann Institute to deliver an
introductory lecture on the physics of supercooled liquids and the
glass transition. Even though I had to talk about equilibrium
behaviour, leaving aside anything connected to aging and
off-equilibrium phenomena, the task looked quite daunting. In
approximately two hours I had to cover an exceptionally vast subject
in front of an unspecialized audience, who explicitly asked me to stay
at an elementary level.  I therefore selected very few concepts that
were in my opinion the keystones in the physics of supercooled
liquids, and I tried to frame them within a brief, but coherent story.
The result was apparently not too bad, and at the end of my talk
the Chairs invited me to turn the lecture into
a review. I was given complete freedom in deciding the plan of the
work, but for one key point: it had to be as elementary, incomplete
and fluent as the original lecture.

The problem is that the field of supercooled liquids is already
embarrassingly rich in reviews and textbooks. References
\cite{angell88,goetze-review,gutzow,debenedetti-review,angell-00, tarjus-01,donth, kob-review,
kob-binder, dyre-review, leuzzi,kive-tarj-08} make just a partial selection. 
Therefore, it seems polite to give a justification of the
apparently needless new effort, and explain what is the new value of
the work, beyond the belief common to all authors to be considerably
better than the predecessors (which is, however, small consolation to
the reader).

I believe that in the literature on supercooled liquids and the glass
transition there is a certain lack of really simple reviews, aimed
at a young and unspecialized audience.  By this I mean reviews that
can be read and understood without any specific training in liquid
theory or disordered systems, but with just a background in
statistical mechanics. This is probably the reason why my lecture at
Weizmann was appreciated.  The aim of the present review is therefore
to span the whole subject in a brief and compact way, at the price of
being utterly inexhaustive and somewhat superficial. The level is
truly elementary. The expert reader is warned that many important
topics are not covered.

Despite the similar pedestrian spirit, the plan of the present work is
quite different from my previous notes, {\it Spin-glass theory for
pedestrians} \cite{cavagna-review}.  In that case, we analyzed in full
mathematical detail a single mean-field spin-glass (the $p$-spin
model), as a paradigm for the entire field.  A similar operation would
be impossible for supercooled liquids, where it is very hard to single
out one paradigmatic system and where no well-established theoretical
framework is present. Therefore, I decided to proceed mainly through
the phenomenology, as on a journey through the various temperature
regimes that are crossed when a liquid is cooled. This journey is
schematically depicted in Fig.1, which can be used as a map throughout
these notes.

After a first part containing some viscoelastic preliminaries, we will
try to understand what are the infamous `certain conditions' under
which crystallization is avoided and a supercooled liquid is formed
below the melting point $T_m$. In this part we will focus on
nucleation theory and in particular on the concept of kinetic
spinodal, namely the temperature below which the crystal nucleation
time becomes shorter than the liquid relaxation time. Below this
temperature the only way to prevent crystallization is to push the
system out of equilibrium.  The kinetic spinodal represents a
metastability limit for the supercooled phase and it is thus essential
to discover under what conditions it can be avoided. This chapter is
important to understand that the existence of supercooled liquids must
not be taken for granted, and that skipping crystallization is more
subtle a process than what is sometimes assumed.

Once the existence of the liquid is secured against unwanted
crystallization, the temperature can be decreased so much as to meet
the dynamic glass transition, $T_g$. At this point the relaxation time
of the system grows so sharply as to exceed our available experimental
time: below the glass transition the system falls inevitably out of
equilibrium, and a glass is formed.  This operational definition may
suggest that $T_g$ is a purely conventional point, whose existence is
solely due to the finite timespan of our experiments.  Moreover, the
structure of liquid configurations close to $T_g$ seems to remain
exactly the same as at higher temperature. This fact too gives the
impression that perhaps nothing crucial is going on at the famous
glass transition, and that the growth of the relaxation time is just a
not-too-relevant {\it quantitative} feature. We will see that this is
not true, and that at least for fragile liquids some really new
physics kicks in at $T_g$. We will focus on those
phenomenological traits that distinguish at a {\it qualitative} level
 a liquid in its fluid phase from a liquid close to the glass
transition. We will find that the two steps relaxation of the
dynamical correlation function is the most insightful landmark of the
proximity to the glass transition.

In order to explain what physical ingredients leads to the glass
transition, we will discuss the crossover from nonactivated to
activated dynamics taking place at the Goldstein's temperature
$T_x$. Above $T_x$ activation is not the main mechanism of diffusion
and dynamics is reasonably described by mode coupling theory. Below
$T_x$, however, the liquid enters in its activated phase, eventually
leading to the dynamic glass transition at $T_g$.  To better
understand the nature of this crossover, we will discuss the
mean-field results obtained within the $p$-spin model and show that
the dynamical transition $T_c$ predicted by both mode coupling and the
$p$-spin is conceptually the same as Goldstein's temperature $T_x$. We
will finally explore the role of saddles of the potential energy and
explain how two steps relaxation can be the result of a system running
out of unstable directions in the phase space.

After this, we will dive into the deeply supercooled phase.  We will
imagine that we can skip the glass transition (we cannot) and keep the
liquid at equilibrium even at very low temperature.  If we do this, we
meet the famous Kauzmann's temperature $T_k$, namely the point where
the entropy of the liquid becomes equal to that of the crystal.
According to some theoretical schemes, at this point there must be a
phase transition, with a true divergence of the relaxation time.
Whether or not such transition really occurs is perhaps not too
important, since anyway the dynamic glass transition prevents us to
reach this point.  However, the physical mechanisms  behind
the phase transition may affect and regulate the behaviour of the system
even at higher temperatures, giving rise to some precursor phenomena
that may be also detected experimentally. For this reason studying the
deeply supercooled phase is not a mere academic exercise.

\begin{figure}
\includegraphics[clip,width=3.7in]{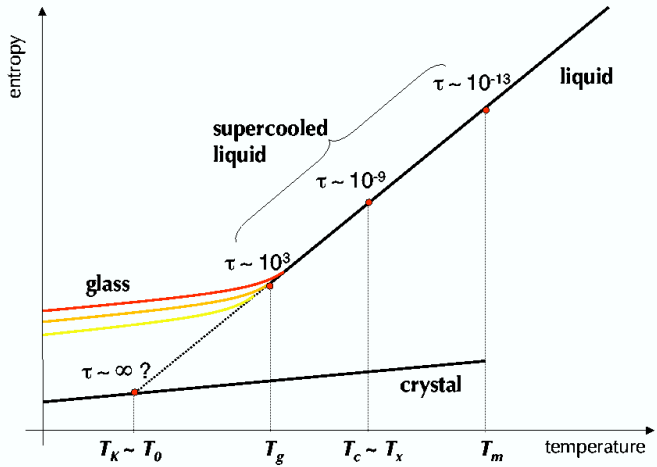}
\caption{ {\bf A pedestrian map of supercooled liquids} - Schematic
representation of the entropy as a function of temperature in a
liquid, from the high-$T$ phase, down to the deeply supercooled phase.
All the relevant temperatures introduced in these notes are marked:
$T_m$ is the melting point, where a first-order phase transition
between liquid and crystal occurs; $T_c$ is the temperature where mode
coupling theory and the $p$-spin model locate a purely dynamic
transition; $T_x$ is Goldstein's crossover temperature from a high-$T$
nonactivated dynamics to a low-$T$ activated one; $T_g$ is the dynamic
glass transition, where the relaxation time exceeds the conventional
experimental time of $10^3$ seconds; the longer the available
experimental time, the lower the temperature where the system falls
out of equilibrium forming a glass (different colours); $T_k$ is
Kauzmann's entropy crisis temperature, where the extrapolated liquid
entropy hits the crystal entropy, and where according to some theories
there is a thermodynamic phase transition; $T_0$ is the temperature
where the Vogel-Fulcher-Tamman fit locates a divergence of the
relaxation time.  Above each temperature we report the approximate
value (in seconds) of the relaxation time.}
\end{figure}

Finally, we will look for a correlation length. Intuition tells us
that wherever there is a large time, there should also be a large
length.  This makes sense: the relaxation time becomes large because
the system needs to rearrange larger and larger correlated regions at
low temperature. A growing lengthscale is a key ingredient of
some important theoretical frameworks of supercooled liquids, like the
Adam-Gibbs  theory and the mosaic theory.  Yet, how to define
this lengthscale in supercooled liquids is not quite clear.  Finding a
correlation length normally requires to define a suitable order
parameter and measure its relative correlation function.  However, how
to do this is  nontrivial in glass-forming liquids, because all
amorphous configurations look the same and it is difficult to detect
correlated regions. Explaining how to uncover the growth of amorphous
order will be the final aim of this notes.

As in my former review on mean-field spin-glasses, I do not cover the
subject of aging and off-equilibrium dynamics. I cannot stress enough
that this is not due to my dislike or disregard of the subject, but,
on the contrary, to the fact that it is such an important and vast
field that it cannot be covered by a single section of a brief review
as this is. I have tried (starting from the title) to use as much as
possible the term 'supercooled liquids' rather than 'glasses'
expressly in order to avoid any confusion between equilibrium and
off-equilibrium physics.

A final word of caution on the theoretical frameworks discussed in
these notes. A shared theory of the glass transition does not exist
and different frameworks clash continuously. As I write, a group of
researchers from all over the world prepares for a workshop in Leiden,
where they will kindly disagree about everything. In this context, I
had two choices: either to present a more or less complete list of all
theories on the market, briefly discussing the pros and cons of all of
them, or to tell a single, coherent, but inevitably partial story, and
discuss what is mainly my personal viewpoint. Not surprisingly, due to
lack of space and overwhelming laziness, I have chosen the second
solution. This was also more true to the spirit of my original
Weizmann lecture. In fact, even the structure I gave to the material
contained in this review reflects my own understanding of the
field. Therefore, even though my view about the fundamental physical
mechanisms ruling supercooled liquids is not entirely a minority one,
the reader should be well aware that this is {\it not} the whole
story.

%%%%%%%%%%%%%%%%%%%%%%%%%%%%%%%%%%%%%%%%%%%%%%%%%%%%%%%%%%%%%%%%%%%%%%%%%%%%%%%
%%%%%%%%%%%%%%%%%%%%%%%%%%%%%%%%%%%%%%%%%%%%%%%%%%%%%%%%%%%%%%%%%%%%%%%%%%%%%%%
%%%%%%%%%%%%%%%%%%%%%%%%%%%%%%%%%%%%%%%%%%%%%%%%%%%%%%%%%%%%%%%%%%%%%%%%%%%%%%%
%%%%%%%%%%%%%%%%%%%%%%%%%%%%%%%%%%%%%%%%%%%%%%%%%%%%%%%%%%%%%%%%%%%%%%%%%%%%%%%
%%%%%%%%%%%%%%%%%%%%%%%%%%%%%%%%%%%%%%%%%%%%%%%%%%%%%%%%%%%%%%%%%%%%%%%%%%%%%%%

\section{Preliminaries}

It seems nice to start a review on supercooled liquids with a
definition of what liquids are. This is particularly important for
deeply supercooled liquids, since in this case the viscosity may be so
large that, from the mechanical point of view, the distinction between
liquid and solid becomes blurred. Viscoelasticity is crucial to
understand the mechanical features of supercooled liquids, and it is
also the ideal setting where to introduce two key tools for the rest
of these notes, namely shear relaxation time and viscosity.  Moreover,
we will introduce the diffusion coefficient, and explain how it is
connected to the viscosity through the Stokes-Einstein relation.
Finally, I will clarify what I intend for `equilibrium supercooled
liquid', as it may seem rather contradictory to define as
`equilibrium' a metastable phase of matter.  As we shall see, however,
there are good reasons to adopt such a terminology.  A good discussion
of the basic concepts of rheology can be found in \cite{ferry}, and in
briefer form in \cite{cates-review}. For a very sharp introduction
about diffusion and the Langevin equation see \cite{zwanzig}.

\subsection{Shear relaxation time and viscosity}

Liquids flow, solids do not. For the level of these notes this may be a good enough definition
of the liquid phase. Yet, let us expand a little. First, we must forget about liquids, and focus on solids.
Consider a cubic solid of side $L$, subjected to a 
small shear. The effect of the shear is to produce a
displacement $u_x$ along the $x$ axis, proportional to the height $y$,
\be
u_x=\gamma \; y \ .
\label{shear}
\ee
When $y=L$, the displacement $u_x$ becomes equal to the maximum shift 
of the top of the cube, let us call it $l$. In this way we have
$\gamma=l/L$. The shear {\it strain} is the $(x,y)$ component of the 
{\it strain tensor}, $u_{xy}$, and it is given by,
\be
u_{xy}=\frac{\partial u_x}{\partial y} \ ,
\ee
and thus,
\be
u_{xy} = \gamma \ .
\ee
The solid responds elastically to such a step shear strain, 
giving rise to a {\it stress} $\sigma_{xy}$ proportional to the 
strain,
\be
\sigma_{xy} = G \, u_{xy} =  G \, \gamma \ .
\label{solid}
\ee
This equation defines the elastic shear modulus $G$ of the solid: the induced stress is proportional to the 
small step-strain and the proportionality factor is the shear modulus. The stress does not decay in time.
Solids do not flow.

Let us apply a similar small shear to a liquid sample: at time $t'$ the liquid is subjected to a step shear
given by (\ref{shear}). Saying that liquids flow is equivalent to say 
that at time $t''>t'$ the stress will be partially relaxed.
Provided that we are at equilibrium, so that time-translation invariance (TTI) holds,
and provided that we are in the linear regime, we can generalize equation (\ref{solid}) to,
\be
\sigma_{xy}= G(t''-t')\; \gamma \ ,
\label{modulus}
\ee
where  $G(t-t')$ is the time-dependent shear modulus (also called linear stress-relaxation function, or 
linear step-strain response function). The time-dependent shear modulus is a decaying function 
of its argument. In a liquid $G(t)\to 0$ for $t\to\infty$, whereas in a solid $G(t)$ decays to a plateau
for infinite times. Therefore, when we say that liquids flow, while solids do not, we must be careful,
since everything depends on the time-scale. Also liquids have a nonzero elastic response to shear 
for short enough times. This was the 
idea of Maxwell, according to which liquids behave (mechanically) like solids at times shorter than a certain
time-scale. This time-scale can be taken as the fundamental definition of shear relaxation time 
$\tau_\mathrm R$. The simplest form of $G(t)$ is provided by the Maxwell model,
\be
G(t)=G_\infty \exp(-t/\tau_\mathrm R) \ ,
\label{maxwell}
\ee
where $G_\infty$ is the infinite frequency (i.e. zero time) shear modulus.
Within the Maxwell model there is just one exponential relaxation time.
In real liquids things are more complicated, and there may be a superposition
of many components, 
\be
G(t)= \sum_i G_\infty^i \; \exp(-t/\tau_i) \ .
\ee
In this case
one has to extract a singly time-scale from $G(t)$ in a more or less arbitrary way
(typically, one focuses on the largest relaxation time).
However, the essence of the problem does not change: the shear relaxation
time of a liquid rules the way the stress-relaxation function $G(t)$
decays. 

For times much smaller than the shear relaxation time the liquid has a
nonzero elastic response. If the liquid undergoes an oscillatory shear
with frequency much larger than $1/\tau_\mathrm R$, it does {\it not}
flow. On the other hand, for times much longer than the relaxation
time, the stress is relaxed.  This may be of little interest at high
temperatures, where $\tau_\mathrm R$ is very small ($\sim 10^{-13}$
sec). However, in the deeply supercooled phase, the relaxation time is
very large ($\sim 10-100$ sec), so that the sample has a nonzero
elastic response on a wide range of times.

Let us consider now an arbitrary time-dependent shear strain $\gamma(t)$, rather
than a step-strain.  In this case we can say that the stress builds up as a
consequence of the many accumulated steps of strain,
\be
\delta \sigma_{xy} = G(t-t')\; \delta \gamma(t') =  G(t-t')\; \dot{\gamma}(t')\; dt' \ .
\ee
If the applied shear is zero for $t<0$ and different from zero for $t>0$, we can 
integrate over time and obtain the total stress,
\be
\sigma_{xy}(t) = \int_{0}^t dt'\ G(t-t')\; \dot{\gamma}(t') \ .
\label{viscoelasticity}
\ee
From this formula it is evident the role of the time-dependent shear modulus $G$ as
a memory kernel. In the particular, but important, case of a constant shear rate $\dot\gamma$,
we get,
\be
\sigma_{xy}(t)= \dot{\gamma} \int_0^t ds \ G(s) \ .
\label{jungle}
\ee
In a normal liquid, the shear relaxation function $G(s)$ decays faster than $1/s$,
so that in the asymptotic regime $t\to\infty$ the system develops a constant stress in response 
to a linear shear strain (i.e. a constant shear rate).
The ratio of shear stress to linear shear strain is the very definition of the
viscosity $\eta$, and thus we have,
\be
\sigma_{xy}= \eta \ \dot{\gamma} 
\quad\quad , \quad\quad
\eta = \int_0^\infty ds \ G(s) \ .
\label{carisma}
\ee
For Maxwell liquids (equation  (\ref{maxwell})) we recover the well-known formula linking viscosity, 
relaxation time and shear modulus,
\be
\eta= G_\infty \, \tau_\mathrm R \ .
\label{magnapompa}
\ee
Let us specify the dimensions here, which will be useful later. The shear modulus has the dimensions
of force per unit surface, and it is normally measured in the c.g.s. system in dyne/cm$^2$.
We conclude that the viscosity has the units of dyne sec/cm$^2$, which is a Poise.

It is interesting to note that there are some soft materials that are
neither liquids, nor solids. When $G(s)$ decays to zero slower than
$1/s$, the shear relaxation function is not integrable for infinite
times: the system is able to relax the shear for long times (in this
sense it flows, so it is not a solid), but it has an infinite
viscosity (so it is neither a liquid). From equation \eqref{jungle} we
see that the response of such systems to a constant shear rate
$\dot\gamma$ is a stress $\sigma_{xy}$ that grows indefinitely in
time.  Reversing the situation, when they are subjected to a constant
stress (as, for example, gravity) these materials do flow, although
slower and slower, that is at a shear rate that vanishes as time goes
by \cite{cates-review}.  Such systems are sometimes called `power law
fluids', since typically $G(t)\sim 1/t^a$ with $a<1$, or `dilatant
compounds'. {\it Silly putty} is a good approximation of what a dilatant
compound is.

\subsection{The diffusion coefficient and the Stokes-Einstein relation}

The last key player in the physics of viscous liquids is the diffusion coefficient $D$.
Let us consider the diffusion equation for the concentration $\rho(x,t)$ of 
a tagged particle in a liquid (for simplicity we stay in one dimension),
\be
\dot\rho(x,t) = D \frac{\partial^2\rho}{\partial x^2} \ ,
\ee
where $D$ is the diffusion coefficient. 
Let us compute the time derivative of the mean square displacement of the particle,
\be
\frac{d}{dt}\langle x^2\rangle = \frac{d}{dt}\int dx\; x^2 \; \rho(x,t)= 2D \ .
\label{msqd}
\ee
from which we derive the relation between mean square displacement and diffusion coefficient,
\be
\langle x^2\rangle = 2 D t \ .
\ee
The LHS of equation \eqref{msqd} can be calculated using directly the particle velocity $v(t)$. 
From,
\be
\langle x(t)^2\rangle = \int_0^t ds \int_0^t du \; \langle v(s) v(u) \rangle \ ,
\ee
we obtain,
\be
\frac{d}{dt}\langle x^2\rangle =2\int_0^t ds \; \langle v(s) v(0) \rangle \ ,
\ee
where we have used time-translation invariance. Therefore, in the long time limit, we have a relation
between diffusion coefficient and velocity-velocity correlation function, which is valid in any dimension,
\be
D=\int_0^t ds \; \langle v(s) v(0) \rangle \ .
\label{velocity}
\ee
The next step is to link diffusion to friction. This can be done through the Langevin equation
for a particle of mass $m$,
\be
m \dot v(t) = - \zeta v(t) + \eta(t) \ ,
\label{langevin}
\ee
where $\zeta$ is the friction coefficient and the noise $\eta(t)$ is delta-correlated in time. 
Equation \eqref{langevin} is a linear, first-order, inhomogeneous differential equation, that can 
be easily solved, giving,
\be
v(t) = e^{-\zeta t/m} v(0) + \int_0^t dt' \; e^{-\zeta(t-t')/m} \; \eta(t')/m  \ .
\label{velocita}
\ee
By computing the long-time limit of the average kinetic energy, $1/2 \; m\langle v^2\rangle$, and by 
using equipartition, we can fix
the balance between amplitude of the noise and temperature, 
\be
\langle \eta(t) \eta(t') \rangle = 2\zeta k_\mathrm B T \delta(t-t') \ ,
\label{sfdt}
\ee
which is known as the static fluctuation-dissipation theorem. This result says that the origin of the noise $\eta$ and
that of the friction $\zeta$ is the same, namely the temperature $T$ \cite{zwanzig}.
By using \eqref{velocita} and \eqref{sfdt}, we can calculate the RHS of \eqref{velocity}, and we obtain,
\be
D=\frac{k_\mathrm B T}{\zeta} \ ,
\label{eisntein}
\ee which is Einstein's relation for the diffusion coefficient. In
order to make a final link with the viscosity $\eta$, we need a
relation between friction coefficient $\zeta$ and $\eta$. This is
provided by Stokes' equation, which we will not derive here (see \cite{stoke}). A large
sphere of radius $a$ moving in a liquid with shear viscosity $\eta$,
undergoes a friction whose coefficient is given by, \be \zeta = C\, a
\, \eta \ ,
\label{stokes}
\ee
where $C$ is a constant depending on the boundary conditions for the fluid on the sphere's surface. Putting together
the last two equations we finally get the Stokes-Einstein relation between diffusion coefficient and viscosity \cite{einstein},
\be
D\eta= \frac{k_\mathrm B T}{C\, a}\ .
\label{stokeseinstein}
\ee
Although this relation strictly holds only for a diffusing sphere much larger than the molecules 
comprising the fluid, it is in fact surprisingly accurate also in 
describing the self-diffusion coefficient $D$ of a molecule surrounded by other molecules of equal size. 
Yet, the Stokes-Einstein relation sometimes fails. This typically happens when the diffusion coefficient remains
finite in situations where the viscosity is either zero (as in superfluid helium), or infinite (as in elastic
crystals). As we shall see, something similar happens in supercooled liquids close to the glass transition, when the diffusion coefficient 
$D$ decreases with decreasing temperature less steeply than $T/\eta$, so that the ration $D\eta/T$ is no 
longer a constant and the Stokes-Einstein relation is violated.

\subsection{Metastable equilibrium?}

In the physics of supercooled liquids and glasses even the terminology
is not universally accepted, and this can give rise to dangerous
ambiguities. Given that it is essential to avoid confusion from the
very beginning, let us specify the terminology used in these
notes.

I use the words `liquid', or `supercooled liquid', whenever
I talk about the {\it equilibrium}, albeit metastable, liquid phase. One may
object that a supercooled liquid is, by definition, out of
equilibrium, since below the melting temperature true thermodynamic
equilibrium is given by the crystal, while the liquid is metastable.
This is true. However, we can experimentally equilibrate a liquid in
its metastable phase, in such a way that time-translation invariance
(and thus the dynamical fluctuation dissipation theorem) holds. In
such situation, no experimental measurement is able to tell us that
the system is metastable. Only an explicit crystallization of the sample
would unveil metastability. Therefore, with a slight abuse of language, I will
call this the `equilibrium' liquid phase, supercooled or not depending on
whether we are below or above the melting temperature. As always in
first-order phase transitions, there is no way to experimentally
detect the presence of the transition temperature, as long as the system
remains equilibrated into one of the two phases.

On the other hand, whenever I will use the word `glass' it will always
be to characterize the {\it off-equilibrium} phase. By this I mean a
phase where time translation invariance is broken, so that two-times
correlation functions do not depends simply on the difference of times
and the fluctuation dissipation theorem does not hold.  Note that this
is not simply a stronger way to be out of equilibrium compared to
metastability; it is a completely different thing.  In principle it is
possible to cool so rapidly a liquid that it goes off-equilibrium even
{\it above} the melting point, where it is not metastable. In an
off-equilibrium glass the relaxation time of the substance is too
large compared to our experimental time, whereas in the equilibrium
metastable phase the crystal nucleation time is larger than our
experimental time, and yet the relaxation time is shorter than the
experimental time.  We shall discuss these two points more in details
later on.

Of course, supercooled liquids and glasses are connected: as we shall
see, the deeper the degree of supercooling of a liquid, the larger the
relaxation time, and the easier is to form a glass. For this reason,
sometimes the words `glass-formers' or `glass-forming liquids' will be
used in alternative to `supercooled liquids'.

%%%%%%%%%%%%%%%%%%%%%%%%%%%%%%%%%%%%%%%%%%%%%%%%%%%%%%%%%%%%%%%%%%%%%%%%%%%%%%%%
%%%%%%%%%%%%%%%%%%%%%%%%%%%%%%%%%%%%%%%%%%%%%%%%%%%%%%%%%%%%%%%%%%%%%%%%%%%%%%%%
%%%%%%%%%%%%%%%%%%%%%%%%%%%%%%%%%%%%%%%%%%%%%%%%%%%%%%%%%%%%%%%%%%%%%%%%%%%%%%%%
%%%%%%%%%%%%%%%%%%%%%%%%%%%%%%%%%%%%%%%%%%%%%%%%%%%%%%%%%%%%%%%%%%%%%%%%%%%%%%%%
%%%%%%%%%%%%%%%%%%%%%%%%%%%%%%%%%%%%%%%%%%%%%%%%%%%%%%%%%%%%%%%%%%%%%%%%%%%%%%%%

\section{Negotiating the crystal}

People working on supercooled liquids and the glass transition
 normally show little interest in the crystal. The reason is obvious:
 we want to study the exciting properties of glasses, not the boring
 crystalline phase. When pressed about the subtleties of
 crystallization, and the risks of disregarding it, the typical glassy
 physicist replies: ``There are systems that do {\it not} have a
 crystalline ground state. In that case, we certainly should not
 care!'', which may be true, but leave us quite anxious about all
 systems that {\it do} have a crystalline ground state.  Therefore,
 despite my full approval of supercooled liquids and glasses, I
 believe that few words must be spent on how crystallization can
 (sometimes) be avoided when cooling a substance. We often read that
 `if we are careful enough', or `if we are fast enough', we can avoid
 crystallization and enter in the supercooled - metastable - phase.
 How does this happen precisely?  Answering this question is what this
 chapter is about.

\subsection{A nucleation primer}

Liquids display a first order phase transition at the melting
temperature $T_m$. Their stable phase at low temperatures is the
crystal, which has a lower free energy density than the liquid. In
order to form a supercooled liquid we therefore have to cool our
sample below $T_m$ avoiding crystallization. A supercooled liquid is
thus a metastable phase. To learn how to skip the crystal, we have
first to learn how it forms within a supercooled liquid. This is the
aim of nucleation theory \cite{turnbull49, turnbull69}. 
In these notes I will only discuss {\it
homogeneous} nucleation, that is the formation of crystal nuclei only
due to thermal fluctuations within the bulk of the sample. In fact,
crystal formation is often assisted by the presence of impurities and
boundaries, and it is known in these cases as {\it heterogeneous}
nucleation. Before we forget about it, note that
heterogeneous nucleation can make the lifetime of a metastable sample
even shorter than what expected from the purely homogeneous theory.

Consider a sample in the liquid phase. Due to thermal fluctuations
there is a nonzero probability that some particles form a nucleus of
the ordered crystalline phase. The crystal has a lower energy
than the liquid, but also a lower entropy. As long as $T>T_m$, the
balance between energy and entropy is still in favour of the entropy
and the liquid phase has a lower free energy than the crystal. In
these conditions forming a crystal nucleus is of no advantage, and the
nucleus will rapidly melt. On the other hand, if $T<T_m$ things
change: now the crystal is thermodynamically favoured, so that the
crystal nucleus formed by thermal fluctuations has a lower free energy
than a liquid droplet with the same number of particles. However,
there is a price we must pay to form the nucleus: this is the free
energy cost due to the mismatch between the two different phases along
the interface between liquid and crystal. This cost is proportional to
the surface of the nucleus.

The total Gibbs free energy change, due to
the formation of the crystal nucleus in $d$ dimensions, is therefore
given by, 
\be \Delta G(R) = \sigma \; R^{d-1} -\delta g \; R^d \ ,
\label{nucleation}
\ee 
where we have disregarded all the dimensionless geometric
prefactors dependent on the shape of the nucleus. Equation (\ref{nucleation}) is
the the pillar of nucleation theory: $\sigma$ is the surface tension, 
i.e. the (positive) free energy cost per unit area in forming the interface 
between the two phases. On the other hand, $\delta g$
is the Gibbs free energy density difference between the metastable and
stable phase, 
\be \delta g = g_\mathrm{metastable} - g_\mathrm{stable} \geq 0
\ .  
\ee 
Note that, by definition, $\delta g =0$ at the melting
temperature $T_m$, and it grows when $T$ goes below $T_m$. In fact,
$\delta g$ is often fitted to a linear form below $T_m$, 
\be 
\delta
g(T)=\frac{\delta h}{\nu} \left( 1-\frac{T}{T_m}\right) \ ,
\label{enthalpy}
\ee
where $\delta h$ is the molar enthalpy of fusion, and $\nu$ is the molar volume of the crystal
\cite{weimberg02}. This is a handy formula, which may however break down well below $T_m$.

Equation (\ref{nucleation}) encodes the competition between the
surface cost, $\sigma$, opposing nucleus formation, and the bulk
thermodynamic drive, $\delta g$, favouring it. For small radii $R$,
the surface term dominates, so that small nuclei are thermodynamically unstable and
melt; on the other hand, for $R$ larger than a critical nucleus size
$R_c$, the volume term takes over, and the nucleus is stable. By
maximising (\ref{nucleation}) we find the critical nucleus $R_c$ and
the free energy barrier to form a stable nucleus, $\Delta G(R=R_c)$, \be
R_c= \frac{\sigma}{\delta g} \ ,
\label{critical}
\ee
\be
\Delta G(R=R_c) = \frac{\sigma^d}{\delta g^{d-1}} \ ,
\label{barrier}
\ee where once again we disregarded dimensionless geometric factors,
irrelevant for the present discussion.  From these relations we
understand that the size of the critical nucleus, and thus the
nucleation barrier, diverge at the melting point, where $\delta g=0$.
This is obvious: at $T=T_m$ the two phases have the same free energy
and there is no thermodynamic advantage in the formation of the
crystal nucleus, whereas there is still a surface disadvantage.
Therefore, the sample always needs a certain degree of supercooling
(i.e. metastability) in order to form stable nuclei, and eventually
collapse to the stable crystal phase. Note that the surface tension is
normally a function weakly dependent on the temperature close to
$T_m$, so that the largest part of the $T$ dependence of $R_c$ and
$\Delta G$ comes from $\delta g(T)$ through equation (\ref{enthalpy}).

It is very important not
to make confusion between $\Delta G$ and $\delta g$: the first one is
the free energy change caused by the nucleus formation, it depends on the
nucleus size $R$, and, once evaluated at the critical size $R_c$,
coincides with the barrier to nucleation; $\delta g$ is the
thermodynamic drive, i.e. the free energy (density)
difference between the two phases and it does not depend on $R$. In
fact, equation (\ref{barrier}) shows that these two quantities are
inversely proportional: when the temperature decreases, entering in
the supercooled phase, the thermodynamic drive to nucleation $\delta
g$ increases, and thus the barrier to crystal nucleation $\Delta
G(R_c)$ is smaller the lower the temperature. Hence, from a 
thermodynamic point of view, the lower $T$, i.e. the larger
$\delta g$, the easier is crystal nucleation.

\subsection{Nucleation time vs. nucleation rate}

After the free energy barrier to nucleation is calculated, it is
common to derive directly the nucleation time, i.e. the time needed to
form a critical nucleus, of size $R_c$.  To do this one simply invokes
the Arrhenius formula that rules activated processes, with $\Delta
G(R_c)$ playing the role of the activation barrier. In this way one
gets for the nucleation time,
\be 
\tau_\mathrm{N} = \tau_0 \ \exp\left(\frac{\Delta
G(R_c)}{k_\mathrm{B}T}\right)= \tau_0 \
\exp\left(\frac{\sigma^d}{k_\mathrm{B}T\ \delta g^{d-1}}\right) \ ,
\label{nuctime}
\ee
where $k_\mathrm{B}$ is Boltzmann's constant and $\tau_0$ is a - far from harmless - 
prefactor.

When we think about it, however, we immediately realize there is
something odd about this formula: where is the sample's volume?  The
stochastic process of nucleus formation is a local one: there is an
equal and independent probability to form a stable nucleus in each
part of the sample. It is as if we are tossing a biased coin (to nucleate or
not to nucleate) in each different position of our system: the larger
the sample, the sooner we will get a critical nucleus. This tells us
that the fundamental quantity of nucleation theory is not the
nucleation time $\tau_\mathrm{N}$, but rather the nucleation rate
$j_\mathrm{N}$, defined as the number of critical nuclei formed per
unit time, {\it per unit volume} \cite{turnbull69}.  The nucleation
rate is a constant for large enough volume $V$, and it is given by,
\be j_\mathrm{N}= j_0 \exp\left(-\frac{\sigma^d}{k_\mathrm{B}T\ \delta
g^{d-1}}\right) \ ,
\label{nucrate}
\ee If the rate per unit volume is constant, then the
time needed to form {\it one} critical nucleus in a sample of volume
$V$ scales as $1/V$, 
\be \tau_\mathrm{N} = \frac{1}{j_\mathrm{N} \; V} \ .  
\ee 
The larger the volume, the smaller the time needed to nucleate the crystal, 
and thus leave the metastable phase. Using small samples is
indeed a well-known experimental trick when trying to avoid nucleation.  
Therefore, the prefactor of the nucleation time $\tau_0$ in (\ref{nuctime}) 
must contain a term inversely proportional to the sample's volume, 
$1/V$. As we shall see, this is just the first of many tricky
ingredients of the infamous nucleation prefactor $\tau_0$.

Although it is important to understand that the nucleation time
depends on the volume of the sample, this fact does not have
momentous effects. First, the volume term $1/V$ is in the prefactor $\tau_0$, that is
{\it outside} the exponential, and it is therefore relatively weak: a
change in temperature of a few degrees close to $T_m$, where $\delta
g$ is almost linear, gives rise to an enormous change in the
nucleation time, and it would take a change in the volume of several
orders of magnitude to compensate it. Secondly, the relationship
$\tau_\mathrm{nuc} \sim 1/V$ must break down at small values of $V$,
either because surface effects become dominant over bulk properties or
simply because the sample size $L$ becomes smaller than the critical
nucleus $R_c$. In this last case taking even smaller volumes decreases
the barrier to nucleation (less particles have to be aggregated in the
nucleus) and the nucleation time decreases too.  We therefore expect
the existence of a minimal volume $v_0$ where the nucleation time is
maximum. We can then set $V=v_0$ in the equations above and
associate to a certain substance a well-defined lifetime of its
supercooled phase at a given temperature.

Note that if we do not fix a reference volume as $v_0$, we end up with the
funny result that all metastable phases have zero lifetime in the
infinite volume limit, since $\tau_\mathrm{nuc}\to 0$ for
$V\to\infty$. If we think about it, this is obvious: if we take an
infinite number of (metastable) diamonds, we have probability one that
at least one of them immediately decays into graphite, whereas most of
the other diamonds will remain stable. If all these diamonds were just
different parts of a single sample, then the immediate nucleation
event would eventually spread throughout the whole sample, making it
collapse into a (very large and very cheap) chunk of stable graphite.

Yet, it would be unwise to use such an argument to imply that no
metastability can exist in the thermodynamic limit. This for at least
two reasons. First, let us remember once again that the $1/V$ factor
in the nucleation time is outside the exponential. We meet
metastable phases with very long lifetimes every day simply because
their volume is too small to make an impact on $\tau_\mathrm{nuc}$,
even though the number of their degrees of freedom is largely beyond
the conventional $10^{23}$ needed to grant the thermodynamic
limit. Discovering that a sample larger than the universe would
effectively be unstable, is not too big a problem.

The second reason is deeper: if a stable nucleus is formed at a
certain position $x$ distant from a position $y$ in the same sample,
the nucleus must {\it grow} from $x$ to $y$, before metastability is
lost at $y$ as well. As we shall see later, the growth of a nucleus
can be slowed down significantly, sometimes even blocked. As long as
the nucleus is far enough from $y$ (beyond few correlation lengths),
any local measurement performed in $y$ will be unable to detect the
loss of metastability taking place in $x$. The underlying reason for
this is that ordinary first-order transitions (unlike second order
ones) are very local in nature, without long range correlations. So,
unless growth is very fast, we need many locally formed nuclei to
loose metastability. We will come back to this point when we will
discuss the kinetic spinodal.

\subsection{Avoiding the crystal}

To understand what is the best route to avoid crystallization we have
now to study how the nucleation time depends on the temperature. For
the time being, let us assume that the prefactor $\tau_0$ does not
depend on $T$. As we shall see this is a wrong assumption, but the
mistake is not essential at the moment. Let us also assume that the
surface tension $\sigma$ does not depend on $T$, which is a more
reasonable hypothesis, especially close to $T_m$. Under these
assumptions, the temperature dependence of the nucleation time in
(\ref{nuctime}) comes from the explicit term $T$ and from $\delta
g(T)$ in the exponential. We recall that $\delta g(T)$ is zero at the
melting point, and it grows when decreasing the temperature, without
however any reason to diverge, not even at zero temperature.  The
nucleation time thus diverges at $T=0$ and at $T=T_m$, and it is
therefore a non-monotonic function of $T$.  This implies that there is
a temperature where the nucleation time reaches a minimum (see Fig.2),
$\tau_\mathrm{min}$.  The
minimum of the nucleation curve is clearly a dangerous zone, where
crystallization is most likely.

Let us cool the system with a certain cooling rate $r$, which is the
temperature variation per unit time, \be r=\frac{dT}{dt} \ .  \ee If
we use a {\it linear} cooling rate, that is if $r(T)$ is a constant,
then we cannot cool slower than a critical value $r_c$, which is
approximately equal to the inverse of the minimum nucleation time
(Fig.2), \be r_c \sim \frac{1}{\tau_\mathrm{min}} \ .  \ee If $r <
r_c$, i.e. if we cool slower than the critical rate, we give to the
system enough time to form a stable nucleus, and thus to start
crystallization. This is why one normally says that to avoid
crystallization we have to cool `fast enough'. Note that cooling
fast a sample may be very nontrivial from the experimental point of
view, so that avoiding crystallization may be not that easy,
especially when $\tau_\mathrm{min}$ is very small and thus $r_c$ very
large.

\begin{figure}
\includegraphics[clip,width=3.4in]{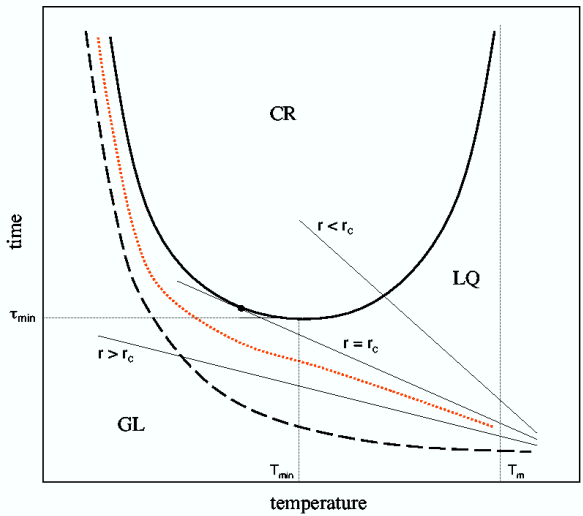}
\caption{ {\bf Nucleation and relaxation time vs. temperature} - The
nucleation time (full line) diverges at the melting point $T_m$ and at
$T=0$, reaching a minimum at $T_\mathrm{min}$.  The relaxation time
(dashed line) is very small close to $T_m$ and it raises sharply as
the temperature lowers. The straight lines represent linear cooling
schedules at different cooling rates.  The slope of these lines is the
inverse of the cooling rate $r$.  In order to keep the liquid (LQ) at
equilibrium we cannot cool too slowly, otherwise we cross the
nucleation time and we get a crystal (CR).  The line that is tangent
to the nucleation time represents the critical linear cooling rate,
i.e. the slowest rate at which we can cool avoiding crystallization.
However, we cannot cool too rapidly either, otherwise we cross the
relaxation time and we go out of equilibrium, obtaining a glass (GL).
The red dotted line represents a nonlinear cooling that is fast close
to the minimum of the nucleation time and slower at lower
temperatures.  }
\end{figure}

If we are not limited to linear cooling schedule, we can do something
smarter: we can cool slowly far from the minimum $T_\mathrm{min}$, and
rapidly just around the minimum.  This nonlinear strategy is indeed
necessary if we wish to keep the sample in the supercooled phase and
not to fall off-equilibrium. As we shall later on in these notes, by
lowering the temperature the relaxation time $\tau_\mathrm R$ of the
supercooled liquid increases dramatically. This means that, to keep
the system at equilibrium, we need to cool it slower and slower, while
lowering the temperature. Such requirement clearly conflicts with the
need of a fast cooling to avoid crystallization. If we stick to a
linear cooling schedule, with $r>r_c$ in order to avoid crystal
nucleation, the sample will inevitably fall out of equilibrium as soon
as the relaxation time exceeds $r_c^{-1}$. On the other hand, if we
fix $r< r_c$ to stay at equilibrium, the sample will crystallize.  We
conclude that there is no way to stay in the supercooled phase at
arbitrary low temperatures by means of a linear cooling schedule: the
sample either drops off-equilibrium, becoming a glass, or it
crystallizes.

If, on the other hand, we  cool nonlinearly, we can cool fast at
higher temperatures, until the minimum of the nucleation curve is
negotiated, and slow down progressively at lower
temperatures to cope with the increasing relaxation time of the
supercooled liquid (Fig.2).

The situation seems therefore under control: we must identify the
minimum of the nucleation curve and negotiate it by using a smart
cooling schedule. In this way we can forget forever about the
crystal. Unfortunately, the solution of the problem is not that easy. First,
$\tau_\mathrm{min}$ and $\tau_\mathrm R$ may be such to make it
experimentally impossible, for a reason or another, to skip
crystallization, yet keeping the system at equilibrium. Second,
measuring the nucleation time, and thus $\tau_\mathrm{min}$, is far
from trivial; as a consequence, there is the concrete possibility that
we do not know the nucleation curve at all. Third, there are cases
where, no matter what we do and know, the metastable liquid is
doomed. This last point seems particularly disturbing, and we
will focus on it.

\subsection{The kinetic spinodal}

The experimental protocol we use to cool our sample is {\it not} the
only factor affecting how deeply we can supercool a liquid. There are
systems that crystallize below a certain temperature,
no matter the cooling schedule we adopt. This temperature is
known as kinetic spinodal, $T_\mathrm{sp}$. The kinetic spinodal is
the metastability limit of the supercooled phase: below this point the
equilibrium supercooled liquid ceases to exist. What is the origin of 
$T_\mathrm{sp}$?

As we have already seen, a supercooled liquid is squeezed in an
uncomfortable time region: if we are too slow when cooling below
$T_m$, the system has enough time to nucleate the crystal; on the
other hand, if we are too fast, the system cannot thermalize, and an
off-equilibrium glass is formed. Thus, to reach the maximum degree of
supercooling we have to keep the sample between the relaxation and the
nucleation curves, a region which narrows at lower temperatures. Yet,
if the qualitative behaviour of relaxation time, $\tau_\mathrm{R}(T)$
vs. nucleation time $\tau_\mathrm{N}(T)$, were always the one depicted
in Fig.2, there would always be a way (at least theoretically) to
supercool the system at arbitrary low temperature. However, this is not
always the case.

\begin{figure}
\includegraphics[clip,width=3.4in]{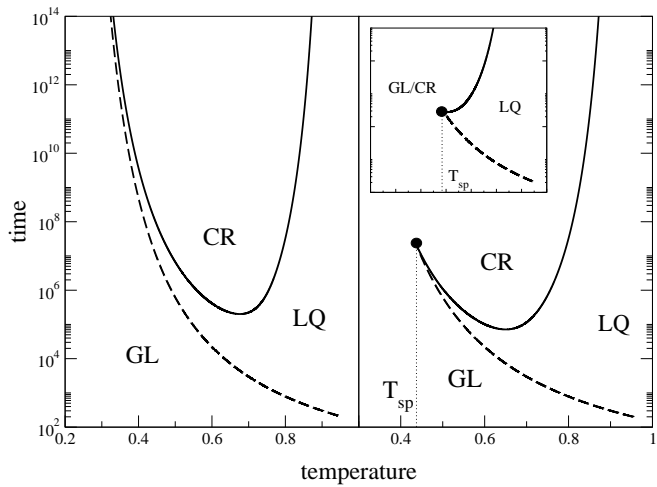}
\caption{ {\bf The kinetic spinodal} - Left panel: If
nucleation time (full line) and relaxation time (dashed line) do not
cross, it is possible to equilibrate the liquid with a nonlinear
cooling at arbitrary low $T$.  Right panel: However, if the relaxation
time becomes larger than the nucleation time (right panel), the system
has a kinetic spinodal $T_\mathrm{sp}$. Below this temperature the
supercooled liquid does not exist, because nucleation starts before
equilibrium can be established. The curves can cross in two different
ways. In the inset of the right panel we see an easy-to-detect
spinodal: the nucleation time is decreasing when it crosses with the
relaxation time. In this case, even a naive extrapolation of the
nucleation and relaxation data clearly indicates that a kinetic
spinodal is incumbent. In the main panel, on the other hand, we see a
sneaky spinodal: both times are increasing in this case. If we only
have data down to $T>T_\mathrm{sp}$, an extrapolation can hardly say
something about the possible existence of $T_\mathrm{sp}$. For this
reason it is in general very difficult to rule out the existence of a
kinetic spinodal. Close to the spinodal the difference between
polycrystal and glass is somewhat blurred \cite{ctls1}.The symbols
CR, LQ and GL indicate respectively the crystalline, the liquid and
the glassy phase. (Reprinted with permission from \cite{attanasi1}; 
copyright of American Physical Society.).
}
\end{figure}

In some materials the relaxation time may exceed the nucleation time
below a certain temperature, as shown in Fig.3 (right panel).  This
temperature is the {\it kinetic spinodal}, which is thus defined by the
relation, \be \tau_\mathrm{N}(T_\mathrm{sp})=
\tau_\mathrm{R}(T_\mathrm{sp}) \ .  \ee If a kinetic spinodal exists,
no equilibrium measurements can be performed on the supercooled sample
below $T_\mathrm{sp}$, because $\tau_\mathrm{N}(T_\mathrm{sp})<
\tau_\mathrm{R}(T_\mathrm{sp})$, so that crystallization starts before
equilibrium is reached in the liquid phase.

The kinetic spinodal therefore marks the metastability limit of the supercooled phase. 
Fig.4 shows the typical time dependence of a physical quantity (in this
case the energy) above and below the kinetic spinodal. Note that 
even above $T_\mathrm{sp}$ there is crystallization if we wait
long enough; however, in this temperature range there is a long
time window between $\tau_\mathrm{R}(T)$ and $\tau_\mathrm{N}(T)$
where equilibrium properties of the supercooled liquid 
phase can be measured. Below $T_\mathrm{sp}$, on
the other hand, this is not possible: there is an overlap of
relaxation and nucleation regimes that makes it impossible to define an
equilibrium phase.

\begin{figure}
\includegraphics[clip,width=3.4in]{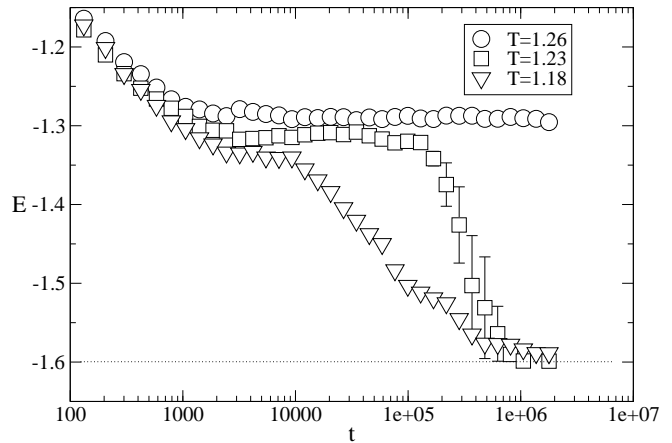}
\caption{ {\bf The time dependence of the energy above and
below the kinetic spinodal} - Numerical simulations of a system with
a liquid-crystal first order transition.  Energy density vs. time at
various temperatures above and below the kinetic spinodal.  Circles:
at this temperature $T>T_\mathrm{sp}$ nucleation time is so large that
crystallization cannot be observed within the maximum experimental
time. Squares: at lower temperature, but still with $T>T_\mathrm{sp}$,
nucleation can be observed within the experimental time
window. However, the energy relaxes to its equilibrium value well
before nucleation occurs, so that the metastable liquid is well
defined. Triangles: for $T< T_\mathrm{sp}$ the two processes of
relaxation and nucleation are no longer well separated, so that we
cannot define the metastable liquid phase in this temperature regime.
(Reprinted with permission from \cite{ctls1}).  }
\end{figure}

It was Kauzmann in 1969 \cite{kauzmann} who first hypothesized the existence 
of a kinetic spinodal in supercooled liquids. He wrote:
``{\it Suppose that when the temperature is lowered a point is
eventually reached at which the free energy barrier to crystal
nucleation becomes reduced to the same height as the barriers to the
simpler motions. At such temperatures the liquid would be expected to
crystallize just as rapidly as it changed its typically liquid
structure to conform to a temperature or pressure change in its
surroundings.}''
Kauzmann introduced the concept 
of kinetic spinodal as an effective
way to avoid the entropy crisis, defined as the point where the
(extrapolated) entropy of the liquid becomes equal to the entropy 
of the solid. We shall discuss this phenomenon in great detail 
later on in these notes.

Let us note that $T_\mathrm{sp}$ is called {\it kinetic} spinodal, and
not simply spinodal, for a reason. The barrier to nucleation is
nonzero at $T_\mathrm{sp}$, and no {\it thermodynamic} loss of
stability occurs here. Despite this fact, the equilibrium phase cannot
be defined any better below $T_\mathrm{sp}$ than if there were a {\it
bona fide} thermodynamic spinodal. Indeed, it is important to
understand that the kinetic spinodal does not depend on the
experimental protocol, but it is rather an intrinsic property of the
sample. Below $T_\mathrm{sp}$ the only equilibrium phase is the
crystal, whereas off-equilibrium glassy and polycrystalline
configurations can be obtained if we cool fast or slow enough,
respectively.

\subsection{Can the prefactor of nucleation save the day?}

Whether or not a particular liquid has a kinetic spinodal that bounds
its region of metastability depends on the material characteristics.
In particular, as we have seen, it depends on the relative $T$-dependence 
of nucleation and relaxation time.
Clearly, those systems whose relaxation time increases more smoothly
when lowering the temperature have a lower danger to present a kinetic
spinodal, whereas a steeply increasing relaxation time makes it more
likely the crossing with the nucleation time, and thus a metastability
limit. In particular, in those systems where relaxation time seems to
become infinite only at zero temperature, there is no {\it a priori}
reason to postulate the existence of a kinetic spinodal: both
$\tau_\mathrm{R}(T)$ and $\tau_\mathrm{N}(T)$ diverge at $T=0$, and
whether they cross or not definitely depends on the details of the
material.

There are systems, however, where the relaxation time increase much
more steeply, and it seems to become {\it very} large well above
$T=0$.  In particular, as we shall see later on, some theories predict
the existence of a temperature $T_k>0$ where the relaxation time
diverges.  If this is true, it seems that $\tau_\mathrm{R}$ is bound
to become larger than the nucleation time $\tau_\mathrm{N}$, which, as
we have seen, only diverges at $T=0$. So, the very existence of a
divergence of the relaxation time at $T_k$ would automatically imply
the existence of a kinetic spinodal, an this in turn would prove that
$T_k$ does not exist! In order to exit from this nasty loop
there is a standard argument, which we report here.

Up to know we have assumed that relaxation and nucleation are
completely independent phenomena. In fact, this is not 
true. The prefactor $\tau_0$ in the nucleation time (\ref{nuctime})
depends on the temperature, although this dependence is quite
complicated and somewhat unclear.  
It has become
customary to approximate the $T$ dependence of $\tau_0$ by using the
self-diffusion coefficient $D$, thus writing \cite{turnbull49, turnbull69},
\be 
\tau_0 \sim 1/D \ .
\label{trick1}
\ee 
Although this equation is definitely {\it not} exact, its meaning is rather 
clear: nucleation proceeds through an aggregation process, and it is
reasonable that the rate of this process is proportional to the
diffusion in the system. By using the Stokes-Einstein relation \eqref{stokeseinstein}, 
one then has, 
\be \tau_0 \sim \eta \sim \tau_\mathrm{R} \ ,
\label{trick2}
\ee 
where we have used the linear relation between shear viscosity and 
relaxation time of normal liquids \eqref{magnapompa}.
From equation \eqref{nuctime} we get,
\be
\tau_\mathrm{N} = \tau_0 \
\exp\left(\frac{\Delta G}{k_\mathrm{B}T}\right)
\sim \tau_\mathrm{R}\exp\left(\frac{\Delta G}{k_\mathrm{B}T}\right)  \ .
\label{beelo}
\ee
This equation implies that the nucleation time is always
{\it larger} than the relaxation time. If this were true,  a kinetic spinodal
would  never exist: nucleation would always be slower than 
relaxation, so that there would always exist a time window where
equilibrium properties could be measured. This is the underlying argument when it
is sometimes said that once a liquid enters in its deeply supercooled
phase, then crystallization is no longer an issue, since it becomes
too difficult. The idea is that when relaxation is too slow, and viscosity
is very high, nucleation, and thus crystallization, is kinetically inhibited.

By killing the kinetic spinodal $T_\mathrm{sp} > T_k$, this argument
saves the existence of the transition at $T_k$. If we accept equation
(\ref{beelo}), then not only the relaxation time, but also the
nucleation time diverges at $T_k$: the two times do not cross, the
kinetic spinodal is avoided and the Kauzmann transition is saved.  Is
this argument satisfying?

Although the prefactor of nucleation definitely increases
by decreasing the temperature, it is hard to claim that this fact 
can completely rule out the existence of a kinetic spinodal.  First, 
there are systems where a kinetic spinodal exist (see \cite{ctls1}, for example),
so that this argument does not work in general.
Secondly, even if equation
(\ref{beelo}) strictly held, it would not be sufficient to grant the
survival of the liquid in a condition where the barrier to nucleation
is very small. In fact, in order to measure the equilibrium properties
of the liquid, we need the nucleation time to be {\it significantly}
larger than the relaxation time. This means that there must be some
orders of magnitudes of difference between the two times, otherwise
the metastable liquid phase is only virtually, and not operationally,
existent. If the exponential is of order $1$, the two times are comparable,
and equilibrium is lost.
Third, in the deeply supercooled phase, where the liquid
becomes very viscous, the Stokes-Einstein relation is known to break
down \cite{rossler-90, cicerone-96}, and typically, 
\be 
1/D \ll \frac{\tau_\mathrm{R}}{T} \ .
\ee 
Even in the extreme case of solids, where the shear viscosity (and thus
$\tau_\mathrm{R}$) is infinite, the diffusion coefficient $D$ remains
nonzero. Nucleation is a local aggregation process that may
(slowly) proceed even when more complex relaxation processes are
completely blocked. Indeed, as we shall see later on, the rearrangements that
grants the ergodicity, and thus true relaxation, to the system are of
cooperative nature at low $T$, whereas nucleation proceeds even at
a single particle level. Thus, the nucleation prefactor remains much
smaller than the relaxation time at low temperatures, and it cannot
grant by itself the condition $\tau_\mathrm{N} \gg \tau_\mathrm{R}$
necessary to skip the kinetic spinodal and guarantee the existence
of the equilibrium liquid. We conclude that, though the prefactor $\tau_0$
can actually depress nucleation in the not-too-cold regime where 
$1/D\sim \tau_\mathrm{R}$, it does not by itself prevent nucleation 
at low $T$. Hence, the prefactor of nucleation cannot save the day.

\subsection{An elastic twist}

There is a  mechanism that can inhibit nucleation in supercooled 
liquids, in some cases so much as to suppress the kinetic spinodal. 
To understand what such mechanism is we have to make a small detour.

Nucleation within a metastable phase is a very
general phenomenon, certainly not limited to liquid-crystal
transitions. 
In particular, when both phases (the stable and the metastable
one) are solids, equation (\ref{nucleation}) is incomplete.
The two solid phases typically
have different specific volumes; to fix ideas let us assume that the stable
phase has a larger volume than the metastable one. This implies that when
a nucleus of the stable phase is formed, the metastable background,
and thus its solid matrix, will be subjected to an elastic strain, due
to the volume misfit. Basically, it is as if we carved a volume
$V_\mathrm M$ from the metastable phase, and tried to replace it with a larger
volume $V_\mathrm S > V_\mathrm M$. Intuition tells us
that we are going to pay something more than the simple surface
mismatch contribution.

Indeed, when the system has an {\it elastic}
response, as in the case of solids, there is an extra energetic price that has to be
paid. Because of the long-range nature of elastic forces, the elastic
price is proportional to the {\it volume} of the nucleus, and it thus
corrects the bare thermodynamic drive $\delta g$ \cite{roi78, cah84, bus03}. 
The correct formula becomes, 
\be 
\Delta G(R) = \sigma \; R^{d-1} - \delta g \; R^d +
E_\mathrm{elastic}\; R^d \ .
\label{elastonuc}
\ee
The elastic term is proportional to the square of the relative 
volume difference between the two phases, $\delta v/v$, and to the
elastic shear modulus, $G_\infty$ \cite{elastoreview},
\be
E_\mathrm{elastic} = k\; \left(\frac{\delta v}{v}\right)^2\; G_\infty \ ,
\ee
where $k$ is a dimensionless constant containing a combination of the bulk and shear moduli.
From (\ref{elastonuc}) we clearly see that as long as $\delta g > E_\mathrm{elastic}$ nucleation
proceeds in the usual way, although it is delayed by an elastic correction to the free energy drive,
\be
\tau_\mathrm{nuc} = 
\tau_0 \ \exp\left(\frac{\sigma^d}{k_\mathrm{B}T\ [ \delta g  -   E_\mathrm{elastic}]^{d-1}}\right) \ .
\label{nuctime_ela}
\ee
Therefore, the effect of the elastic contribution when $\delta g > E_\mathrm{elastic}$, 
is simply to increase the nucleation time.
On the other hand, when 
$\delta g < E_\mathrm{elastic}$, there
is no volume advantage in forming the nucleus, since the free energy gain is 
smaller than the elastic cost. In such situation nucleation is  totally
suppressed. In particular, if $\delta g < E_\mathrm{elastic}$ at all temperatures,
the only way for the system to collapse in the stable phase is 
to perform a so-called zeroth-order phase transition, where the entire sample jumps to the
stable phase, in order not to create any elastic strain \cite{bus03}.

As thrilling as all this may sound, the reader may nevertheless ask: Who
cares?  What this has to do with the liquid-crystal transition?
Liquids are not solids.  Why, then, should we bother about elastic
effects?  The answer is that at low temperatures, viscous liquids
{\it do have} a nonzero elastic response on time scales that may be
relevant for nucleation.  As we have seen in Section II.A, a
liquid is described by a  time-dependent shear modulus $G(t)$ that decays 
to zero only for times significantly larger than the relaxation time (equations
(\ref{modulus}) and (\ref{maxwell})). For $t\gg \tau_\mathrm R$ 
the liquid is able to relax the stress, $G(t)\sim 0$, and there is
no elastic contribution, while for $t\ll\tau_\mathrm R$ the liquid responds 
like a solid, with finite shear modulus, $G(t)\sim G_\infty$.

At high temperatures, close to $T_m$, nucleation is slow and
relaxation is fast, so that $\tau_\mathrm N \gg \tau_\mathrm R$.  In
this regime, thus, $G(\tau_\mathrm N)\sim 0$ and there is no elastic
contribution.  On the other hand, at low temperatures nucleation
occurs on time scales that may be comparable to the relaxation time (see Fig.2),
$\tau_\mathrm N \sim \tau_\mathrm R$ and thus $G(\tau_\mathrm N)\neq
0$
\footnote{In fact, the correct time scale to be plugged into the time-dependent shear 
modulus $G(t)$ is the time for the {\it formation} of the nucleus, $t_f$, which is
considerably smaller than the nucleation time, $\tau_\mathrm N$. In fact, 
this last quantity is roughly speaking the time between two nucleation events: during
much of this time nothing happens; while $t_f$ is the time needed
to actually build the nucleus once the process has started. Because
$t_f \ll \tau_\mathrm N$, we have $G(t_f)\gg G(\tau_\mathrm N)$ and the 
elastic argument holds even stronger. See \cite{attanasi2}.}. 
In this case when a nucleus is formed the background liquid
undergoes an elastic strain due to the volume mismatch of the newly
formed crystal nucleus (the specific volumes of the liquid and crystal
are typically different). As a result, an elastic price, proportional
to the nucleus' volume, must be paid \cite{zonko02, zonko03}.

The elastic response depresses the thermodynamic drive to nucleation
and increases the nucleation time (equations \eqref{elastonuc} and
\eqref{nuctime_ela}), therefore $G(\tau_\mathrm N)$ decreases and the
elastic response is in turn depressed. Therefore, finding
$\tau_\mathrm N$ becomes a self-consistent problem.  It can be proved
\cite{attanasi1, attanasi2} that the final effect of this
self-consistent elastic mechanism is encoded in the following
dimensionless parameter, \be \lambda = \left(\frac{\delta
v}{v}\right)^2 \frac{G_\infty}{\delta h/\nu} \ , \ee where we remind
that $\delta h/\nu$ is the molar enthalpy of fusion per molar volume
(see equation (\ref{enthalpy})), which is the drive to nucleation. The
ratio $G_\infty/\delta h$ thus encapsulates the competition between
elastic loss and thermodynamic gain in forming the nucleus.  In those
liquids where $\lambda$ is small, elasticity is weak compared to the
thermodynamic drive to nucleation, so that nucleation is only
moderately depressed, and a kinetic spinodal may still exist. On the
other hand, when the parameter $\lambda$ is large, elastic effects
become so strong that the kinetic spinodal can be completely
suppressed \cite{attanasi1, attanasi2}.

We conclude that some supercooled liquids, those with a large enough
elastic response, may survive in their metastable phase even when the
relaxation time becomes very large, because the kinetic spinodal is
suppressed by elastic effects. Of course, the absence of a kinetic
spinodal does not mean that these systems cannot crystallize: we still
have to be very careful when cooling. However, knowing that, at least
in principle, there are systems where the metastable phase is
well-defined, no matter how low is $T$, is somewhat a consolation.

\subsection{Nucleation vs. growth}

The curve of the nucleation time vs. temperature is
quite instructive from a theoretical point of view, but it is not of
great help experimentally. What we can measure in an experiment is
crystallization, rather than nucleation.  In particular, we can
measure the time $t_x$ the system takes to develop a substantial
amount of crystalline order. This amount must be large enough to be
detected experimentally and see that crystallization has started at 
that particular temperature $T$.  The function $t_x(T)$ obtained in this way
is often called Time-Temperature-Transformation (TTT) curve, and it is
the only experimentally available tool to check how likely is crystal
formation at a given temperature \cite{uhlmann72}. The TTT curve, as the nucleation
curve, displays a minimum, indicating that at this point
crystallization is most likely.  In the experimental context the TTT
curve is normally plotted in a temperature vs. time representation, so
that this minimum looks in fact like the 'nose' of the curve, and this
is the word normally used for it.

One may argue that the TTT curve is just the experimental counterpart
of the nucleation curve: they both tell us how likely crystallization
is, they both have a minimum, they both diverge at $T_m$ and $T=0$. In
fact, up to know we made little difference between nucleation and
crystallization.  However, nucleation is a different phenomenon from
crystallization, and the two curves have a fundamentally different
meaning.

In order to form a substantial amount of crystal, which can
be detected experimentally, the system has to first form some stable
nuclei and then to grow them until they invade a large  part of
the sample. The first process is {\it nucleation}, the second is {\it
growth}.  These two processes are different, and crystallization is
given by the concourse of the two of them. Nucleation is dominated by
the competition between surface tension and free energy
(potentially renormalized by an elastic factor). 
Growth, on the other hand, is largely
dominated by the viscosity of the background liquid in which the
nucleus must expand. Moreover, growth may be slowed down because
different mismatched crystallites come in contact with each other,
giving rise to a very slow process of crystal domain growth (or
coarsening) \cite{bray}. Therefore, growth is ruled mostly 
by relaxation mechanisms. As a result, crystallization proceeds at the pace
of the slowest process, nucleation or growth.

Close to the melting point the viscosity of the liquid is very small,
whereas nucleation time is very large. Therefore, it is
always nucleation to slow down crystallization near $T_m$ and to make
the TTT curve diverge here.  Moreover, if there is no kinetic spinodal, i.e.
if nucleation is always significantly slower than relaxation, then as
slow as growth may be, nucleus formation is typically even slower, and
thus nucleation is the real bottleneck of crystallization. In this case
the TTT curve gives the same qualitative physical information as the 
nucleation curve (Fig.5 - left panel).

\begin{figure}
\includegraphics[clip,width=3.4in]{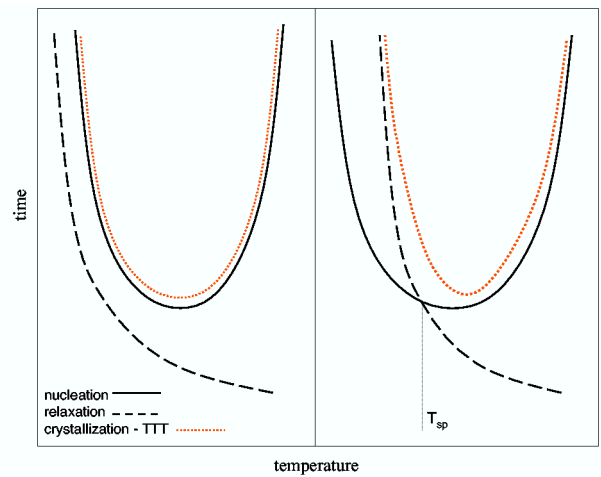}
\caption{ {\bf Crystallization curve vs. nucleation
 curve} - Left panel: when there is no spinodal, the bottleneck of
 crystallization is nucleation at all temperatures, so that the
 TTT-crystallization curve is basically the same as the nucleation
 curve. Right panel: when a spinodal is present, the right branch of
 the TTT curve is still given by nucleation, while the right branch is
 dominated by slow growth. In this case nucleation and crystallization
 (TTT) curves have different meanings.  }
\end{figure}

However, there is one important case where nucleation is fast, 
and yet crystallization is slow, being hampered by slow growth. 
This happens in presence of a kinetic spinodal. In this case we have to be 
very careful in the interpretation of the TTT curve (Fig.5 - right panel). 
Nucleation and relaxation times intersect at the spinodal point 
$T_\mathrm{sp}$. Below $T_\mathrm{sp}$
nucleation is fast, whereas growth is slowed down by the large
relaxation time. This means that while the right branch of the TTT
curve is still dominated by nucleation, the left branch, the one for 
$T<T_\mathrm{sp}$, is dominated by growth. Thus, a metastability limit 
makes the meaning of the crystallization (TTT) and nucleation curves quite different
\cite{ctls1,ctls2}.

Strictly speaking, below the kinetic spinodal the equilibrium
supercooled liquid does not exist, and therefore it is formally not possible to
speak about the liquid relaxation time. Therefore, when we say that
below $T_\mathrm{sp}$ the growth of a nucleus is slowed down by the
large relaxation time, we are quite sloppy, although the statement is
reasonable.  We mean that the local relaxation of the liquid
background where the nucleus is expanding is comparable to the
extrapolation of the relaxation time of the liquid below the spinodal.

The final message of this section is the following: when trying to
avoid crystallization, we must make the best use of the TTT curve,
which is the only experimentally available tool; therefore, we have to
cool rapidly around the nose, and cross our fingers. However, there is
no way we can detect the existence of a kinetic spinodal by simply
inspecting the TTT curve, as it looks always the same (a
divergence at $T_m$ plus a nose), both when there is and when there is
not a metastability limit. The difference is that 
when a kinetic spinodal is present the left branch 
of the TTT curve is ruled by growth, rather than
nucleation (Fig.5); but this is something
the TTT curve itself is not going to tell us.  As a consequence, we
have always to be very careful in ruling out the possibility that our sample 
has crystallized. Even after we have negotiated the nose of the TTT curve,
there is always the risk that nucleation starts, but growth is so slow
that it keeps the crystallites' size below the experimentally detectable 
threshold. In that case our sample is a sneaky polycrystal, i.e. 
an off-equilibrium system: however slow growth may be, it is bound to 
drive away the system from its supercooled equilibrium condition.
Therefore, when a spinodal is present (Fig.5 - right panel), we can only study the supercooled 
liquid in the regime of equilibrium arrested nucleation (right branch of the TTT 
curve), not in the regime of off-equilibrium arrested growth (left branch of the TTT curve) 
\cite{ctls1,ctls2}.

Even though off-equilibrium glasses are not our main concern in these notes,
it is worthwhile saying that such off-equilibrium polycrystals, as those obtained 
from a widespread nucleation arrested by a very slow growth, raise some conceptual
issues about the nature of the glassy phase. The question is whether or not it is
always possible to identify an order parameter that clearly distinguish these sneaky
polycrystals from {\it bona fide} amorphous (i.e. glassy) configurations.
In general, it is not at all easy to clear up the difference between the two: there are 
systems where it exists a grey zone where it proves very hard to {\it qualitatively} 
distinguish a highly disperse polycrystal from a truly amorphous glass 
\cite{ctls1, ctls2}. This said, for the rest of these notes we will make the simplifying assumption
that {\it bona fide} amorphous minima, qualitatively well separated from the crystal, 
can always be defined.

%%%%%%%%%%%%%%%%%%%%%%%%%%%%%%%%%%%%%%%%%%%%%%%%%%%%%%%%%%%%%%%%%%%%%%%%%%%%%%
%%%%%%%%%%%%%%%%%%%%%%%%%%%%%%%%%%%%%%%%%%%%%%%%%%%%%%%%%%%%%%%%%%%%%%%%%%%%%%
%%%%%%%%%%%%%%%%%%%%%%%%%%%%%%%%%%%%%%%%%%%%%%%%%%%%%%%%%%%%%%%%%%%%%%%%%%%%%%
%%%%%%%%%%%%%%%%%%%%%%%%%%%%%%%%%%%%%%%%%%%%%%%%%%%%%%%%%%%%%%%%%%%%%%%%%%%%%%
%%%%%%%%%%%%%%%%%%%%%%%%%%%%%%%%%%%%%%%%%%%%%%%%%%%%%%%%%%%%%%%%%%%%%%%%%%%%%%
%%%%%%%%%%%%%%%%%%%%%%%%%%%%%%%%%%%%%%%%%%%%%%%%%%%%%%%%%%%%%%%%%%%%%%%%%%%%%%
%%%%%%%%%%%%%%%%%%%%%%%%%%%%%%%%%%%%%%%%%%%%%%%%%%%%%%%%%%%%%%%%%%%%%%%%%%%%%%

\section{The glass transition and thereabouts}

The glass transition is one of the most interesting open problems 
in condensed matter physics. There is first of all an issue of conceptual definition,
since the glass transition is not a 'transition' at all. But more importantly,
there is the problem of whether or not a liquid close to the glass
transition is in any respect qualitatively different from a normal liquid.
In other words: does the glass transition mark a fundamental change in 
the physical properties of the liquid, or is it a mere conventional 
point? I will try to convince the reader that a supercooled liquid close to 
the glass transition is indeed different from a high temperature liquid at a
qualitative level, and not simply because its relaxation time is much larger
than normal. 

Throughout these notes I will always use temperature, rather than density,
as the control parameter triggering glassy behaviour. For a thorough comparison
of the lowering $T$ vs. increasing $\rho$ effects on glass-forming liquids see \cite{tarjus-0,tarjus-rhot}.
Excellent brief accounts of the glass transition and related phenomena are given in \cite{angell88} and \cite{kob-review}.

\begin{figure}
\includegraphics[clip,width=3.5in]{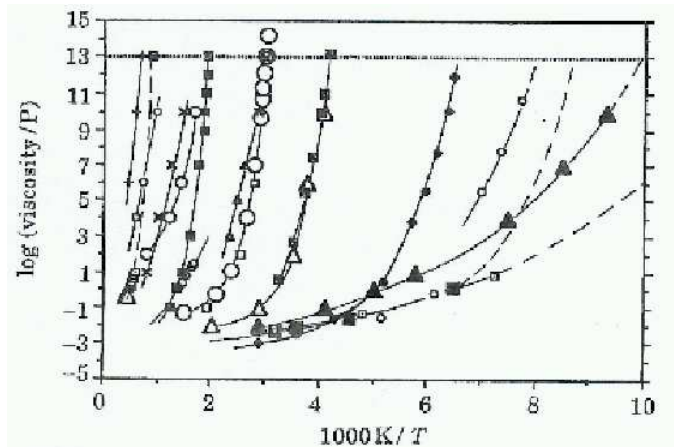}
\caption{ {\bf The growth of the viscosity close to the glass
transition} - Logarithm of the viscosity of several substances as a
function of the inverse temperature.  The horizontal line marks the
value $\eta=10^{13}$ Poise, which conventionally defines the dynamic
glass transition.  (Reprinted with permission from
\cite{angell-poole}; copyright of Societa' Italiana di Fisica.)  }
\end{figure}

\subsection{Going off-equilibrium: the dynamic glass transition}

Once we have learned how to defend our supercooled liquid sample from
crystallization, we turn to face another enemy, the glass. When we
cool a liquid significantly below the melting point, the shear
relaxation time and the viscosity grow sharply. This growth
is very impressive and very general (see Fig.6). Data 
show that viscosity  can increase by up to $14$ decades in a relatively
narrow range of temperature  \cite{angell-poole}. This sharpness is very different
from what happens above the melting point, where the viscosity 
increases in a much smoother way. This dramatic growth is common 
to many liquids with very different microscopic structures, 
including polymers.

By decreasing $T$ we are therefore bound to hit a temperature where
the relaxation time exceeds the time experimentally available to our
measurement.  Below such temperature it is impossible to equilibrate
the system: the sample is out of equilibrium on the time scale of our
experiment, and we have formed a {\it glass}. We give a provisional
definition of {\it dynamic glass transition} (or kinetic glass
transition, or simply glass transition) as the temperature $T_g$ where
the relaxation time exceeds the experimental time, \be \tau_\mathrm R
(T< T_g) > t_\mathrm{exp} \ .
\label{tglass}
\ee
What happens from the experimental point of view at the glass transition?
The most important practical consequence of going off equilibrium is that
we do not give the system enough time to properly explore the phase space, and,
in so doing, we sharply cut the number of degrees of freedom accessible to the system.
This causes a sharp drop (up to a factor $2$) of the constant pressure specific heat $c_p$
at $T_g$ \cite{angell88}. A schematic view of the typical behaviour of $c_p(T)$ is reported in Fig.7. 
The experimental time is smaller than the ergodicity time, i.e. the time needed by
the system to explore a representative fraction of the phase space. In this dynamic sense, we can say that
the system is no longer ergodic.

\begin{figure}
\includegraphics[clip,width=3.7in]{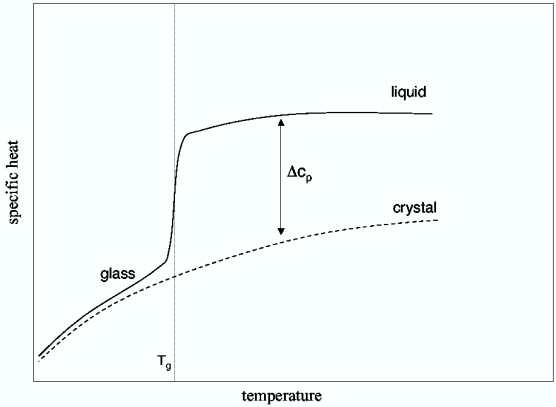}
\caption{ 
{\bf Specific heat vs. temperature at the dynamic glass transition} - 
The specific heat drops at the dynamic glass transition to approximately the same value it has in the crystal phase.
This is because below $T_g$ we are not giving the system enough time to be ergodic. Roughly speaking,
a glass is stuck in a single potential energy minimum for a long time, so that it looses all the configurational 
degrees of freedom.
}
\end{figure}

This phenomenon becomes all the more clear when we notice that
the specific heat below $T_g$ drops to a value very close to that of the 
crystalline phase. In a crystal the motion of all particles consists of
vibrations around their (ordered) equilibrium positions, without any kind of rearrangement.
Ergodicity is broken and the system is confined to one (absolute) energy minimum in the 
phase space. The behaviour of the specific heat thus suggests that also in 
a glass at low temperature particles vibrate around their (disordered) 
equilibrium positions, with almost no structural rearrangement. Ergodicity is dynamically
broken and the glass is confined to one (local) energy minimum in the 
phase space. For this reason, the specific heat is approximately the same in the crystal 
as in the low temperature glass. 

Even though the view of a glass stuck in a local minimum is good enough to
understand the behaviour of static quantities as the specific heat, it is unfortunately 
far too simplistic if we want to understand the dynamic properties of the off-equilibrium phase.
This is not our focus, but we nevertheless must be a bit more precise here.
A glass is something  more complicated than a system vibrating around an amorphous 
minimum of the energy. Were this simple picture true, the glass would be at a 
(broken-ergodicity) equilibrium within this minimum,
as it happens to the crystal. However, a glass is drastically {\it out} of equilibrium.
Even though one-time quantities (as the volume or the energy) may look almost constant 
in the long time limit, two-time quantities (as the dynamic correlation function) show a 
stark off-equilibrium behaviour, in that they depend explicitly on both times, rather 
than on their difference. In other words, the properties of the system depend on the time
elapsed from the instant the system was cooled below $T_g$. This is aging. The reasons for this
behaviour are complex and go beyond the scope of these notes.
Very nice introductions on aging and 
off-equilibrium dynamics can be found in \cite{leticia-review} and \cite{biroli-aging}.
Here, let us be content in knowing that on short time-scales the most conspicuous effect
of going off-equilibrium below $T_g$ is indeed to remain stuck in an amorphous 
configuration, where only vibrations contribute to the specific heat, and for this
reason $c_p$ has a drop.

There is, of course, a crucial difference between an off-equilibrium glass and a crystal:
ergodicity breaking in a glass is a purely dynamical accident, whereas in a crystal 
is a truly thermodynamic phenomenon. On the time scale
over which we observe a glass, ergodicity cannot be restored and equilibrium is lost.
In such a situation, in principle, we should not measure the specific heat at all, which
is an equilibrium concept. If we stubbornly do that anyway, we inevitably record a loss
in the number of degrees of freedom that the system is dynamically prevented to use. 
If we waited enough time to equilibrate the system, 
we would see the curve $c_p(T)$ modify drastically, and the specific heat 
drop would disappear. In a crystal, on the the contrary, the system is confined in a limited 
portion of the phase space forever, since energy 
barriers are infinite. In other words: a crystal is a {\it bona fine} broken ergodicity 
equilibrium phase of the system, whereas an off-equilibrium glass is not.

To conclude this section, imagine that a very patient researcher give
us (as a gift) a liquid sample thermalized at a temperature where we
would not be able to equilibrate the system, that is at $T$ smaller
than our experimental glass transition $T_g$.  The sample now is at
equilibrium at very low temperature and we can start doing our
equilibrium measurements. The energy will have the right value, and so
the volume. But do we get the right specific heat? The answer is no.
In fact, we would still give a very bad off-equilibrium experimental
estimate of $c_p$.  The reason is that, even though our system is
initially into an equilibrium configuration (that is a configuration
that has been correctly drawn with the Boltzmann-Gibbs distribution),
in order to perform an equilibrium measurement of the specific heat we
need to let the configuration explore ergodically the phase
space. But we do not have enough time to do so, otherwise we could
have equilibrated the sample ourselves! Of course, there are some
quantities (like the volume, or the energy) whose measurement would
coincide with the right equilibrium value, but this is small
consolation.  Most of the interesting observables, that is those
involving the fluctuations, need the system to be truly ergodic in
order to be measured correctly.  Thus, few equilibrium configurations
do not solve the problem at all.

This apparently pointless issue is in fact relevant in the field of
numerical simulations: there are powerful algorithms that use a
non-physical dynamics to produce very low $T$ configurations
equilibrated according to the Boltzmann-Gibbs distribution.  One may
be tempted then to use these equilibrium configurations as a starting
point to run a simulation with a more physical dynamics (as, for
example, molecular dynamics) and measure the specific
heat. This, however, cannot be done, since, as we have seen, it would
be impossible to make the system explore ergodically the phase space,
and to exploit all the relevant degrees of freedom.

\subsection{Is $T_g$ a meaningful concept?}

According to \eqref{tglass}, the temperature where the glass
transition occurs depends on how long is our available experimental
time $t_\mathrm{exp}$, and thus on the cooling rate. Different values
of $t_\mathrm{exp}$ yield different values of $T_g$, as shown in
Fig.1. This really looks as an odd definition of a `transition': what
is the point in fixing a transition at a temperature whose value
depends on the experimental protocol, and in particular on the cooling
rate?

Truly enough, if the value of $T_g$ for a certain substance
depended {\it strongly} on the experimental time, then it would be 
pointless to define a glass transition. However, this 
is not what happens. In many systems the increase of the relaxation time
is so sharp to make it very hard to move significantly the position of 
$T_g$ even by a substantial change in the cooling rate. The 
reason for this is that, whatever is the true underlying physical mechanism, 
the increase of the relaxation time when decreasing the temperature is {\it at least} 
exponential, and often sharper than that.
This means that a change of order $1$ in $t_\mathrm{exp}$ causes at best a logarithmic change in
the glass transition temperature. 
To fix ideas, let us assume that the shear relaxation time has a simple
Arrhenius dependence on the temperature, so that at $T=T_g$ we have,
\be
t_\mathrm{exp} = \tau_0 \exp\left(\frac{\Delta}{T_g}\right) \ .
\label{arr}
\ee
By differentiating both sides of \eqref{arr}, we obtain,
\be
d T_g = -\left(\frac{\tau_0 \Delta}{T_g^2}\right) \exp\left(-\frac{\Delta}{T_g}\right) d t_\mathrm{exp}  \ ,
\label{ucciouccio}
\ee
which clearly shows that a change of $t_\mathrm{exp}$ is exponentially damped,
so to give a very small change of the glass transition temperature $T_g$.
This fact is also expressed by the Bartenev-Ritland phenomenological
equation connecting $T_g$ to the cooling rate $r$ \cite{gutzow},
\be
\frac{1}{T_g} = a - b \log (r) \quad \quad ,\quad\quad r=\frac{dT}{dt}
\label{ritland}
\ee
where $a$ and $b$ are two constants. Equation \eqref{ritland} is just a
trivial consequence of \eqref{ucciouccio}.

Equation \eqref{arr} is the expected behaviour for a system where the
mechanism of relaxation is ruled by barrier crossing and where the
barrier $\Delta$ does not have any dependence on the temperature.  As
we shall see, one of the most interesting features of glass-formers is
that the increase of the relaxation time is often {\it steeper} than
purely Arrhenius, i.e. it is super-Arrhenius.  In this case, the
dependence of $T_g$ on $t_\mathrm{exp}$ is even weaker than in
\eqref{ucciouccio}, and the point we wanted to make is thus even
stronger: the dependence of the glass transition on the experimental time
is quite weak in general. For all practical purposes, when the
temperature arrives around a certain value, the system falls out of
equilibrium, and there is very little we can do to move $T_g$ much
further down. For this reason it makes sense to define the dynamic
glass transition, and it is more so the sharper is the slowing down of
the substance. When the relaxation time (and the viscosity) increase
by 12-14 decades under a temperature change of a factor $3$, we get
the strong feeling that something significant is going on there, and
we want to mark this point.

For these reasons, it is sensible to give a general definition of the kinetic glass transition 
by fixing a {\it conventional} value for the maximum experimental time we are prepared to wait
in order to equilibrate a liquid. This was first suggested in \cite{laughlin} and 
 fixed at $100-1000$ seconds, so that we can write,
\be
\tau_\mathrm R (T_g) \sim 10^2-10^3 \  \mathrm{sec} \ .
\label{glasstau}
\ee
By using the relation \eqref{magnapompa} between shear viscosity, relaxation time and 
infinity frequency shear modulus, and using the standard value for $G_\infty\sim 10^{10}-10^{11}$ dyne/cm$^2$, 
equation \eqref{glasstau} is equivalent to define  $T_g$ through the following classic relation,
\be
\eta(T_g) \sim 10^{13} \  \mathrm{Poise}
\label{glasseta}
\ee
At the melting point a liquid's viscosity seldom exceeds the value $\eta\sim 10^{-2}-10^{-3}$ Poise,
so we see that at the glass transition the viscosity is pretty huge.

Of course, if we decrease by several orders of magnitude our
experimental available time, than we {\it do} move upward
significantly the glass transition. This is exactly what happens in
numerical simulations of liquids, which, compared to real experiments,
can cover a much smaller time interval. For this reason, the
definition of 'glass transition' in numerical simulations is ruled by
the CPU time, rather than the simulated 'real' time of the liquid, and
$T_g$ is significantly higher than in realistic experiments.

The conclusion of this section is that, yes, the glass transition is a
meaningful concept, at least from a practical point of view.  Even
though it is not a sharply defined transition, the increase in the
relaxation time is so steep, that for all practical purposes it makes
sense to mark a point where this happens.  As we shall see in the next
section, for some systems $T_g$ not only makes practical sense, but it
also cries for a physical explanation.

\subsection{Fragile vs. strong liquids}

The conventional definition of glass transition given in \eqref{glasseta}  allows us to 
report in a compact way the viscosity data of many different substances in a single plot. 
The benchmark for the increase of the relaxation time (and thus of the viscosity) is the 
Arrhenius law \eqref{arr}. For this reason data are normally plotted as a function of $1/T$, 
as done in Fig.6. In this way, however, there is a spread of the data
due to the different values of $T_g$ of the various systems. To balance this we can plot 
the logarithm of the viscosity as a function of $T_g/T$, in such a way that all curves have the same value
(that is $10^{13}$ Poise) at the point $T_g/T=1$. This kind of graph, conventionally called
Angell's plot \cite{laughlin, angellplot}, 
is shown in Fig.8, and it is quite instructive. 

We see that different substances
may have quite different behaviours. Some systems fall on a straight line, meaning that the
simple Arrhenius behaviour is a reasonable description. Other systems, on the contrary, are
very far from an Arrhenius behaviour and display a smooth growth at high temperatures,
which becomes however steeper and steeper as the temperature is lowered. These two extremes
cases have names \cite{angellplot}: the almost purely Arrhenius systems are called {\it strong} 
liquids, whereas those that have a sharper super-Arrhenius behaviour, displaying the largest
deviation from a straight line in the $T_g/T$ representation, are called {\it fragile} 
liquids. Archetypical strong liquids are SiO$_2$ (window glass) and GeO$_2$,
whereas o-terphenyl and toluene are two champions of the fragile class.

\begin{figure}
\includegraphics[clip,width=3.4in]{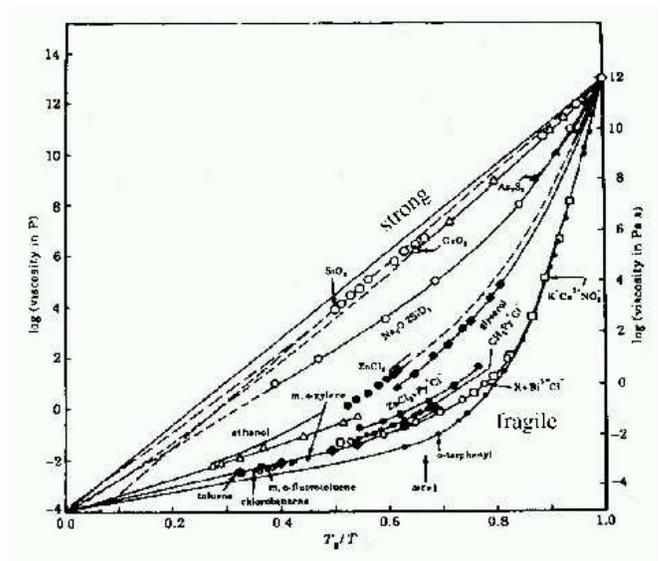}
\caption{ {\bf Angell's plot} - 
Logarithm of the viscosity vs. rescaled inverse temperature for many substances. In this 
representation a purely Arrhenius behaviour shows up as a straight line. This is typical
of strong glass-formers. On the other hand, sharper super-Arrhenius increase of $\eta$
correspond to fragile glass-formers.
(Reprinted with permission from \cite{angell-poole}; copyright of Societa' Italiana di Fisica).
 }
\end{figure}

Of course, we have to be careful with the strong-fragile
classification, because, as Fig.8 shows clearly, there is a whole
spectrum of behaviours in between the two extrema. Moreover, even in
the most fragile cases, it is clear that a piece-wise Arrhenius fit,
with two different values of the barrier $\Delta$ at high and low
temperatures, will fit very reasonably the data (as can be appreciated
also by eye).  The only important point we have to grasp here is that
the increase of the viscosity by lowering $T$ is not equally sharp in
all systems. For strong-like liquids the increase is {\it relatively}
smooth, in the sense that it is ``only'' exponential. For strong
systems it seems that at $T_g$ really nothing particular happens,
apart the viscosity hitting $10^{13}$ P; in this case, one may argue
that the definition of $T_g$ has purely practical implications, and
nothing more.

On the other hand, in strongly fragile liquids the increase is sharper
the closer we get to $T_g$. Fragility can be easily quantified by
measuring the slope of $\log(\eta)$ vs. $T_g/T$ at the point
$T_g/T=1$: the larger this quantity, the sharper the growth of the
viscosity, the larger the fragility.  Fragile behaviour suggests that
something new does happens close to $T_g$, and therefore that the
glass transition is a more fundamental quantity for fragile systems
than for strong ones.  For this reason, fragile liquids are the most
interesting glass-formers, and many of the questions we ask about
glassy physics are in fact pertinent only for this kind of systems. If
a strong glass-former may be interpreted simply as a very very viscous
liquid, this seems to be inappropriate for fragile liquids, which
seems to require a deeper explanation.

We have provided an unambiguous definition of the glass
transition, equation \eqref{glasstau}, and we have reasons to believe
that the physics of fragile liquids becomes tricky close
to the glass transition.  Yet, we must admit that we are still
somewhat unsatisfied: the $T_g$ we introduced is indeed well-defined,
but entirely conventional.  It seems a completely
anthropocentric concept: give that our lifespan is $~10^2$ rather than
$~10^{30}$ years, and given that that we get bored at doing
experiments longer than $1$ day, we give a different name to samples
whose relaxation time exceeds a certain threshold, and we call them
{\it glass}. Even though in very fragile liquids one feels
that $T_g$ has a deeper meaning, from the formal point of view there is
little or nothing upon which to base this judgment.  Fragility is a
{\it quantitative} feature, not a {\it qualitative} one. What about the whole
zoology of intermediate systems? What is $T_g$ for them?  The problem
is that, from what we have seen up to now, nothing {\it qualitatively}
remarkable happens at $T_g$, apart from our inability to keep the
sample at equilibrium.  Let us reformulate the problem in the
following way: if our lifespan were $~10^{30}$ years, and we were able
to equilibrate a supercooled liquid well below $T_g$ as defined in
\eqref{glasstau}, would we notice any qualitative difference compared to 
the high $T$ phenomenology?  Fortunately for the popularity of the
field, the answer to this question is: yes.

\subsection{Supercooled liquids are structurally unexciting}

Before we see how to answer positively to the question above, it is
important to point out that standard structural observables 
{\it cannot} distinguish the deeply supercooled phase at a 
qualitative level. The static structure of the particles in a 
supercooled liquid close to $T_g$, and even in a glass below $T_g$, 
is virtually indistinguishable from that of a liquid at temperatures 
well above $T_g$. As far as structure is concerned, a glass looks exactly the same as a liquid.
Let us see in more detail what `structurally' means here.

The simplest way to give a structural characterization of a homogeneous
and isotropic liquid 
is to consider the radial distribution function, $g(r)$ \cite{hansen,allen}. 
This quantity tells us what is the probability to find a particle
at distance $r$ from a certain focal particle. Its formal definition is the following,
\be
g(r) = \frac{1}{N} \frac{1}{4\pi r^2 \rho} \langle \sum_{i}^N 
\sum_{j\neq i}^N\delta(r-r_{ij}) \rangle \ ,
\label{gidierre}
\ee
where $N$ is the total number of particles, $\rho$ is the density and,
\be
\quad r_{ij} = || \vec x_i - \vec x_j ||   \ .
\ee
From this relation we get,
\be
\int_0^\infty dr \; 4\pi r^2  \rho g(r) = N-1\sim N  \ .
\ee
This shows that the mean local density at distance $r$ from a focal particle is 
equal to $\rho \, g(r)$. 
From its definition it is clear that $g(r)$ must
go to $1$ for $r\to\infty$, whereas it has a nontrivial structure
at finite values of $r$. 

The radial distribution function $g(r)$ is very good at distinguishing
different phases (gas, liquid, crystal) of a particle system: the
higher the degree of order in the system, the more structured in term
of peaks is the $g(r)$. In a liquid, the typical shape of this function
is showed in Fig.9: at small $r$, $g(r)$ is  zero, due to the
short-range repulsion that prevents particles from getting too close
to each other; at larger $r$ the function steeply rises in
correspondence to the first layer (or shell) of particles around the
focal one; at even larger $r$ there are some weaker, although still
well defined, peaks corresponding to the various shells around the
focal particle. In the liquid there is no long range order, so that
the peaks are weaker and weaker the larger $r$.  In a crystal, on the
other hand, the peaks are very sharp, and do not decay, because long
range order sets in. On the contrary, in a gas there is only the drop
of probability at very low $r$ due to the hard core of the
particles, and no peaks at all, since there is no structure.

\begin{figure}
\includegraphics[clip,width=4.4in]{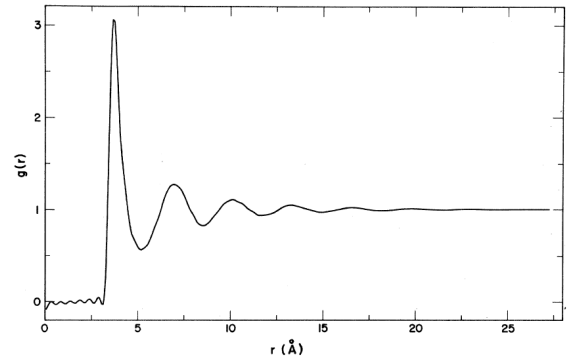}
\caption{ {\bf The radial distribution function} - The radial
distribution function $g(r)$ in liquid Argon at relatively high
temperature.  The various peaks correspond to various shells around
each particle.  (Reprinted with permission from \cite{yarnell-73});
copyright of American Physical Society.  }
\end{figure}

From the experimental point of view the easiest thing to measure is 
the static structure factor $S(q)$, which is experimentally accessible from
inelastic neutron scattering \cite{hansen}. This quantity
is related to the radial distribution function $g(r)$ by a simple
Fourier integral \cite{hansen, allen},
\be
S(q)= 1+4\pi\rho \int_0^\infty dr \, r^2 \frac{\sin qr}{qr}\, (g(r)-1)  \ ,
\label{strutto}
\ee
and it provides in momentum space $q$ the same kind of structural information as 
$g(r)$.

Do these structural quantities show anything peculiar near $T_g$?
Not at all. The static structure factor of a deeply supercooled
liquid, and even of an off-equilibrium glass, is virtually
indistinguishable from that of a high $T$ liquid \cite{price, ernst-91, blaade,
leheny}.  This result is clearly shown in Fig.10 \cite{gleim}.  As the
temperature is lowered, there is a very small modification of the
peaks structure, in particular the peaks become somewhat sharper.
In the same temperature range, on the other hand, the increase of
the relaxation time is very significant (Fig.10). In general, the relaxation
time $\tau_\mathrm R$ increases by $12-14$ orders of magnitude at
$T_g$. Yet, when this happens the structure does not show anything
much more relevant going on than what is shown in Fig.10.
We conclude that it is impossible to use the structure factor, or any
other standard structural quantity, to understand whether or not the
sample is close to the glass transition. 

\begin{figure}
\includegraphics[clip,width=3.4 in]{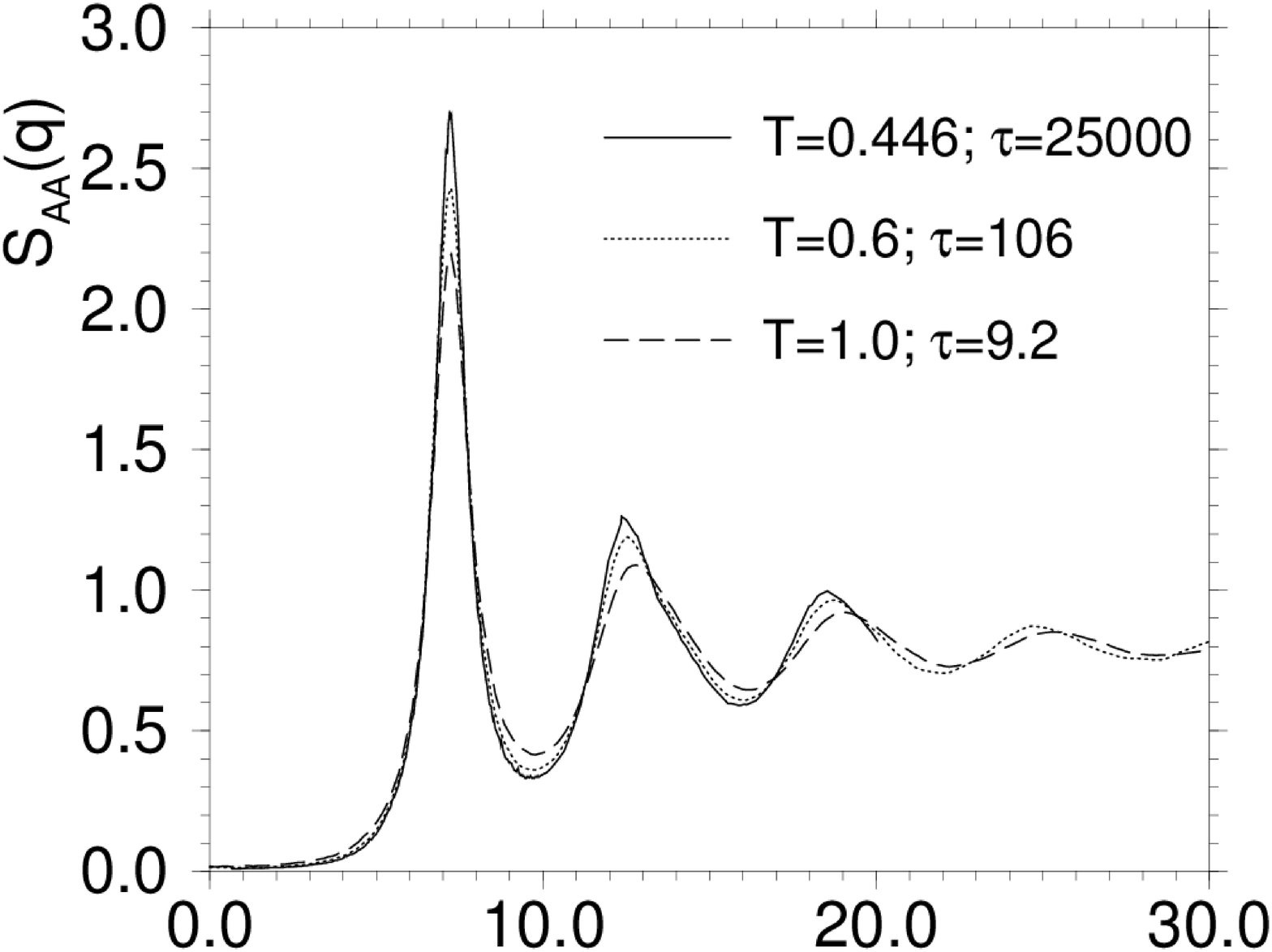}
\caption{ {\bf The static structure factor} - 
The static structure factor $S(q)$ in a Lennard-Jones liquid at three different temperatures.
The relaxation time $\tau_\mathrm R$ increases by almost $4$ orders of magnitude, and yet the 
structure factor shows no particular change.
(Reprinted with permission from \cite{kob-review}).
 }
\end{figure}

The weak modification of the structure factor with temperature also
implies that any lengthscale that can be reasonably extracted from
$S(q)$ or $g(r)$ shows a depressingly weak dependence on $T$ near the
glass transition \cite{price, ernst-91, blaade, leheny}.  This is
surprising. The common wisdom is that the presence of a sharply
increasing (or diverging) relaxation time should be associated to a
sharply increasing (or diverging) correlation length. The wisdom comes
mainly form the theory of critical phenomena \cite{cardy}, and the
argument is basically that a large relaxation time derives from the
need to rearrange larger and larger correlated regions. Even though
common indeed, such wisdom deserve some carefulness anyway.
First, we cannot stress too much the (obvious) fact that in a system with
a {\it finite} number of degrees of freedom, the relaxation time must
be equally {\it finite}. A real divergence can only occur in the thermodynamic 
limit and only in presence of a phase transition. Still, the time can be
very large, typically because of the presence of large energy or free 
energy barriers to relaxation. In general, a barrier could arise from an
external potential, and in this case there is no need to invoke
the existence a large length scale. However, in all interesting
systems barriers almost invariably arise from the internal interaction among the 
degrees of freedom. In this case, increasing barriers (and thus increasing 
relaxation times), are indeed due to the increasing number of degrees of 
freedom that must be rearranged together to relax the system. Hence: large time, 
large length.

This argument makes sense, but of course in order to detect a sharply growing
correlation length, we need to identify a suitable correlation
function. This is the crucial point: in very viscous liquids and
glasses we are completely lost when it comes to identify the correct
order parameter. Structural correlation functions are not up to the
job and fail to provide any exciting characterization of the viscous
and glassy phase. This is unfortunately true for all other standard
static correlation functions.

The reason for this can be either because we are using the wrong
correlation function, or simply because there is {\it no} sharply
growing lengthscale (the common wisdom could be wrong!).  In
fact, we will see at the end of these notes that the first hypothesis
is the correct one.  For now, we simply note that the standard
structural observables are not good enough to say something about the
deeply supercooled phase. Structure remains qualitatively the same
going through the glass transition.  Supercooled liquids and glasses
are structurally unexciting.

\subsection{Equilibrium fingerprint of glassiness: two steps relaxation}

The failure of a standard static approach to find a signature of the
glass transition is, in fact, not surprising. The very definition of the
glass transition is purely dynamic in nature, and therefore if
something new happens close to $T_g$, the dynamics, rather than the
static structure, should detect it.  In particular, the viscosity, which
marks the onset of glassiness, is the integral over time of a dynamic
correlation functions, namely the shear relaxation function, see
equation \eqref{carisma}. The same is true for the diffusion
coefficient, related to the time integral of the velocity-velocity
correlation function \eqref{velocity}. An integral wraps up an entire
function into a single number, thus loosing a lot of information. Hence, it seems
a good idea to check the dynamic correlation functions, rather than
their integrals, to see whether they show some qualitative signature
of $T_g$.

Let us consider, in full generality, the dynamic correlation function,
\be
C(t_1,t_2) = \frac{1}{N}\sum_{k=1}^N \langle \varphi_k(t_1) \varphi_k(t_2) \rangle  \ ,
\ee
where $\varphi_k(t)$ is a generic quantity relative to particle $k$, observed at time $t$.
If the system is at equilibrium,
then time translation invariance (TTI) holds, and the correlation function only depends
on the difference of times, $t=t_2-t_1$, so that $C(t_1,t_2) = C(t)$, and we can write,
\be
C(t)=\frac{1}{N}\sum_{k=1}^N \langle \varphi_k(t) \varphi_k(0) \rangle  \ .
\label{dyncorr}
\ee 
In liquids, a typical choice for $\varphi_k(t)$ is the Fourier transform of the density fluctuations 
of a tagged particle $k$,  $\delta\rho_k({\bf q},t)=\exp[-i{\bf q}\cdot{\bf r}_k(t)]$, at fixed 
momentum $\bf q$. In this case, the dynamic correlation function $C(t)$ coincides with the incoherent intermediate 
scattering function $F_s(q,t)$ \cite{hansen}, which is normally measured in 
experiments. In systems other than liquids, $\varphi_k(t)$ can be any meaningful observable 
carrying a real space label (be it particle or spin). 

The correlation function $C(t)$ measures how quickly correlations within the system decay in time. At high temperatures
we expect a very short-time ballistic regime (for Newtonian dynamics), where particles move freely with no 
mutual interactions, followed by a dissipative regime, described by 
a normal exponential relaxation,
\be
C(t)=C_0\,\exp(-t/\tau)  \ .
\ee
In principle the relaxation time $\tau$ depends on the particular 
observable $\varphi$. However, it is natural to expect that at high $T$ there is
only one intrinsic time scale in the system, for example the shear relaxation time $\tau_\mathrm R$,
and that all the other time scales are a simple rescaling of it.

We already know that by lowering the temperature the relaxation time
$\tau_\mathrm R$ grows very sharply, so that the decay of $C(t)$ is increasingly slower
approaching $T_g$. Therefore, from a quantitative point of view, the dynamic correlation
function differs significantly from the structural correlation function, which
shows no dramatic temperature dependence close to $T_g$.  However, were the sharp increase
of $\tau$ the only effect of lowering $T$, there would be no
qualitative signature of approaching glassiness. But this is
not what happens. Fig.11 shows the typical behaviour
of the dynamic correlation function at various temperatures, down to a
temperature that is above, but close to $T_g$. What the figure shows is that the
qualitative shape of $C(t)$ changes significantly approaching $T_g$: in a
log-time representation, a plateau is formed at low temperature, so
that, overall, the decay is no longer purely exponential \cite{kob-prl-94,kob-corr}. 
We call this kind of decay {\it two steps relaxation}.

\begin{figure}
\includegraphics[clip,width=3.4in]{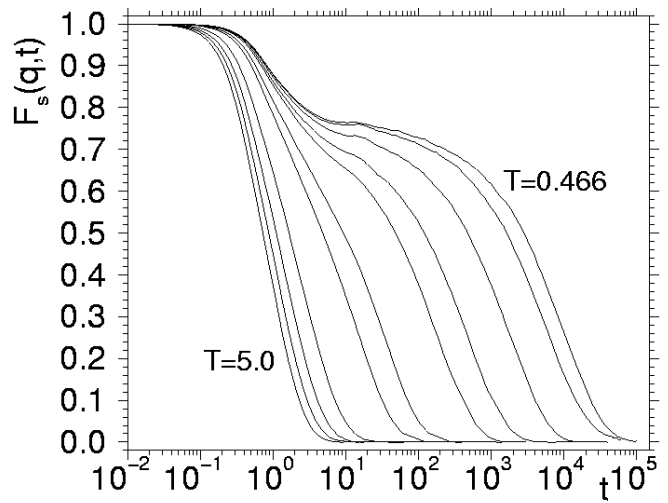}
\caption{ {\bf Two steps relaxation}
The dynamic correlation function $C(t)$  in a Lennard-Jones
system. In this case $C(t)$ is the incoherent intermediate scattering function $F_s(q,t)$,
evaluated at the value of $q$ where the static structure factor has the main peak.
 At high temperatures the decay is exponential, but when the temperature get 
close to $T_g$ a plateau is formed and relaxation proceeds in two steps.
(Reprinted with permission from \cite{kob-corr}; copyright of American Physical Society).
 }
\end{figure}

Two steps relaxation is the qualitative fingerprint of approaching
glassiness. By performing a purely equilibrium
measurement, we have a clear way to say whether or not our sample is
close to the dynamic glass transition, $T_g$. The relaxation time
increases (more or less sharply) at low temperatures, but it is
only from the nonexponential relaxation of the dynamic correlation
function, and in particular from the formation of a plateau, that we
can say that a system is approaching the glass transition.  Of course,
the precise temperature window where this non-exponential behaviour
kicks in depends on the system (and much less strongly on the
observable $\varphi$), but, on balance, it is fair to say that whenever the
dynamic correlation function develops a plateau, the glassy
phase is not far down in temperature, even though the relaxation time
(and the viscosity) may still be significantly lower than their $T_g$
values.  

The nonexponential relaxation and the plateau of the
dynamical correlation function appears above the glass transition
$T_g$, as a qualitative precursor of it.  We therefore see that the
glass transition $T_g$ {\it does} have some physical significance,
after all, because something qualitatively new shows up in the
equilibrium properties of a liquid close to $T_g$.

Let us describe more carefully the shape of the correlation function at low $T$ (Fig.11).
The time-scale over which the correlation function arrives to the plateau does not
depend very much on $T$, whereas the length of the 
plateau, i.e. the time needed to leave the plateau, becomes larger the
lower the temperature. This second phenomenon is the main contribution to the sharp
increase of any reasonably defined relaxation time $\tau_\mathrm R$,
may it be the point where $C(t)$ reaches a certain arbitrary
threshold, for example, 
\be C(t> \tau_\mathrm R)\leq 0.1 \ , 
\ee 
or a time-scale proportional to the integral of $C(t)$, 
\be \tau_\mathrm
R\sim \frac{1}{C(0)}\int_0^\infty dt\;C(t) \ .  
\ee 
The very shape of
$C(t)$, however, strongly suggests that describing its decay in terms
of a single times scale is unwise.  Indeed, unlike at high $T$,
where a simple exponential decay $\exp(-t/\tau_\mathrm R)$ fits well the data,
the plateau structure of the low $T$ correlation function cries for (at least) 
a two time-scales description. Roughly speaking, we can say that there is a fast process
related to the approach to the plateau, and a slow process related
to the decay from the plateau. The fast process is weakly dependent of $T$,
while the slow process depends strongly on the temperature. 
Conventionally, these two processes are called, respectively, $\beta$ (fast) and $\alpha$ (slow)
relaxation. 

Given this structure of the correlation function, when 
one mentions the relaxation time it would be better to specify 
what time one is talking about. Throughout these notes, by
relaxation time $\tau_\mathrm R$ we intend the $\alpha$ relaxation 
time, i.e. the time of the longest relaxation processes.
Of course, the plateau is the common reference frame for both processes, 
so that the late $\beta$ relaxation overlaps with the early $\alpha$ relaxation. 
Nevertheless, the structure of $C(t)$ clearly indicates that
there is a separation of time-scales, which is sharper the lower the temperature.
This separation of times scales is the qualitative landmark of glassiness.

The appearance of two steps relaxation in the dynamic correlation function 
occurs gradually on approaching the glassy phase. For this reason
it is not possible to sharply define a temperature where the 
departure from the high-$T$ exponential behaviour firstly appears.
Under this respect, we may say that the transition from exponential to two steps relaxation,
and thus from fluid to glassy behaviour, is continuous. 
However, under another important respect the transition can be considered as discontinuous.
Indeed, the height of the plateau, i.e. the value $C(t)$ has at the plateau,
is already different from zero when the plateau appears. 

We can make this 
observation more concrete in the following way. Imagine we are at a temperature $T$ where the
plateau is well defined. By using some reasonable fitting method, we can measure the height of the plateau, 
let us call it $C^\star$. This quantity is sometimes called {\it nonergodicity parameter} in 
the literature. 
If we now raise the temperature we observe that $C^\star$ decreases when $T$ increases \cite{sastry-00}.
This dependence, however, is rather weak (Fig.11). For this reason the value of $C^\star$ is still well above
zero when we reach the temperature regime where there is no plateau anymore. In this regime
the nonergodicity parameter $C^\star$ is ill-defined and we cannot assign a value to it. 
Conversely, if we decrease, rather than increase, the temperature, we observe that as soon as
the plateau exists, $C^\star$ has a nonzero value. In this sense the transition from the 
fluid to the glassy phase is discontinuous. We will go back to this crucial point when discussing 
mode coupling theory and spin-glasses.

\subsection{The cage}

What is the origin of two steps relaxation?
To have a first clue we turn to a different dynamical observable:
the mean square displacement (MSD) of a tagged particle,
\be
\langle r^2(t)\rangle =\frac{1}{N}\sum_i \langle || \vec x_i(t) - \vec x_i(0)||^2\rangle \ .
\ee
We have already met this quantity in the Section II.B  when discussing 
diffusion. We expect the MSD
to have an early regime where $\langle r^2(t)\rangle\sim t^2$, when particles move 
ballistically without many collisions, followed by a diffusive regime, where
$\langle r^2(t)\rangle\sim t$, dominated by collisions. In fact, as we have seen, the diffusion
coefficient $D$ is given by the ratio between MSD and time in the asymptotic regime.

In a deeply supercooled liquid, however, when we plot the MSD as a
function of the logarithm of time, the ballistic and diffusive regimes
are separated by a plateau, whose length increases on lowering the
temperature (Fig.12). The similarity with the dynamical correlation
function (Fig.11) is obvious, but the advantage now is that the
immediate physical meaning of the MSD provides a simple interpretation
of what is going on.  In the time region of the plateau the MSD
increases very little with time. The tagged particle is definitely
beyond the ballistic regime, but despite its many collisions with the
other particles something prevents it from a standard diffusive
motion, and the particle remains confined in a small region of
space. If we look at the actual value of the MSD in correspondence of
the plateau, we discover that it is quite small, well below the
(square) interparticle distance \cite{kob-msd}.

\begin{figure}
\includegraphics[clip,width=3.2in]{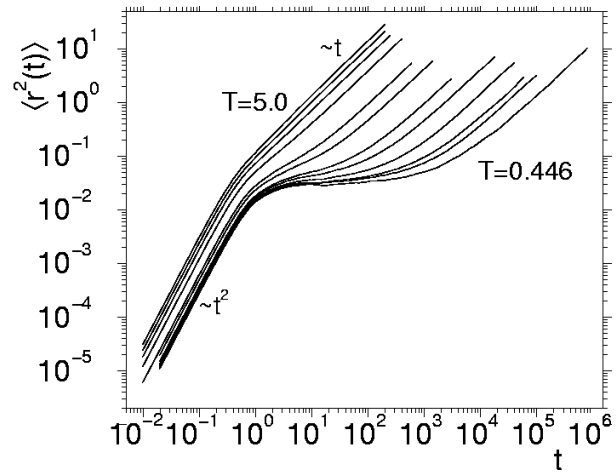}
\caption{ {\bf The mean square displacement} -
The mean square displacement (MSD) as a function of the logarithm of time in a Lennard-Jones
system. At high temperature there is a crossover from ballistic transport to diffusion.
At low temperatures this crossover is interrupted by a plateau similar to what happens
for the dynamic correlation function.
(Reprinted with permission from \cite{kob-msd}; copyright of American Physical Society).
}
\end{figure}

We can explain these facts by hypothesizing that, on the time scale of
the plateau, the particle cannot exit the {\it cage} formed by its
neighbours. Under this interpretation, the plateau corresponds to the
vibrations of the tagged particle within the cage. If we wait long
enough, though, the particle finds its way out of the cage, and a
standard diffusive dynamics sets in. Beware: we are watching things in
log-time. Therefore, after the particle has got out of the cage, it is
effectively out of {\it any} cage! The time needed to get out of the
cage is longer the lower the temperature, and it corresponds to the
$\alpha$ relaxation process, whereas the $\beta$ relaxation is given
by the particle vibrations within their local cages.

This seems a reasonable interpretation, and by similarity it also extends to the plateau in the dynamic
correlation function. Few comments, however, are in order. First, the cage explanation is in fact not so
much of an explanation: particles have neighbouring particles at all temperatures; what we would like to 
understand is why below a certain temperature, i.e. close to glassiness, the cage suddenly becomes stiff.
Moreover, how the particle gets out of the cage after a certain time, and why this time increases by decreasing 
the temperature? 

That of the cage is little more than a description of what is going on, but it is nevertheless a very
useful one. For example, it immediately suggests that to break a stiff cage, and
thus to exit the plateau and restore ergodicity,  particles must in some way rearrange.
This can be done by either finding a rare `hole' through the many energy barriers giving rise to the cage 
(a process that may be entropically costly, but energetically harmless), 
or by climbing up these barriers (energetically costly, but entropically harmless). 
We will get back later to these two alternatives, but the example shows that the cage interpretation 
should always be kept in mind.

Note that the size of the cage, i.e. the value of the MSD at the plateau, changes very little with 
the temperature, and it does not approach zero when raising $T$.
As soon as we start seeing the cage (i.e. a plateau of the MSD), its physical size is already 
different from zero. Because of this, the transition from high-$T$ simple diffusion to low-$T$ two
steps diffusion is discontinuous for the MSD exactly as it was for the dynamic correlation function. 

The cage description of two steps relaxation is deeply rooted in real space:
it is only concerned about what the particles do in the real physical space, as opposed to phase space arguments. 
On one hand this is good, since real physical systems live in real space, and what happens to them is ultimately 
related to real space phenomena. On the other hand, this is somewhat puzzling, because there are mean-field 
systems that have no real space structure at all, but where the plateau of the dynamical correlation 
function is anyway present at low temperatures. In these cases the cage effect cannot be the right interpretation, 
since there is  no real space, no local neighbours and no cage: each degree of freedom (particle) interacts with all the others, 
and a phase space description is the only one we are left with.
These observations do {\it not} imply that real space arguments are incorrect. Rather, the presence of
mean-field model showing two steps relaxation suggests that there may be an alternative 
explanation of this crucial landmark of glassiness. As we shall see in the next chapter, this alternative 
explanation can be given in terms of what the system does in phase space.

A final warning. Even though we described decaging as a single particle event (the particle `gets out
of the cage'), this is a simplistic view. As we shall see later on, dynamical
(as well as static) relaxation is achieved through cooperative behaviour. Many particles must be
dynamically correlated in order to unstuck from their local vibrational positions. Hence, one should rather say 
that it is the cage that collectively breaks up to free the particles, before another cage is formed. However,
at this stage of our study we still do not know what are the details of such dynamical cooperativity, so we ask
the reader to simply keep this remark in the back of her/his mind whenever we will use the naive
short-cut of a particle getting out of the cage. Things may become clearer when we will study the 
dynamical correlation length.

\subsection{Stretched exponential, dynamical heterogeneities and Stokes-Einstein violation}

One may think that the departure of the dynamic correlation function from exponential relaxation is solely due to the 
break-up of the decay into two steps; this would imply that the late relaxation, i.e. the $\alpha$ decay
from the plateau is exponential. However, this is not the case: even the late relaxation 
is nonexponential. In particular, it is found that the Kohlraush-Williams-Watts stretched exponential form
\cite{kohl, watts},
\be
C(t)= C_0 \ \exp\left[-(t/\tau)^\beta\right]  \quad , \quad  \beta<1
\label{kww}
\ee
fits reasonably well the data. The interesting point is that the exponent $\beta$ decreases
when the temperature is decreased, marking a larger and larger deviation from a standard
exponential relaxation, the deeper we get into the supercooled phase \cite{angell-00}. On the other hand,
at higher $T$ the exponent $\beta$ approaches $1$, and at the same time the whole two steps
structure of the dynamical correlation function disappears, and the relaxation goes back
to simple exponential. 

Regarding the origin of nonexponential relaxation we can put forward
two hypotheses, let us call them the {\it heterogeneous} and the {\it
homogeneous} explanation.  According to the first, the relaxation of
the entire system is stretched because different regions have
significantly different relaxation times, and in this sense dynamics
is heterogeneous. When we measure the global relaxation time, we are
typically averaging over many different regions, each one with a
purely exponential relaxation ruled by a different relaxation time,
and from this heterogeneous average one gets a global nonexponential
decay.  On the other hand, according to the homogeneous explanation,
relaxation is equally nonexponential in all regions, so that the
stretched nature of the correlation function is not due to an
heterogeneous spatial average, but it is an intrinsic phenomenon also at the
local level, due to the disordered environment each particle sees
around itself.

These two views are not necessarily contradictory. In fact, there are
reasons to believe that they are both valid to some degree. 
There are mean-field systems (as the $p$-spin model) where
two steps relaxation and stretched decay of the dynamic correlation
function are present at low $T$ just as in supercooled liquids. In
these systems, due to their mean-field nature (each spin interacts equally
with all other spins) there is no space structure, so that there cannot
be any {\it spatial} etherogeneities. Still, these mean-field systems are spin-glass
models, where there is quenched disorder in the interactions between
the degrees of freedom; it is therefore reasonable that the relaxation
becomes nonexponential at low $T$, even without the aid of
spatially heterogeneous dynamics. In fact, even in mean-field 
systems, at low $T$, different spins have very different flipping 
frequencies \cite{ricci-03,ricci-03b}.

On the other hand, it is now well established 
both numerically \cite{perera-96, donati-97,doliwa-98,donati-98} and
experimentally \cite{vidal-00, kegel-00} that close to the glass
transition the dynamics of real supercooled liquids {\it is}
heterogeneous. It is possible to directly observe domains few
nanometers away from each other with significantly different mobility
and relaxation times.  Therefore, in real liquids heterogeneous
dynamics is likely to contribute substantially to the formation of an
overall nonexponential relaxation. Given that heterogeneous dynamics
has implications beyond the stretched exponential relaxation, it is
worth considering it more carefully. The literature on this subject is
vast; see \cite{sillescu-99} and \cite{ediger-00} for experimental
reviews and \cite{glotzer-00} for a more numerically oriented one.

If we take one snapshot of our system, we see nothing impressive close
to $T_g$. We have already noted this fact, structural quantities are
unexciting. Let us now consider {\it two} subsequent snapshots, taken
at two instants of time separated by an interval $t$. We can now
measure how much each particle moved in this time interval.  If $t$ is
very short, still in the ballistic regime, we do not expect great
variations of the particles mobility, because interaction has still
not kicked in to make things interesting; similarly, if $t$ is very
large, much larger than the structural relaxation time $\tau_\mathrm
R$, then we are averaging over a temporal window so large that the
time average will be equal to the ensemble average (remember: we are
at equilibrium!); but all particles are statistically the same, so we
cannot observe differences once we take the ensemble average, and thus
each particle will again have a similar mobility. On the other hand,
if $t$ has an intermediate value, long enough to monitor particles
interaction, but short enough not to restore statistical homogeneity,
we see something very different: there are particles with mobility
significantly higher, and lower, than the average; moreover, particles
with higher mobility tend to cluster together in groups. This means
that across the system there are different patches of very fast and
very slow particles. A liquid close to $T_g$ is dynamically
heterogeneous, another qualitative landmark of incoming glassiness in
a system.

Few remarks. First, it is important to appreciate the fact that the
time interval $t$ must have an intermediate value, in order to observe
an heterogeneous dynamics. Such intermediate value is close to the
plateau regime of the dynamic relaxation function $C(t)$
\cite{perera-96, donati-97,doliwa-98,donati-98}.  This observation
strongly suggests that the nonexponential relaxation of the dynamical
correlation function, whose main feature is indeed the plateau, is linked
to heterogeneous dynamics.  Each patch has a different value of the
local relaxation time, and all locally exponential relaxations add up
to a stretched exponential one. Second, when we raise the temperature,
the dynamical correlation function and the MSD loose their plateau
structure, the ballistic regime is directly followed by the diffusive
one, and dynamics becomes homogeneous, irrespective of the value $t$
has. Third, we must not forget that these patches, or clusters, are
dynamical, not structural. We could not detect them by using {\it one}
snapshot of the system. We need two. Mobility fluctuations becomes
very large at low $T$, even though nothing similar happens at the
structural level.

Heterogeneous dynamics is believed to be at the basis of another important phenomenon in low $T$ supercooled liquids, namely 
the violation of Stokes-Einstein (SE) relation between diffusion coefficient and viscosity \eqref{stokeseinstein},
\be
D \sim \frac{T}{\eta}  \ .
\label{se2}
\ee
Close to the glass transition $T_g$, the diffusion coefficient $D$ may be several orders of magnitude 
larger than $T/\eta$ \cite{rossler-90, fuja-92, chang-94, cicerone-96}. In particular, the ratio $T/(D\eta)$, 
far from remaining constant as the SE relation wold require, decreases sharply when the temperature 
is lowered close to $T_g$ \cite{rossler-90, fuja-92}. 
This means that $D$ and $1/\eta$ have different functional dependence on $T$.
Hence, in the temperature regime where structural relaxation becomes very sluggish, 
self-diffusion is still relatively vital. Why is that?

Several authors have linked the SE violation to heterogeneous dynamics. In particular, 
Cicerone and Ediger provide in \cite{cicerone-96} a strikingly simple argument, which 
we report here. The core of the argument is the following: if the dynamics is heterogeneous, then 
diffusion is dominated by the fastest clusters, whereas structural relaxation is dominated by 
the slowest ones. Thus, as soon as mobility fluctuations become relevant, diffusion and relaxation
decouple, and the SE relation is violated. Let us see how this happens. First, remember that 
in general we can assume $\tau_\mathrm R\sim \eta$, so that at fixed temperature $T$ the SE
relation reduces to,
\be
D \sim \frac{1}{\tau_\mathrm R}  \ .
\ee
It is therefore a decoupling between diffusion coefficient and relaxation time that we are after. Now imagine an 
oversimplified heterogeneous dynamics: there are just two kinds of clusters, fast and slow ones, with relaxation 
time and diffusion coefficient respectively $\tau_f \ll \tau_s$ and $D_f \gg D_s$. In each cluster the SE relation is
obeyed, so that $D_f\sim 1/\tau_f$ and $D_s\sim 1/\tau_s$. Let us also assume
that fast and slow clusters are equally numerous across the system. Under these hypotheses, the global relaxation 
time measured in an experiment is given by,
\be
\tau_\mathrm R = \frac{\tau_f+\tau_s}{2}  \sim \tau_s/2  \ ,
\ee
whereas a measurement of the diffusion coefficient gives,
\be
D =\frac{D_f+D_s}{2}\sim D_f/2  \ .
\ee
We conclude that,
\be
D \gg  \frac{1}{\tau_\mathrm R}  \ ,
\ee
so that the SE relation is violated: diffusion is enhanced compared to structural relaxation. 
The crucial point in the argument is that when we measure the diffusion coefficient we
are actually measuring a square displacement, because we use the diffusion formula $\langle x^2\rangle = 2 D t$;
on the other hand, in measuring $\tau_\mathrm R$ we are measuring a time, which scales like $1/D$. Thus, when
the distribution of these two quantities are broad, i.e. when the dynamics is heterogeneous, their average value 
are no longer linked by the SE relation that holds locally. The argument is somewhat oversimplified, but it gives a rough idea of what 
is going on in a supercooled liquid.

The violation of SE relation has been also put in direct connection with the 
nonexponential decay of the dynamic correlation function. More precisely, by testing the SE relation in various substances  
at the glass transition, it has been concluded \cite{cicerone-96}  that the stretched exponent $\beta$ of \eqref{kww}  is smaller the 
larger is the measured deviation from the SE relation.
Furthermore, it has been shown that the SE equation is restored  when the diameter of the probe 
used to measure the diffusion coefficient increases \cite{cicerone-96}. This makes sense in terms of heterogeneous dynamics:
when the probe is large enough, i.e. larger than the size of the patches, we are averaging over many 
heterogeneities, thus washing out their influence.
These results strongly supports the link between stretched exponential, violation of the SE equation and 
heterogeneous dynamics.

Even though mobility clusters do not necessarily imply cooperativity, their very existence strongly suggests that particles 
in these patches are in fact correlated in some way, and thus that the cluster size is a very good candidate for a correlation length, 
even though of a dynamical nature. Indeed, empirical data show that on lowering the temperature mobility 
clusters grow in size \cite{donati-97,donati-98}. In the final chapter of these notes, when we will look for a 
growing correlation length in supercooled liquids,  we will see in detail how this dynamical lengthscale can be  defined.

%%%%%%%%%%%%%%%%%%%%%%%%%%%%%%%%%%%%%%%%%%%%%%%%%%%%%%%%%%%%%%%%%%%%%%%%%%%%%%%%%%%%%%%%%%%%%%%%%%%%%%%%
%%%%%%%%%%%%%%%%%%%%%%%%%%%%%%%%%%%%%%%%%%%%%%%%%%%%%%%%%%%%%%%%%%%%%%%%%%%%%%%%%%%%%%%%%%%%%%%%%%%%%%%%
%%%%%%%%%%%%%%%%%%%%%%%%%%%%%%%%%%%%%%%%%%%%%%%%%%%%%%%%%%%%%%%%%%%%%%%%%%%%%%%%%%%%%%%%%%%%%%%%%%%%%%%%
%%%%%%%%%%%%%%%%%%%%%%%%%%%%%%%%%%%%%%%%%%%%%%%%%%%%%%%%%%%%%%%%%%%%%%%%%%%%%%%%%%%%%%%%%%%%%%%%%%%%%%%%
%%%%%%%%%%%%%%%%%%%%%%%%%%%%%%%%%%%%%%%%%%%%%%%%%%%%%%%%%%%%%%%%%%%%%%%%%%%%%%%%%%%%%%%%%%%%%%%%%%%%%%%%
%%%%%%%%%%%%%%%%%%%%%%%%%%%%%%%%%%%%%%%%%%%%%%%%%%%%%%%%%%%%%%%%%%%%%%%%%%%%%%%%%%%%%%%%%%%%%%%%%%%%%%%%
%%%%%%%%%%%%%%%%%%%%%%%%%%%%%%%%%%%%%%%%%%%%%%%%%%%%%%%%%%%%%%%%%%%%%%%%%%%%%%%%%%%%%%%%%%%%%%%%%%%%%%%%
%%%%%%%%%%%%%%%%%%%%%%%%%%%%%%%%%%%%%%%%%%%%%%%%%%%%%%%%%%%%%%%%%%%%%%%%%%%%%%%%%%%%%%%%%%%%%%%%%%%%%%%%
%%%%%%%%%%%%%%%%%%%%%%%%%%%%%%%%%%%%%%%%%%%%%%%%%%%%%%%%%%%%%%%%%%%%%%%%%%%%%%%%%%%%%%%%%%%%%%%%%%%%%%%%

\section{Theoretical views on the glass transition}

As I said in the introduction, the quest for a theory of the glass transition
is far from being over. Different theoretical frameworks provide different interpretations
of the phenomenology, and it is hard to write
about this subject without having in mind our own ideas. 
Here I will try to present a coherent perspective, rather than to
enumerate all different approaches. The result will
inevitably be partial.

\subsection{Goldstein's energy landscape scenario}

Goldstein's picture of the equilibrium dynamics of a deeply
supercooled liquid is so simple, that it may seem
trivial today, although it was definitely not when it was formulated
back in 1969 \cite{gold}.  The very fact that his description seems so natural to
us is proof of how vast the influence of his work has been on
the understanding of this field.

Goldstein put the emphasis on the evolution of the system in the phase
space, i.e. the space of all the configurational degrees of
freedom. For example, for a monatomic liquid in three dimensions,
this is the space of all $3N$ coordinates of the particles.  Over this
space it is defined the total potential energy of the system, and the
surface of this function is often called potential energy
landscape. Each different configuration is represented by a point in
the phase space, and the dynamics of the system can be thought of as the
motion of this point over the potential energy landscape. 

The local minima of the potential energy correspond to locally stable
configurations of the particle system. One of them is of course the
crystal, and this will be the absolute minimum, i.e. the ground
state. Moreover, there will be all the minima obtained by introducing
defects and dislocations into the crystal; we can see these as
excitations over the crystalline ground state.
But apart from these crystalline, polycrystalline and defected crystal 
minima, there will be many local minima 
corresponding to particles arrangement that are completely lacking long-range crystalline
order. These are amorphous, or glassy, minima, and have a potential
energy that is extensively larger than the crystal one. For the rest
of these notes we will not deal with crystalline and polycrystalline
minima and only be interested in amorphous local minima.

Goldstein's idea is that at low enough temperatures a supercooled
liquid explores the phase space mainly through activated jumps between
different amorphous minima, separated by potential energy
barriers. It is important to understand that the system does not do
these jumps in the attempt of reaching lower energy minima: we are
considering a system at equilibrium (albeit metastable with respect to the crystal), 
so that the
level of average potential energy where the system lives is
constant in time. The system passes from one minimum to another of similar
energy (on average), and in so doing it is ergodic and in equilibrium.

Even though the idea of a system globally jumping from one minimum
to another is clear enough, we should not forget that the true
dynamics of the system takes place in real space.  Goldstein himself
is very keen in giving a real space interpretation of the hopping
process in phase space. The jump over a barrier leading the whole system from a
minimum to another corresponds to the local rearrangement of a
relatively small number $n$ of particles localised in a limited region
of space. Particles far from where the rearrangement takes
place are very weakly affected by this process, even though in a
phase space description it is the point representing the global system
that performs the transition.

It is indeed very important to appreciate  the {\it local} nature of 
the hopping processes in Goldstein's scenario. The potential energy barrier separating 
two minima in the phase space is proportional to the number of particles
participating in the rearrangement, and thus it scales as some power of $n$.
We will discuss thoroughly later on about how the barrier grows with $n$,
but the important point to understand now is that $n$ is sub-extensive, 
that is negligible compared to the size $N$ of the entire
system. Therefore, also the barrier is sub-extensive, and thanks to this it can be surpassed by the system,
which is equipped of a thermal energy $k_\mathrm B T$ of order $1$. On the
contrary, in a fully connected mean-field model, each particle (or spin, or any other degree
of freedom) interacts with all other particles, so that exiting a local
energy minimum requires changing $N$ degrees of freedom. In this case the barrier is 
extensive and it cannot be surpassed. For this reason, once a mean-field system is
trapped within a local minimum it cannot escape and ergodicity is broken.

Moreover, in a large system the fact that each event of rearrangement (jump over a barrier) 
is localized in space implies that several independent molecular rearrangements
will be occurring in different regions at the same time: however small is the transition rate
per unit volume, in a very large system there will always be a nonzero probability to have
a rearrangement happening somewhere at a given instant of time (we met a similar argument 
when discussing nucleation).
This, however, may seem a bit counterintuitive: a system that is always on top of a barrier
does not exactly match our naive understanding of activated barrier crossing, where one
is supposed to wait a long time between different jumps.

This confusion arises from the difficulty in visualizing 
an activated event performed by a small number of degrees of freedom embedded 
in a multidimensional space (the phase space), where most of the other degrees
of freedom are unaffected by the transition. Once again, it is essential to keep in mind 
the local nature of the activated jump. Goldstein writes a very illuminating sentence about this \cite{gold}:
{\it ``The system will always be in the process of transition, rather than some
of the time, but always near a minimum, in the sense that a sudden cooling will 
drop it into a minimum with relatively small changes of most of the coordinates''.}
The last words are the most important: even though in principle it
is the entire system that performs the transition, in fact it is only 
a very small subpart of it that crosses the barrier. If we focus our attention on 
this subpart, we indeed recover our intuitive idea of activated dynamics.

A further reason why it is important to appreciate the local nature of
the jump from one minimum to another is that the temperature dependence of 
the number $n$ of particles involved in the rearrangement is
a natural candidate to explain the super-Arrhenius behaviour of fragile glass-formers.
We said above that $n$ is small, so that the barrier is of order $1$ and
it is comparable to the $k_\mathrm B T$. This was however rather vague.
As we shall see in Chapter VII, the aim of most theories of
the deeply supercooled phase is to estimate $n$ and
to explain why and how $n$ increases on lowering $T$.

When the temperature is increased, it is clear that Goldstein's
scenario must break down at the point where thermal energy
becomes comparable and even higher than the typical potential energy
barriers. Under this condition the theory of activation itself is a
bad approximation of the true dynamics, since activation only holds when the
barrier is much larger than $k_\mathrm B T$ \cite{chandra}. In this regime the
liquid becomes very fluid, and local rearrangements will be
no longer ruled by thermal activation. Therefore, the break down of
Goldstein's scenario marks a conceptually useful border, that is 
the temperature separating a low-$T$
activated and viscous regime, from a high-$T$ nonactivated and fluid
regime. Let us call this temperature $T_x$. In his paper Goldstein
provides three independent estimates of what
should be the value of the shear relaxation time at $T_x$, concluding that \cite{gold},
\be
\tau_\mathrm R (T_x) \sim 10^{-9} \ \mathrm{sec}  \ .
\ee
Other estimates \cite{binder-00} give $\tau_\mathrm R (T_x)\sim 10^{-8}$ sec. 
Both figures are well below the conventional glass transition value,
\be
\tau_\mathrm R (T_g)\sim 10^2-10^3 \ \mathrm{sec}  \ ,
\ee
but they are quite larger than that of simple liquids at their melting point, 
which is of order,
\be
\tau_\mathrm R (T_m)\sim 10^{-13} \ \mathrm{sec} \ .
\ee 
We conclude that,  
\be
T_g < T_x  < T_m   \ .
\label{panino}
\ee
Let us not be fooled by the apparent similarity of the relaxation times at $T_m$ and $T_x$, 
compared to its value at $T_g$. We must remember that the increase of $\tau_\mathrm R$ is very 
sharp. For this reason the two temperature intervals $[T_g:T_x]$ and $[T_x:T_m]$ are comparable, 
even though in time they are wildly different. In particular, in very fragile systems, where the raise of
the relaxation time is super-Arrhenius, $T_x$ and $T_g$ can
be quite close. Goldstein's  temperature $T_x$ is therefore a precursor of the glass transition.

One of the arguments used by Goldstein in his derivation of $\tau(T_x)$ is worth
to be mentioned explicitly. We have stressed above that each activation event takes place
locally, and that particles that are far only participate as a background.  This background
is however crucial, since it constitutes the rigid matrix that provides the potential energy barrier
resisting the rearrangement. As long as the temperature is well below $T_x$ different events in different 
positions will be independent, each one of them surrounded by a rigid background. 
However, when the temperature is raised the activation rate increases, so that 
there will be more and more rearrangements taking place simultaneously and rather close to each other.
In this case, different activation events will no longer be independent. As a consequence,
the view of a rigid background breaks down. To locate this temperature Goldstein  compares the 
time to complete a rearrangement with the shear relaxation time, which is a measure of the time 
the background matrix can retain its rigidity. In this way, using some rather arbitrary, but sensible assumptions, 
he finds $\tau(T_x) \sim 10^{-9}$ seconds.

Goldstein's scenario provides a natural description in terms
of two time scales below $T_x$: a short time due to the vibrational relaxation within a potential 
energy minimum, and a long time relaxation due to the transition between different minima.
The second time scale is much longer than the first one, since the hopping time is given by activation and
it becomes exponentially large when lowering the temperature. The low-$T$ separation of time scales
has been indeed confirmed via numerical simulations and it has been used as an explicit way to locate $T_x$ \cite{sastry-00}.

This separation of time scales seems to fit very well the two steps
relaxation of the dynamical correlation function. In particular, one may interpret
the short vibrational time scale and the long activated time scale of Goldstein's description 
respectively with the $\beta$ (fast) and $\alpha$ (slow) relaxation of the dynamic correlation 
function.  Under this interpretation, vibrations in the phase space around a single potential energy 
minimum correspond to vibrations of particles within their cages, whereas crossing of a potential 
energy barrier corresponds to the local decaging of some particles. By accumulating many decaging
events, i.e. many activated rearrangements, the system decorrelates and the dynamic correlation function 
leaves the plateau, giving rise to the $\alpha$ relaxation.

This interpretation is correct at low temperatures, that is in the regime where Goldstein's scenario holds, 
for $T<T_x$.  However, two steps relaxation is present in general also {\it above} $T_x$. This is a very important point: 
a plateau structure in the 
dynamical correlation function and in the mean square displacements is already observed in the temperature regime where 
$\tau_\mathrm R(T) < \tau_\mathrm R(T_x)$, a regime where activated dynamics is {\it not} the main mechanism of diffusion. 
This fact proves that vibrations around minima and activated barrier crossing cannot be the right interpretation of 
two steps relaxation at all temperatures, and in particular it is not the right one for $T>T_x$.
The way particles get out of their cage, that is the way the dynamic correlation function leaves the 
plateau, must change from activated rearrangement below $T_x$, to something else above $T_x$.
We will see in the next sections what this `something else' is.

\subsection{Mode coupling and the infamous $p$-spin}

Placing two rather extensive subjects as Mode Coupling Theory (MCT)
and mean-field Spin-Glass Theory (SGT) in the same  section is
perhaps the most pedestrian choice of this entire review.
Mode coupling theory and mean-field spin-glass theory have been
developed completely independently, within rather different communities.  MCT was
formulated in the attempt to describe at a quantitative
level the dynamical properties of supercooled liquids. SGT, on the
other hand, was originally developed to study the properties of a class 
systems (spin-glasses), which had apparently little in common with
liquids. Just to mention one thing, spin-glasses have quenched disorder, 
liquids do not. However, at some point the SG community stumbled upon a
class of models that showed a phenomenology strikingly similar to that of
supercooled liquids.  The $p$-spin model is the paradigm of such class.
It was proved that this particular model was described
by a set of dynamical equations formally identical to those obtained by MCT. 
Moreover, and most importantly,
these equations reproduced the two steps relaxation typical of
supercooled liquids, even though none of the two theories made use of
activation.  Therefore, the reason for putting together these two theoretical 
frameworks is simple: both MCT and
SGT provide a physical mechanism able to explain the two steps
relaxation of the dynamical correlation function {\it without}
resorting to thermal activation.

I will not give any detail about MCT and SGT, because the literature
on both MCT and SGT is already  extensive.  Apart from the
original papers on MCT \cite{bengzelius, leuth}, useful reviews can be
found in \cite{goetze-review, sjogren, kob-review, kob-binder}. On the
other hand, classic reviews on SGT can be found
in \cite{binder-young,virasoro, hertz, young}. For a more specific, yet
elementary, account of the $p$-spin model and of its similarities with
supercooled liquids see \cite{cavagna-review}. For the quantitative
connection between $p$-spin model and MCT dynamical equations see
\cite{bouchaud-96}.

By using the Zwanzig-Mori formalism \cite{zwanzig-61, mori}, MCT
aims to write a set of self-consistent equations for the dynamical
correlation function of the density fluctuations.  
This program is rather complicated, and MCT needs to make 
a number of simplifying approximations. Even though these approaximations
are the defining core of the theory \cite{bengzelius, leuth}, it is unfortunately,
not that easy to explain the physical intuition behind them 
(for a diagrammatic approach see \cite{bouchaud-96}).
Here we simply note that the approximations adopted made it possible 
for MCT to formulate a set of dynamical equations that can be handled numerically, and 
even analytically at the qualitative level \cite{goetze-review}.
The crucial point is that the input of the MCT equations is
given only by the static observables, and in
particular the static structure factor $S(q)$ \eqref{strutto}. 
This is somewhat surprising, since we have seen
that structural quantities do not show anything peculiar close to
$T_g$: how can the MCT equations be anymore exciting, having the
boring structural quantities as an input? 

The answer is the very
nonlinear form of the interaction vertex in the MCT equations, which
gives rise to a sharp feedback between static structure and dynamics:
even a tiny change in
the structural properties causes a steep slow down of the dynamical
relaxation \cite{goetze-review}.
This feedback in the MCT equations has a striking consequence: at low
temperature the MCT dynamical correlation function displays
two steps relaxation, developing a plateau whose length increases when
the temperature is decreased. This is qualitatively the same as 
in real supercooled liquids.  Moreover, the behaviour of
the nonergodicity parameter $C^\star$, i.e. the height of the plateau,
is as discontinuous as in liquids: as soon as the plateau can be
defined, its value is already different from zero.  The qualitative
agreement between MCT and real liquids regarding the shape of the
dynamic correlation function is an argument in favour of the
theory. But it is not the only one.

MCT predicts in a quantitative fashion the way the correlation
function should arrive to and depart from the plateau, and thus
it gives a precise description of both the $\beta$ (fast) and
$\alpha$ (slow) relaxation. These prediction agrees rather well with 
the experimental and the numerical evidence \cite{kob-prl-94, kob-msd, kob-corr, gotze-99}. 
This is remarkable, and sometimes overlooked: in the
general quest for a theory able to explain the {\it long} time
dynamics, or $\alpha$ relaxation, most of the theories simply forget
about the {\it short} time $\beta$ relaxation, which is however such a
conspicuous feature of glassy relaxation: as we have seen, the only
qualitative fingerprint of glassiness is two steps relaxation, not
simply the sharp growth of the relaxation time.

MCT thus captures the qualitative feature of two steps relaxation and it also gives some quantitative
predictions about the approach and departure from the plateau, which are (depending on the system) 
in more or less good agreement with the data. However, there is also a down side.
The theory predicts that the length of the plateau, and thus the $\alpha$ relaxation time
{\it diverges} at a finite temperature, $T_c$. The predicted divergence is a power law,
\be
\tau_\mathrm R = \frac{1}{(T-T_c)^\gamma}  \ .
\label{powerlaw}
\ee
If the divergence located by MCT were at very low temperature, well below $T_g$, we could not 
completely rule out its existence. However, this is not the case, and there is in fact overwhelming evidence that
such divergence is not present in experimental data, but it is rather an artefact of the theory.

Whenever the relaxation data are fitted to a power law, as in \eqref{powerlaw}, the resulting
$T_c$ is invariably and significantly larger than $T_g$, so that the $\alpha$ relaxation time
is {\it not} infinite at $T_c$. In fact, even in those systems where MCT is most successful what happens
is the following: the data follow rather well a power law, with a certain 
critical temperature $T_c$. However, the closer one gets to $T_c$, the larger the discrepancy 
between data and fit is, until right at $T_c$ such discrepancy is infinite \cite{angell88}.
Unfortunately, this means that
whenever we want to perform a MCT fit to a data-set we are in troubles. Imagine we include 
in the fit data in the range $T_1\leq T \leq T_2$, and get a certain $T_c$, which is by construction lower than $T_1$.
If we now perform again the fit using the data in the interval $T_c \leq T\leq T_2$ we are bound to obtain 
a lower critical temperature $T_c'$, since at the former $T_c$ the fit will try to stick to the data,
rather than locating a divergence. 

Why MCT predicts an inexistent divergence?
There is a common consensus that this happens because MCT in its original 
form does not take into account activated barrier crossings. 
The relaxation mechanism of MCT is something different from activation,
even though it is not easy to understand its nature by a mere inspection of the MCT equations.
Below $T_c$ the MCT dynamical mechanism is completely stuck, so that the theory cannot 
go beyond $T_c$ and it therefore locates a divergence here.
A real liquid, however, can switch from the MCT mechanism to activated barrier crossing, thus keeping
the relaxation time finite, although sharply increasing with lowering the temperature.
When we include low $T$ data in 
a MCT power law fit, we are effectively asking the theory to work in an activated regime, where it is not supposed 
to work

This considerations suggest a key connection with Goldstein's scenario. As we have seen above, 
activated dynamics only makes sense at temperatures low enough, where
potential energy barriers are substantially larger than the thermal energy, i.e. for
$T<T_x$. Above $T_x$ Goldstein's scenario breaks down because activation is 
no longer the main mechanism of relaxation.  On the other hand, MCT breaks down 
below $T_c$ because activation  becomes the main mechanism of relaxation.
This strongly suggests the identification,
\be
T_x \sim T_c  \ .
\label{goldmct}
\ee
The two theoretical frameworks, Goldstein's and MCT, therefore joins at this temperature, 
having mirroring validity regimes. Relation \eqref{goldmct} has been verified by several
investigations \cite{sokolov-98, sastry-00}. In fact, the observation done in \cite{angell88} 
that the shear relaxation time at $T_c$ is often of order $10^{-9}$ sec, i.e. the same as 
Goldstein's estimate for $\tau_\mathrm R(T_x)$, suggested relation \eqref{goldmct}, together 
with the theoretical scheme we have just discussed. For a less schematic discussion of
the crossover from non-activated to activated dynamics taking place at $T_x\sim T_c$
and of the break-down of MCT see \cite{burmer-04,bagchi-08}.

The identification of the MCT transition temperature with 
Goldstein's crossover temperature has an important implication: even though one 
knows there is no real divergence, it is nevertheless very useful to work out $T_c$ (typically 
via a power law fit of the data) as a reference temperature for a system 
approaching glassiness. In fact, from what we have discussed above one may conclude that 
the definition of $T_x\sim T_c$ is more fundamental than that of $T_g$, which depends on 
an arbitrarily fixed time or viscosity scale (see equation \eqref{glasseta}). For example,
the position of the dynamic glass transition $T_g$ is effectively very different in experiments
and numerical simulations, since in the latter we can reach much shorter time scales; on 
the other hand, the temperature where a crossover from nonactivated to activated dynamics
takes place is independent from the experimental time, so that $T_x\sim T_c$ is conceptually a 
better defined quantity. In particular, while in real experiments $T_g$ can be quite 
below $T_c$, in numerical simulations, where $T_g$ is 
pushed at much higher temperatures due to the shorter available time, the two
temperatures are typically  much closer. For this reason, it is somewhat customary
to use $T_c$ as a landmark of impeding glassiness in numerical simulations.

The scenario depicted above, and in particular equation
\eqref{goldmct}, relies on the hypothesis that $T_c$ is the limit of
validity of MCT: the theory does not include activated events, and
therefore at the temperature where there is a crossover from
non-activated (MCT) to activated (Goldstein) dynamics, MCT locates a
divergence.  However, we did not provide any solid evidence in favour
of this hypothesis, apart from saying that there is common consensus. 
In particular, given the complete lack of details
in our exposition of MCT, it is far from obvious that the theory's
failure at low $T$ is the result of disregarding activation. One could
argue that the spurious divergence at $T_c$ just means that MCT is
wrong, end of story. To investigate this point we take a rather indirect 
road, and dive into the physics of spin-glass.

After the explosion of spin-glass physics in the early 80s, a great deal of discussion 
started about the possible analogies between glass-forming liquids and spin-glasses. 
However, there was one conceptual problem: supercooled liquids display a sort of discontinuous transition, whereas
spin-glass models analyzed in the early days had a continuous nature. We have seen above 
what `discontinuous' means in the context of the dynamical correlation function of a liquid close to 
$T_g$: when the plateau develops at low $T$, its height is already quite different from zero. The discontinuous
nature of the structural glass transition is something more profound than the behaviour of the dynamic correlation 
function; it is a trait that pops out in very diverse contexts, including off-equilibrium behaviour. 
For example, when the 
system goes out of equilibrium at $T_g$ the specific heat jumps discontinuously to a smaller (crytal-like) value (Fig.7),
rather than simply changing slope. This discontinuous nature of glass-forming liquids clashed with
early spin-glass models, which had a continuous behaviour of all the corresponding observables.
In particular, the nonergodicity parameter was zero at the (spin) glass transition \cite{virasoro}.

Quite soon, however, it was discovered a new class of mean-field spin-glass models that displayed a discontinuous 
glass transition. These models included the random energy model \cite{derrida-81}, the $p$-spin model 
\cite{gross-84}, and the Pott's spin-glass model \cite{gross-85}. In the following we will use the 
$p$-spin models as a representative of this entire class of systems. 
It was Kirkpatrick and Wolynes who first noted in \cite{kirk-0} the close analogy between the
discontinuous transition in this class of spin-glasses and that of glass-forming liquids. They also noted the similarity
between dynamic spin-glass theory and mode coupling theory. In a subsequent series of remarkable papers 
\cite{kirk-1,kirk-2,kirk-chi4, kirk-3}, Kirkpatrick, Thirumalai and Wolynes reached a 
deep understanding of this class of spin-glasses and of its connections with structural glasses.

The Hamiltonian of the $p$-spin model for $p=3$ is the following,
\be
H=-\sum_{i, j, k=1}^N J_{ijk}\; \sigma_i \sigma_j\sigma_k  \ .
\label{pspin}
\ee
The degrees of freedom $\sigma$ are spins (they may be real or integer variables) 
and the couplings $J$ are quenched random variables \cite{cavagna-review}. 
The model is mean-field because all spins interact with all other spins: the couplings $J$ do not decay with the distance between 
spins. This is therefore a model where there is no underlying lattice, nor space structure at all,
where each spin is equally close (or distant) to every other spin (such mean-field models are also called
fully connected or infinite dimensional models; for this reason, normal, non-mean-field systems are sometimes 
called finite dimensional systems). We wrote explicitly Hamiltonian \eqref{pspin} just to make it clear how far is this model
from a real liquid, whose Hamiltonian looks something like,
\be
H=\sum_{i, j=1}^N V(||\vec r_i -\vec r_j||)  \ ,
\ee
where the degrees of freedom are the positions $\vec r_i$ of the particles and 
$V(r)$ is a pair-wise interaction potential. 

It therefore comes somewhat as a surprise 
to find some impressive analogies between the phenomenology of the $p$-spin model and that
of real supercooled liquids. Thanks to its mean-field nature, the Langevin dynamics of model
\eqref{pspin} can be studied analytically \cite{kirk-2}, 
and in particular one can write a self-consistent equation 
for the correlation function,
\be
C(t)= \frac{1}{N}\sum_i \langle \sigma_i(t_0)\sigma_i(t_0+t)\rangle \ ,
\label{corrspin}
\ee
which at equilibrium only depends on the time-difference $t$. This function is large when the two 
configurations $\sigma(t_0)$ and $\sigma(t_0+t)$ are very similar, and it decays to zero when they
becomes completely decorrelated. What is found from the equations is that at low temperatures the
correlation function of the $p$-spin model develops a plateau, giving rise to the (now familiar) two steps
relaxation pattern. As in real liquids and in MCT, the length of the plateau increases when decreasing the temperature
and the formation of the plateau has a discontinuous nature.
Unlike real liquids, however, but similar to MCT, the length of the plateau (and thus the relaxation time $\tau$) 
diverges as a power law at a finite temperature $T_c$, called the {\it dynamical transition} in the spin-glass literature 
\footnote{The dynamical transition is sometimes indicated with $T_d$ in the spin-glass literature.
Here we call it $T_c$ in analogy with the MCT transition, in order to keep as small as possible the number of 
different labels for the same concept.}.

The interesting thing is that no thermodynamic (static) divergence,
nor anomaly takes place at $T_c$: the specific heat, free energy and
so on show nothing noteworthy around $T_c$: from a standard
thermodynamic treatment of the model, it is impossible to detect the
dynamical transition. What happens at $T_c$ is that the system remains
dynamically trapped within metastable states. Due to the fully
connected interaction, in a mean-field model metastable states are
surrounded by infinite free energy barriers that the system cannot
surpass by using thermal activation \cite{cavagna-review}. For this reason in
mean-field metastable states have an infinite lifetime and can be
sharply defined.  The system is therefore forbidden to restore
ergodicity, and a true dynamical divergence occurs. Of course,
metastable states are irrelevant from a thermodynamic point of view,
since their individual weight is negligible. For this reason $T_c$
goes undetected by a standard thermodynamic investigation. I strongly
encourage the reader to see the original papers by Kirkpatrick, Thirumalai
and Wolynes about the nature of the dynamic transition at $T_c$, and
in particular the illuminating discussion at the end of Ref.\cite{kirk-chi4}.

The fact that two steps relaxation in the $p$-spin models ends up in a
true dynamical transition at $T_c$ is  quite similar to what
happens at the corresponding $T_c$ in the context of MCT.  Moreover,
this divergence is described by a power law both in MCT and in the
$p$-spin.  All these coincidences are not fortuitous. In fact, the
dynamical equations for the correlation function analytically found in
the $p$-spin model are formally identical to those formulated by
MCT \cite{kirk-2,bouchaud-96}.  This is what really brought together for the
first time the supercooled liquids and spin-glass communities. 
MCT writes a set of dynamical equations that are formally 
identical to those describing the exact dynamics of a mean-field
model, where, by definition, activation is forbidden. 
Even though far from a theorem, it seems reasonable
to conclude that in doing its approximations MCT lost the ability to
describe activated events. For this reason MCT predicts a
divergence at $T_c$, which is formally completely equivalent to $T_c$
in the $p$-spin. Following the same line of thought, the nonactivated
relaxation mechanism of MCT and  that ruling the dynamics of the $p$-spin 
above $T_c$ are probably similar.

At this point one may ask: given that the $p$-spin model is so similar
to MCT, why do we need to bother about it? After all, our dish was
already full enough! The reason why it is useful to consider both
these theoretical frameworks is that they
originate from completely different starting points, and this helps
us a lot in understanding what is going on. MCT is an approximated
theory of a real system staged in real space; the cage effect is the
pivotal interpretation of two steps relaxation in terms of real
particles. The correlation function leaves the plateau when particles get 
out of their cage. The $p$-spin model, on the other
hand, has no real space structure: no
distance, no neighbouring spins (particles), and thus no cage,
decaging, and all that. Any reasonable interpretation of the dynamical
slowing down for the $p$-spin model must be staged in phase space, not in
real space: activation is forbidden by construction, not because of an
approximation.  

Despite these differences, the phenomenology and the
formal structure of the dynamical equations of the $p$-spin model are
identical to MCT. It is clear that understanding what is the dynamical
relaxation mechanism in the $p$-spin above $T_c$, can be very helpful
to clarify what is the origin of two steps relaxation and decaging in
MCT, and perhaps in real liquids too. Indeed, whereas MCT is somewhat the
pinnacle of the analytic effort in real liquids, and it is very hard
to go beyond it, the big advantage of the $p$-spin is that, being a
mean-field model, many calculations can be performed exactly, and its
physical behaviour can be understood in great detail.

\subsection{From minima to saddles}

To understand the origin of the dynamical singularity in the $p$-spin
model we have to investigate the topology of the potential energy
surface visited by the system at a certain temperature $T$.  
Roughly speaking, we shall see that Goldstein's activated scenario
corresponds to a motion from minimum to minimum in the phase space,
whereas MCT/$p$-spin nonactivated dynamics corresponds to a motion
from saddle to saddle. Most of the following concepts are treated in
more detail in \cite{cavagna-review}, where all the relevant
references may be found.

Let us start below the dynamical transition, $T<T_c$. In the $p$-spin model 
we can (analytically) equilibrate a configuration in this temperature region,
and compute the dynamical correlation function \cite{burioni}. If we
do this what we find is that for large times $C(t)$ approaches a
plateau and it never leaves it. In other words, after a brief $\beta$
relaxation, the correlation function does not show any $\alpha$
relaxation, since the $\alpha$ relaxation time is infinite. This is
due to the fact that ergodicity is broken, and the system never leaves
the metastable state where it has been initially equilibrated. The
system is at equilibrium within this particular state,
but it cannot jump out, because free energy barriers are
infinite. Thus, ergodicity is broken and there is no long term
relaxation.

The nice thing about the $p$-spin model is that the topological
properties of the minima trapping the dynamics for $T<T_c$ can be 
worked out very precisely \cite{cavagna-review}. By topological properties we mean
essentially two things: the {\it height} of the minimum in the
landscape, i.e. the energy density $E$ of the bottom of the well, and
the {\it shape} of the minimum in the $N$-dimensional phase space,
where $N$ is the number of degrees of freedom.  In a minimum, by
definition, the gradient of the potential (i.e. the force) is zero,
and thus it is the matrix of the second derivatives, the so-called
Hessian, that contains much of the information about the shape. The
$N$ eigenvalues of the Hessian give the curvature along the $N$
directions (eigenvectors) out of the minimum. By definition  of minimum, 
the eigenvalues are all positive, but their values tells us how soft
or stiff is the minimum: a large eigenvalue $\lambda_i$ means that the
energy climbs very rapidly along direction $i$, whereas a small (but
positive) eigenvalue $\lambda_i$ tells us that direction $i$ is almost
flat (at least at the second order of the expansion). Such soft
directions are called {\it marginal} when the eigenvalue is 
zero \footnote{Each continuous symmetry of the system, i.e. each
continuous transformation that leaves the energy invariant, generates
a null (marginal) eigenvalue of the Hessian. For example, when
translational invariance holds in $d=3$ there are $3$ zero eigenvalues
of the Hessian, corresponding to the three directions along which the
system can be moved without changing the energy. Clearly, these
directions correspond to trivial global movements of all degrees of
freedom, and do not contribute to the relaxation of the system. To fix
ideas, it is better to assume that all continuous symmetries are
broken, and that the Hessian has no trivial zero modes.}. 

A compact way to assess the eigenvalues of the Hessian is to define the
spectrum, i.e. the fraction of the eigenvalues equal to a particular
value $\lambda$, 
\be 
\rho(\lambda) = \frac{1}{N}\sum_{i=1}^N
\delta(\lambda-\lambda_i) \ .  
\ee 
The question now is: what is the typical spectrum
$\rho(\lambda)$ of the minima trapping the dynamics for $T<T_c$?

The spectrum $\rho(\lambda)$ can be computed exactly in the $p$-spin model 
\cite{cavagna-review}. More
precisely, we can compute the average spectrum of all minima with a
given bare energy $E$, that is $\rho(\lambda;E)$, and see how it
changes with $E$. It can be shown that at each $E$ the spectrum has a
compact support (Fig.13), with a
minimum eigenvalue $\lambda_\mathrm{min}$, the so-called lower band edge
\footnote{The spectrum of the Hessian in the $p$-spin model is a
semicircle. This is nothing else than the Wigner law for the spectrum
of Gaussian random matrices. The reason for this is that the random
couplings $J_{ijk}$ are Gaussian in this model, and this reflects
on the properties of the Hessian matrix.}.

\begin{figure}
\includegraphics[clip,width=3.8in]{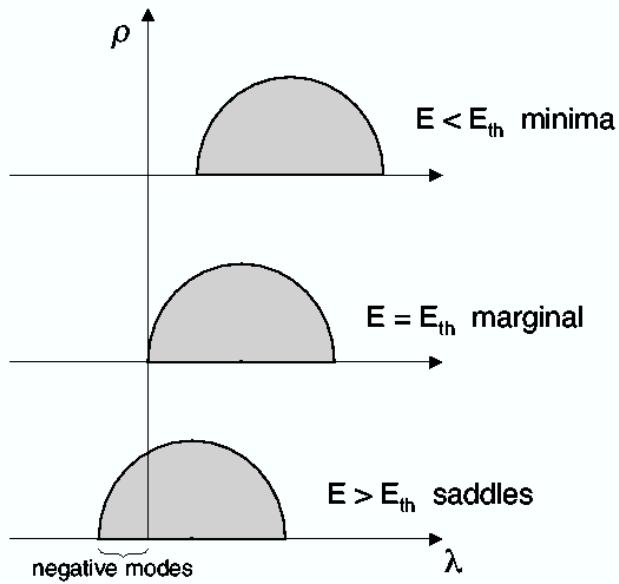}
\caption{ {\bf The spectrum of the Hessian} -
In the $p$-spin model the spectrum $\rho(\lambda; E)$ of the Hessian can be computed exactly. This
quantity gives the distribution of the eigenvalues of the stationary points with energy $E$. For $E<E_\mathrm{th}$
the spectrum is entirely contained in the positive semi-axis. As $E$ increases the spectrum moves to the
left, until for $E=E_\mathrm{th}$ the lower band edge touches the zero. This means that threshold
minima are marginal. For larger energies the spectrum has a finite part in the negative semi-axis,
i.e. the typical stationary point has a finite fraction of negative eigenvalues, and it is thus
an unstable saddle.}
\end{figure}

The crucial thing is that, when
the energy $E$ increases the whole spectrum shifts to the left,
i.e. towards lower values of $\lambda$ (Fig.13). In other words, by
increasing $E$ the lower band edge $\lambda_\mathrm{min}$ approaches zero,
meaning that higher energy minima are softer than lower energy
ones. At a certain energy, called {\it threshold energy}
$E_\mathrm{th}$, the lower band edge of the spectrum touches zero, 
\be
\lambda_\mathrm{min}(E_\mathrm{th}) = 0 \ .  
\ee 
Therefore, threshold
minima have a marginal direction, i.e. a flat direction along which a
second order displacement costs no energy. The important point is that
the spectrum is continuous, so this marginal direction is not
isolated: many eigenvalues are very small, even though not precisely
zero, implying that many directions within a threshold state are {\it
almost} flat. 
Furthermore, in the energy regime $E>E_\mathrm{th}$ the spectrum has a
support partly contained in the negative semiaxis (Fig.13). This
means that a nonzero fraction of eigenvalues of the Hessian is
negative: the typical stationary point with bare energy $E$ above the
threshold is thus an unstable {\it saddle} \cite{cgp-98,cgp-01}.

The interpretation of this result is the following: minima in the
$p$-spin model are less trapping the larger is their energy, until at
the threshold energy they open up along the marginal direction and
turn into something different, namely unstable saddle points. 
Therefore, at the threshold
energy a topological transition occurs between a region of the
landscape mainly populated by stable minima ($E<E_\mathrm{th}$), and a
region mainly populated by unstable saddles ($E>E_\mathrm{th}$).  We
stress once again that for the $p$-spin model (for which MCT dynamical 
equations are exact) this topological transition is
found analytically, without any approximation.

What the topological transition has to do with the real dynamics of
the system, and in particular with the dynamical transition at $T_c$?
After all, there is nothing dynamic nor thermal in the energy
landscape: it is the same at all temperatures and times. However,
the properties of the landscape depend on the energy level $E$ at 
which we explore it, and this in turn depends on the temperature. 
We must only be careful to a point: when we consider a minimum with energy $E$, 
this is the energy density of the bottom of the well; but at finite temperature the system vibrates around 
the minimum, so that the average potential energy density of the system 
will be larger than $E$ by factor roughly equal to $1/2 \; k_B T$. We will refer to
$E$ as the {\it bare} energy of the system. At low temperature, the full average energy $\langle H\rangle$
is equal to the bare energy plus thermal fluctuations,
\be
\langle H \rangle \sim E +\frac{1}{2} k_B T   \ .
\label{bare}
\ee
The bare energy $E$ is a better label than $\langle H\rangle$ because it is more directly connected 
to the topological properties of the stationary point. By $E(T)$ we indicate the bare energy of the typical minima
trapping the system at temperature $T<T_c$. In the $p$-spin we can compute exactly the curve $E(T)$: 
as expected, the bare energy grows with the temperature, meaning that the system is trapped by higher energy 
minima the higher the temperature. If this is not particularly exciting, the following result is,
\be 
E(T_c) = E_\mathrm{th} \ .
\label{tatanka}
\ee 
The bare energy of the minima visited by the system at
the dynamical transition $T_c$ is exactly equal to the threshold energy. 
Moreover, it is possible to prove that
for $T>T_c$ the closest stationary point to an equilibrium
configuration is not a minimum, but a saddle; the fraction 
of unstable (negative) eigenvalues of this saddle goes to
zero for $T\to T_c^+$ \cite{cgp-01}.

The nature of the dynamical transition in the $p$-spin model 
is now clear: at equilibrium for $T<T_c$ the system is
trapped within minima of the energy, with all the eigenvalues of the
Hessian larger than zero; the system never escapes, and the dynamical
correlation function $C(t)$ does not decay to zero; the plateau and
the $\alpha$ relaxation time have infinite length; ergodicity is
broken.  For $T>T_c$, on the contrary, the typical stationary point is
a saddle, with a finite fraction of negative eigenvalues; saddles do not trap
the dynamics for infinite times, since sooner or later, by thermal
fluctuations the system finds the negative modes and leaves the
saddle; this is {\it not} an activated mechanism; the dynamic
correlation function decays to zero and the $\alpha$ relaxation time
is finite; ergodicity is not broken. 
Even though the dynamical transition at 
$T_c$ is the most visible effect of the border between these two
regimes, the fundamental transition takes place at $E_\mathrm{th}$: the
thermal transition is just a manifestation of the topological one.

Note that in the $p$-spin model
for $T<T_c$ it is possible to sharply define the
nonergodicity parameter $C^\star(T)$ as the infinite time limit of the correlation 
function once the system is thermalized within one state,
\be
C^\star(T) = \lim_{t\to\infty} C(t,T)  \ .
\ee
The discontinuous nature of the transition amounts to say that,
\be
\lim_{T\to T_c^-} C^\star(T) \neq 0  \ .
\ee

The topological transition between minima and saddles not only gives
us an explanation of the ergodicity breaking at $T_c$, but it also
clarifies the dynamical behaviour slightly {\it above} the
transition. 
In this regime, as we have seen, the dynamical correlation
function is characterised by two steps relaxation, with a plateau
whose length increases when $T_c$ is approached from above. We now
know that this plateau is not due to local minima trapping the system:
in this temperature regime unstable saddles, and not minima, are the
stationary points most relevant for the dynamics \cite{cgp-01}. However, for $T\sim
T_c$, and thus $E\sim E_\mathrm{th}$, these saddles have very few
negative eigenvalues; moreover, due to the fact that the spectrum is
continuous, these negative eigenvalues are also very close to zero. In
this condition, it is entropically very difficult for the system to
find the way out of the saddle, and this is the fundamental reason why
the plateau forms \cite{single-saddle}.

The two steps relaxation can be interpreted in the following
way. First, the system relaxes exponentially along the many positive
eigenvalues, $\lambda_i>0$, that is along those directions with the
steepest positive curvature, giving rise to the $\beta$ relaxation and
the approach to the plateau. In this early phase there is little
difference with the relaxation within a minimum, since only the
positive modes dominate; given that we are close to the threshold, the
positive eigenvalues are the overwhelming majority.  For later times,
the effect of the few, almost marginal, negative modes becomes
important, and the system can slowly leave the saddle along this sort
of flat tunnel \cite{laloux}. 
This is the $\alpha$ relaxation: it does
not correspond to an activated event, nor to a jump over a potential
energy barrier; rather, it is the time the system needs to fully
exploit the negative modes that will lead it out of the saddle, thus
restoring ergodicity. The closer the temperature is to $T_c$, the
closer the bare energy is to $E_\mathrm{th}$, and the smaller the
fraction of negative eigenvalues of the Hessian.  This is the reason
why the $\alpha$ relaxation time (and thus the plateau) increases when
approaching $T_c$: it is a direct consequence of the fact that the
fraction of negative eigenvalues vanishes at the threshold.

The $p$-spin model therefore provides in natural way a nonactivated
mechanism of diffusion, whose slowing down is responsible for the
appearance of the two steps relaxation in the correlation function. We
recall that the exact dynamical equations of this model for $T>T_c$
are formally identical to those of MCT in the same temperature regime. 
Given that  MCT describes
supercooled liquids reasonably well for $T>T_c$, it is natural then to ask whether the nonactivated
diffusion mechanism of the $p$-spin model above
$T_c$ is valid also in supercooled liquids. Is the
decreasing number of unstable directions of saddle points responsible
for the two steps relaxation, and the corresponding slowing down, in
the dynamics of supercooled liquids?

\subsection{Topological origin of the glass transition}

Saddles of the potential energy are present also in finite dimensional
systems as real liquids, of course.  However, after Goldtein's paper
in 1969 \cite{gold} numerical investigations focused for a long time
only on potential energy minima. In particular, given an equilibrium
configuration of the system, it is possible by following the gradient
of the potential energy, to associate a unique local minimum to the
initial equilibrium configuration \cite{stillinger-weber}. This local
minimum is also called {\it inherent structure} in the literature. The
partition of the phase space into basins of the local minima is well
defined and unique, and it suites perfectly Goldstein's scenario of an
activated dynamics between different basins. This was the reason why
local potential energy minima have largely dominated the analysis of
supercooled liquids for quite a while. It has been indeed thanks to
such numerical investigations that it has been possible to validate
Goldstein's picture \cite{sastry-00} and to understand many important
properties of the dynamics of supercooled liquids at low temperatures
\cite{angell-95, stillinger-95, sastry-98, wales-04, sciortino-05}.

However, we know that for $T>T_x\sim T_c$ the description of the
dynamics in terms of minima is not appropriate, since activation is no
longer the main mechanism of diffusion in this regime. We also know
that two steps relaxation is anyway present  for $T>T_c$, and that
MCT reproduces it. Finally, we have seen that the $p$-spin
model gives a neat framework to explain nonactivated two steps
relaxation: the cornerstone of this framework is the topological
transition between minima and saddles taking place in the energy
landscape. All these considerations suggest
that in order to better understand the crossover between nonactivated
to activated dynamics at $T_x$ we need to study the properties of saddles 
of the potential energy.

A first attempt to exploit unstable modes of the potential energy to
study the diffusion properties of supercooled liquids was done by the
Instantaneous Normal Mode (INM) approach \cite{inm}.  In this context,
one would compute the fraction of negative eigenvalues of the Hessian
averaged over all the instantaneous equilibrium configuration visited
by the system.  The extrapolated temperature where the fraction of
negative modes goes to zero should give an estimate of the glass
transition.  Despite the great interest it stirred, the INM method is
affected by a big problem: many of the negative modes visited by an
instantaneous equilibrium configuration have nothing to do with
diffusion, so that the INM negative modes do not correlate very well
to the slowing down of the dynamics. What one should do is to compute
the negative modes in correspondence of a saddle point, whereas the
INM focuses on instantaneous equilibrium configurations that, due to
thermal fluctuations, are quite far from any stationary point.  For
this very reason we considered the bare energy, $E$, rather than the
average energy, $\langle H\rangle$, in the $p$-spin. Therefore, to go
beyond both Goldstein's picture and the INM approach, we must study
the properties of unstable saddles in supercooled liquids. 
The first discussion of the role of saddles in supercooled liquids was 
done in \cite{cavagna-99}, building on the study of saddles carried out 
for the $p$-spin model in \cite{cgp-98} and more in general on the 
topological approach to glassy dynamics discussed in \cite{laloux}.

Finding unstable saddles numerically is less trivial than finding minima, but it
can be done. The real problem, however, is that there is no unique
way to associate to an equilibrium configuration a
saddle. This is unlike what happens with minima, where the gradient descent
method uniquely partition the phase space into basins.  In the case of
saddles, the mapping between equilibrium configuration and saddle
depends on the particular algorithm one uses \cite{grigera-06}, so that
there is not a natural way to partition the phase space into
basins of unstable stationary points.  This is not small a problem,
because it means that when we say sentences like,
`the typical saddle visited by the system at temperature $T$', we are
in fact using an ill-defined concept.

There are two ways to
face this problem. First, we can stick to one particular mapping (algorithm), and plot all
saddle properties as a function of the temperature $T$ of the equilibrium 
configurations used to generate the saddles \cite{angelani-00}. This 
first possibility is a natural generalization of the INM approach, 
with saddles in place of instantaneous configurations. The second possibility 
is to use a method that relies as little as possible on the mapping between
equilibrium configurations and saddles \cite{broderix-00}. The difference 
between the two methods is critically discussed in \cite{grigera-06}.
Here, we will explore the second method, which is the one that most directly
connects to the $p$-spin scenario.

The fundamental quantity in the $p$-spin is the threshold energy $E_\mathrm{th}$,
marking the border between minima and saddles. To define
$E_\mathrm{th}$ we do not need to mess with configurations at
thermal equilibrium: the threshold energy is a purely geometric quantity,
which describes the topology of the potential energy
surface. To find the threshold in a realistic model we can sample
saddles using {\it any} reasonable algorithm and for each saddle measure its
bare energy density $E$ and its density $k$ of unstable directions, namely
the number of negative eigenvalues of the Hessian divided by the
number of degrees of freedom. If we plot these two numbers on the
$(E,k)$ plane for each saddle, we obtain the graph in Fig.14:
stationary points are not found anywhere in the $(E,k)$
plane, but they are concentrated around a backbone, which forms an
approximately linear, monotonously increasing function \cite{broderix-00,grigera-02,grigera-06}. 
This is very reasonable: the larger the energy of the saddle, the more unstable
this is. 

The interesting point is that the temperature at which we sample
saddles (as well as the algorithm we use to do this), only changes the part
of this backbone that we uncover, but not the backbone itself, which 
is a property of the landscape, independent of the temperature. Besides, this backbone
is very weakly dependent on the sampling method: it has indeed been confirmed that 
different numerical algorithms used to sample saddles give rise to the same backbone in
the $(E,k)$ plane \cite{grigera-06}. The same, however, is not true if we use the generalization of
the INM approach mentioned above: when we compute the average 
index $\bar k$ as a function of the temperature $T$ at which we are sampling 
saddles, we obtain a curve that strongly depends on the sampling algorithm
\cite{grigera-06}. This fact seems a good reason to prefer the `geometric'
$(E,k)$ approach, with respect to the `thermal' $\bar k (T)$ approach.

\begin{figure}
\includegraphics[clip,width=3.4in]{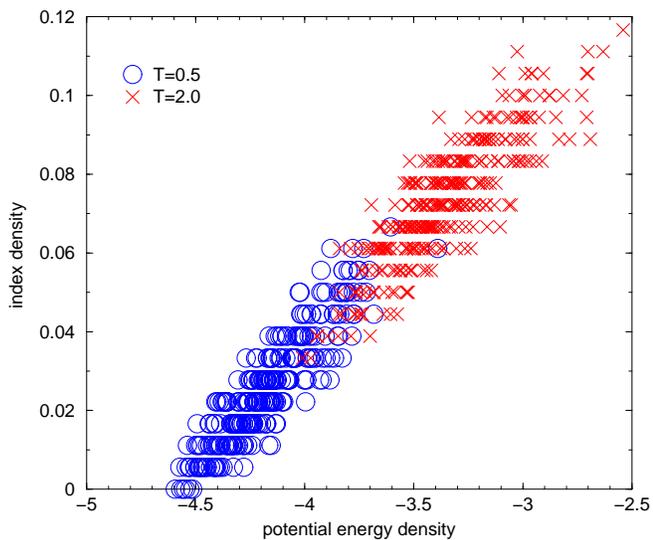}
\caption{ {\bf Instability index vs. energy of saddles} - For each
sampled saddle one measures its energy density $E$ and the number of
negative eigenvalues of the Hessian, divided by $N$, i.e. the
instability index (density) $k$. These two values are reported for
each saddle in the $(E,k)$ plane in the figure. The scatter plot
cluster around a backbone curve $k(E)$. Changing the temperature
amounts to change the region of the curve we are sampling, but it does
not change the curve itself. The system is a Lennard-Jones binary mixture.
 (Reprinted with permission from
\cite{broderix-00}; copyright of American Physical Society).  }
\end{figure}

\begin{figure}
\includegraphics[clip,width=3.4in]{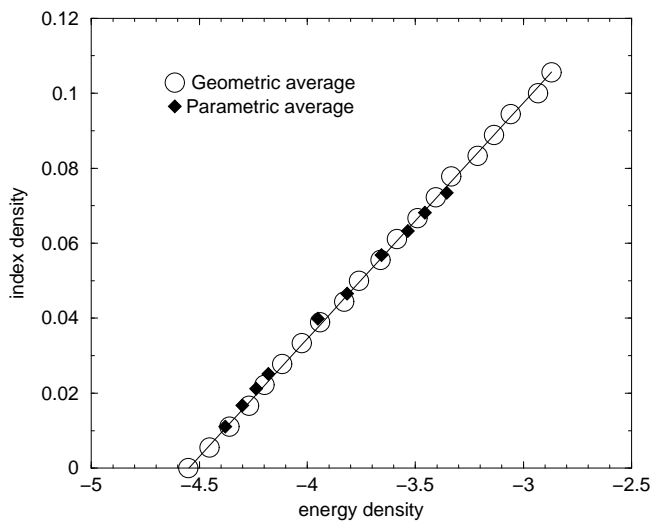}
\caption{ {\bf Average instability index vs. energy of saddles} - 
In order to identify a threshold energy one must average the curve in Fig.14. This can be done in different,
although at the end equivalent, ways (see \cite{broderix-00} for details). The point where $k(E)$ goes
to zero identifies the threshold energy. The system is a Lennard-Jones binary mixture.
(Reprinted with permission from \cite{broderix-00}; copyright of American Physical Society).  }
\end{figure}

When we average the energy of all stationary points with the same
instability index $k$, we obtain the curve in Fig.15 \cite{broderix-00,grigera-02,grigera-06}. 
This curve
identifies a threshold energy, $E_\mathrm{th}$, as the point where
$k(E)$ goes to zero \footnote{We have to be careful here. As we mentioned, by changing 
the temperature at which we are sampling saddles, we simply 
highlight different portions of the scatter plot in the $k(E)$
curve. However, if we thermalize the system at low
temperature, any saddle-searching algorithm is bound to find
mainly minima ($k=0$); if the temperature is {\it very} low, these
minima will have lower and lower energy, moving the point at
$k=0$ closer to the ground state, which is {\it different} from the threshold.
For this reason, the $k=0$ point of the $k(E)$ curve is biased
by the lowest temperature used for the sampling. This is not true for 
the part of the curve at larger $k$. We conclude that, in general,
the threshold energy is better obtained by the extrapolation 
of the $k>0$ part of the curve, rather than by its
endpoint at $k=0$.}, 
\be 
k(E_\mathrm{th})=0 \ .  
\ee 
From equation \eqref{tatanka} we know that in the $p$-spin model the dynamical
transition corresponds to the temperature where the bare equilibrium
energy reaches the threshold. To test this conclusion in real liquids we
must define in some way the bare potential energy. In accordance with
equation \eqref{bare}, this can be done by subtracting from the
average potential energy $\langle U \rangle$ at temperature $T$, the
contribution of the thermal vibrations, 
\be 
E(T)= \langle U \rangle -\frac{3}{2} k_\mathrm B T 
\label{il3}
\ee 
The bare energy $E(T)$ can thus be easily
measured numerically, and this measure is completely independent
from the determination of the threshold
\footnote{The reader should not get confused by the difference between
the factors $1/2 \  k_B T$ in \eqref{bare} and $3/2 \ k_B T$ in \eqref{il3}.
Each spin in the $p$-spin model has $1$ degree of freedom, so that in the energy per spin,
equipartition gives a factor $1/2 \ k_B T$. On the other hand, in simple
monoatomic liquids each particle has $3$ translational degrees of freedom,
so that the energy per particle has a factor $3/2 \ k_B T$. Moreover, the potential energy 
$U$ for a liquid has exactly the same role as the hamiltonian $H$ for the $p$-spin.}. 
Inspired by equation
\eqref{tatanka}, we can now define the temperature $T_\mathrm{th}$
where the bare energy crosses the threshold energy \cite{broderix-00, grigera-02}, 
\be
E(T_\mathrm{th}) = E_\mathrm{th} \ , 
\label{titozuro}
\ee 
and check whether $T_\mathrm{th}$ is in any way relevant for the system. 
We  know that, unlike in the $p$-spin model, in a realistic liquid
there is no true dynamic transition. However, we also know that mode
coupling theory predicts a sharp transition at $T_c$, as a consequence
of the fact that below this temperature activated phenomena,
i.e. jumping between different minima, becomes the main mechanism of
diffusion. This is the reason why $T_c$ has been identified with the
crossover temperature of Goldstein, $T_x$: below $T_c$ the
dynamics is dominated by activated jumps between minima; above
$T_c$ a new, nonactivated mechanism kicks in, and the analysis
of this chapter suggests that this may be a saddle ruled mechanism.

The temperature $T_\mathrm{th}$ defined in \eqref{titozuro} is the thermal
counterpart of the topological transition between minima and saddles
at $E_\mathrm{th}$, and thus, in order for this saddle ruled scenario
to be coherent, it is necessary that the following relation holds, 
\be
T_\mathrm{th} \sim T_c  \ . 
\label{ulala}
\ee 
This relationship between threshold temperature and Mode Coupling
temperature has been tested in a series of numerical investigations,
either in the geometric framework discussed here \cite{broderix-00, 
grigera-02}, or in a thermal framework, where $\bar k (T)$
is computed \cite{angelani-00, charu-01, charu-02}. In all cases the result of
the test has been positive: the threshold temperature is within few
percents equal to the mode coupling temperature. We must note that
$T_c$ is determined by measuring dynamical observables, whereas
$T_\mathrm{th}$ has a completely different (topological) origin. The
fact that they are so close indeed supports the view that the
dynamical crossover at the glass transition is in fact the most
visible effect of a more fundamental topological transition, taking
place in the phase space.

The scenario we have presented in these last sections gives a topological
interpretation of the dynamic crossover from nonactivated to activated
dynamics that takes place at $T_x\sim T_c$. This is the point where the systems
stops visiting saddles, and thus runs out of unstable eigenvalues, which 
are equivalent to nonactivated diffusive modes. Below the crossover, the system 
spends most part of the time around minima, rather than around saddles, 
and an activated Goldstein-like diffusion mechanism kicks in.

Yet, we know that $T_x > T_g$ and we also know that in the temperature 
region $T_g < T < T_x$ fragile systems have a steep super-Arrhenius increase of the relaxation time.
What can we say about this regime? Very little in this context. The topological approach
is rooted in phase space, and knows nothing about the size of energy barriers, which is
determined by the number of particles that really make the transition in real space.
Numerical studies in fragile glass-formers \cite{broderix-00, grigera-02} show that at $T_x$, where 
the topological transition takes place, the size of the barriers is already quite large, typically,
\be
\Delta U(T_x)  \sim 10 \; k_B T  \ .
\ee
This observation indicates that at the temperature $T_x$ where for first time the system must
use activation, this mechanism is already very slow, because barriers are
already quite large. This explains the sharp increase of $\tau_\mathrm R$ on approaching 
$T_x$. In particular, the different size of $\Delta U$ at $T_x$
can rationalize the difference between fragile (large $\Delta U(T_x)$) and 
strong (small $\Delta U(T_x)$) systems \cite{cavagna-99}.
However, in order to understand why there is super-Arrhenius behaviour
below $T_x$ one must understand why and how $\Delta U$ increases when $T$ decreases. Both the 
topological approach and MCT have nothing to say about this, because they cannot go below $T_x$.
To continue in our exploration we therefore have to go to the deeply supercooled phase.

%%%%%%%%%%%%%%%%%%%%%%%%%%%%%%%%%%%%%%%%%%%%%%%%%%%%%%%%%%%%%%%%%%%%%%%%%%%%%%%%%%%%%%%%%%%%%%%
%%%%%%%%%%%%%%%%%%%%%%%%%%%%%%%%%%%%%%%%%%%%%%%%%%%%%%%%%%%%%%%%%%%%%%%%%%%%%%%%%%%%%%%%%%%%%%%
%%%%%%%%%%%%%%%%%%%%%%%%%%%%%%%%%%%%%%%%%%%%%%%%%%%%%%%%%%%%%%%%%%%%%%%%%%%%%%%%%%%%%%%%%%%%%%%
%%%%%%%%%%%%%%%%%%%%%%%%%%%%%%%%%%%%%%%%%%%%%%%%%%%%%%%%%%%%%%%%%%%%%%%%%%%%%%%%%%%%%%%%%%%%%%%
%%%%%%%%%%%%%%%%%%%%%%%%%%%%%%%%%%%%%%%%%%%%%%%%%%%%%%%%%%%%%%%%%%%%%%%%%%%%%%%%%%%%%%%%%%%%%%%
%%%%%%%%%%%%%%%%%%%%%%%%%%%%%%%%%%%%%%%%%%%%%%%%%%%%%%%%%%%%%%%%%%%%%%%%%%%%%%%%%%%%%%%%%%%%%%%
%%%%%%%%%%%%%%%%%%%%%%%%%%%%%%%%%%%%%%%%%%%%%%%%%%%%%%%%%%%%%%%%%%%%%%%%%%%%%%%%%%%%%%%%%%%%%%%

\section{Going deeply supercooled}

One of the main conclusions of the former chapter is that at the Goldstein's
temperature $T_x$ there is a crossover from high-$T$ nonactivated
 to low-$T$ activated dynamics, where barrier crossing
becomes the main mechanism of diffusion. This slows down dramatically
the dynamics, so much that at $T_g$ the relaxation time exceeds the
experimental time and the dynamic glass transition occurs. Clearly
$T_g < T_x$, but whether or not the two temperatures are close to each
other depends on how large the potential energy barriers are at $T_x$,
when activation kicks in, and to what extent they increase with lowering
the temperature. 

At this stage of our study we do not have a clear idea about the precise nature
of these potential energy barriers. Up to now we just assumed that
they exist: there are minima in the potential energy landscape, and in
order to visit different minima in the attempt of being ergodic the
system will have to cross the barriers connecting these minima. Now
this explanation is no longer sufficient. Once we have understood that 
activation is the central diffusion mechanism at low temperatures, 
it is essential to get a clearer understanding about barriers.

In order to do this we must first focus our attention on the empirical evidence
in the low $T$ regime. Moreover, even though, by definition, we have no equilibrium
data below $T_g$, the extrapolation  down in the
$T<T_g$ region is very interesting. Indeed, it was such an
extrapolation that stimulated a rather famous observation by Kauzmann
in 1948.

\subsection{Entropy crisis}

Let us quote the remarkable paper that 
Kauzmann wrote in 1948 \cite{kauzmann48}:
{\it ``The vitreous or glassy state of liquids evidently only exists
because experiments performed by mortal beings must of necessity be of
limited duration. It is interesting to speculate on the behaviour which
liquids would show at very low temperatures if enough time could be
allowed in the thermodynamic measurement to avoid vitrification.''}

Basically, Kauzmann says that it seems restrictive not to investigate what happens
at low temperatures solely because we do not live long enough to make 
an equilibrium measurement. What we can do, therefore, is to 
make an {\it extrapolation} of equilibrium data, and cross our fingers.
This is what Kauzmann did, by performing a low-$T$ extrapolation of various
thermodynamic quantities. In particular, Kauzmann was interested in the 
metastable nature of the liquid phase with respect to the crystal, and
thus he plotted the liquid-crystal difference of quantities such as 
enthalpy, entropy, free energy and specific volume.

Let us focus our attention on the entropy. The liquid entropy decreases much 
more rapidly than the crystal one. This is obvious, since the derivative 
of the entropy is the specific heat,
\be
\frac{dS}{dT}=\frac{c_p(T)}{T}  
\ee
and the specific heat of the liquid is larger than that of the crystal.
We therefore conclude that the difference between
liquid and crystal, also called {\it excess entropy} of the liquid, 
\be 
\Delta S(T) = S_\mathrm{LQ}(T) - S_\mathrm{CR}(T)  \ ,
\ee
decreases when the temperature is decreased.
If we normalize this quantity by its melting point value, i.e. if we plot 
$\Delta S(T)/ \Delta S(T_m)$ vs. $T/T_m$ we can inspect many different liquids 
on the same diagram (Fig.16 is the original plot of Kauzmann's paper). 
What immediately catches our attention is the fact that 
for some systems the extrapolated excess entropy seems to vanish at finite 
temperature. Abiding to the common convention, we will 
call this the Kauzmann temperature, and indicate it as $T_k$ (we shall see later that  
this is in fact a very poor conventional choice). 
Thus, if we trust the extrapolation, we conclude that
there is a temperature $T_k$ such that,
\be
\Delta S(T_k) = 0  \ .
\ee
As we have seen, extrapolations suggest that $T_k$ is 
finite for many substances. This is equivalent to say
that the entropy of the supercooled liquid becomes lower than the entropy of the crystal 
for $T<T_k$. This phenomenon is known as {\it entropy crisis} or {\it Kauzmann's paradox}.
If we think about it, a metastable liquid with less entropy than its relative 
stable crystal is sort of counterintuitive. Even though there is no fundamental law of nature forbidding
this fact, it certainly stirs our curiosity about the significance of $T_k$: is there something
funny going on at this temperature?

\begin{figure}
\includegraphics[clip,width=3.4in]{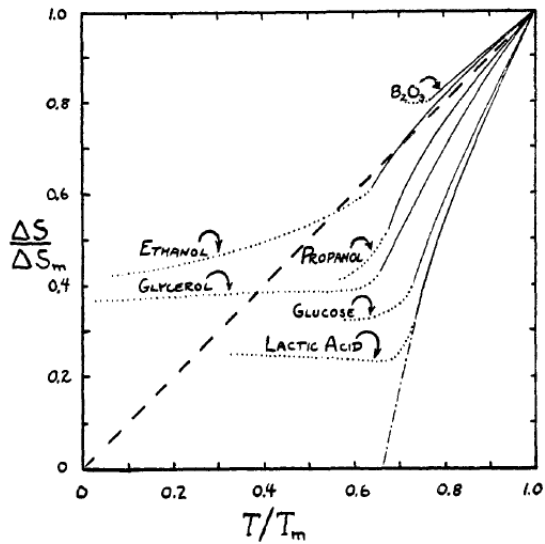}
\caption{ 
{\bf Kauzmann entropy crisis} - 
Kauzmann made a low-$T$ extrapolation of the excess entropy of many substances, 
normalized by their melting point value. He noted that in several cases the excess entropy
seems to vanish at temperature larger than zero. 
(Reprinted with permission from \cite{kauzmann48}). }
\end{figure}

The interesting point is that several other thermodynamic observables
have an extrapolation qualitatively similar to that of the entropy:
they all indicate a point where their liquid-crystal difference
vanishes. However, the temperatures where this happens are not the same
for different observables; for example, the point where the extrapolation of
the entropy difference vanishes does not coincide with the point where
the extrapolation of the enthalpy difference vanishes.  Most importantly,
Kauzmann noted that the liquid-crystal free-energy difference $\Delta
F$ {\it does not} decrease when lowering the temperature, so that any (reasonable) extrapolation does not indicate that
the free energies of the two phases converge to each other. From this
observations, Kauzmann immediately ruled out the most naive
interpretation of $T_k$, i.e. that it could be the locus of a
continuous transition from the liquid to the crystal phase. This
would require converging free-energies and similar values of $T_k$ for 
different observables, which is not the case.

An alternative way out of the conundrum posed by $T_k$ is to argue
that the extrapolation is just wrong. In particular, the curve of $\Delta S$
vs. $T$ could become flat at some temperature below $T_g$ (note that the
flattening of $\Delta S$ depicted by Kauzmann in Fig.16 - dotted lines - does not refer to
this, but to the dynamic glass transition).  If
this were the case, $\Delta S$ would not vanish, and there would be no
$T_k$. This is quite an effective way to cool down all the commotion
about the Kauzmann temperature: the extrapolation is, indeed, just an
extrapolation, so we cannot trust any result coming from it. Also
this objection, however, is not completely reassuring. First, some
empirical observations, both experimental and numerical, performed at rather 
lower temperatures, arrive quite close to $T_k$ without showing 
any evidence of an `elbow' of the equilibrium $\Delta S(T)$ curve \cite{angell88}. 
Second, such a flattening of $\Delta S$
would be for all practical purposes identical to what happens at the
dynamic glass transition $T_g$: the entropy freezes at a certain level,
and the specific heat drops to its crystalline value. Kauzmann
rightly noted that this would be weird, since a {\it dynamic} glass
transition certainly has nothing to do with the genuine {\it thermodynamic}
behaviour we were after when we made the entropy extrapolation.

Kauzmann found a more effective (and elegant) way to get rid of $T_k$
and of all the conceptual problems related to it: in his view,
the extrapolation of the equilibrium curves of a supercooled liquid is {\it always}
interrupted by a kinetic spinodal at a temperature
$T_\mathrm{sp}>T_k$. In fact, Kauzmann has been the first one to
describe in \cite{kauzmann48} the kinetic spinodal in clear terms.  Below $T_\mathrm{sp}$
the supercooled liquid simply does not exist, and no meaningful
extrapolation can be performed. We have extensively described the
mechanism behind the kinetic spinodal in Section III.D: at a certain point
the barriers to relaxation becomes larger than the barriers to
crystallization; below this point the liquid phase is no longer experimentally
well defined. No liquid, no extrapolation, no entropy crisis. Problem solved.

Thus, Kauzmann did not believe that $T_k$ existed at all. In fact, he
did not even give a name to such temperature. On the other hand, he
did give a name to the kinetic spinodal, and he call it $T_k$. At this
point, the utterly confused reader may ask: why do we indicate with
$T_k$ (where the {\it k} is after Kauzmann) a temperature that
Kauzmann rejected  (i.e. the entropy crisis
point), while we call $T_\mathrm{sp}$ the temperature
Kauzmann christened as $T_k$ (i.e. the kinetic spinodal)?  The reason is
that Kauzmann's proposal of a kinetic spinodal never enjoyed much fortune, whereas the
concept of entropy crisis (and of what would be going on at $T_k$)
stirred an enormous interest within the glassy community and it was very influential
in the development of later theories. Because of this, irrespective of
Kauzmann's own preferences, the name $T_k$ stuck to the entropy crisis
temperature. Reluctantly, we abide to this (now well established) convention,
and use the symbol $T_\mathrm{sp}$ for the kinetic spinodal.

How stands today Kauzmann's proposal of a kinetic spinodal as a
resolution of the entropy crisis problem? In my opinion, it is a very
sensible idea. In fact, it is unclear why such idea has been
largely disregarded by the community. As we stressed in Chapter III, the
kinetic spinodal is a very sneaky enemy of the supercooled liquid, and
it is very hard to rule it out a priori in a given
substance. Therefore, in many cases the extrapolation discussed above
is probably interrupted at $T_\mathrm{sp}$, and no $T_k$ is present. On
the other hand, we have seen that elastic effects may stabilize
significantly the supercooled liquid phase, in some cases so much as to
kill the kinetic spinodal. Therefore, we do not see
particularly strong reasons to believe that a kinetic spinodal must {\it
always} be present in all supercooled liquids. In several systems,
where elastic effects are strong, the entropy crisis is therefore still an issue.

There is a second reason why Kauzmann's way out of the problem is not fully
satisfying, which is  more profound. The sharp approach of the
liquid entropy to the crystal value, seems an interesting fact by itself, 
something which cries for an explanation, 
irrespective of whether or not the metastable liquid 
phase actually survives down to $T_k$ or not.
This same observation is also a reasonable reply to those who object
that the entropy crisis is irrelevant because we will never be
able to experimentally thermalize a liquid down to $T_k$. The {\it
incumbent} entropy crisis at $T_k$ is important, even though
we cannot equilibrate at $T_k$, because it may have
consequences on the physics of the system even in the temperature
range where we can equilibrate the system.
To better appreciate this point, we must go deeper into the
meaning of the excess entropy $\Delta S$ and the physical implications of
its decrease. To do this we must go back to Goldstein's
description of the low-$T$ dynamics of a supercooled liquid. As we
shall see, this deeper insight will also suggest a new interpretation
of what happens at $T_k$.

\subsection{Configurational entropy}

Within Goldstein's picture, activation is the main mechanism of diffusion
and thus at very low $T$ the system spends a long time vibrating
within an amorphous minimum of the potential energy, with rare jumps
between different minima. The lower the temperature, the more accurate
is this description. A rather reasonable consequence of Goldstein's description is that the entropy
of an equilibrium liquid, $S_\mathrm{LQ}$, can be split into two parts: a {\it vibrational} 
contribution corresponding to the intra-minimum short-time vibrational dynamics, 
and a {\it configurational} contribution given by the large number of
different amorphous minima. We formalize this hypothesis by writing,
\be
S_\mathrm{LQ} = S_\mathrm{vib} + S_\mathrm{c}  \ .
\label{splatto}
\ee
The configurational entropy $S_c$ is different
from zero only if the number of minima is exponentially large in 
the size of the system. In fact, $S_c$ is defined by the relation, 
\be
S_\mathrm{c}=\frac{1}{N} \log \; \cal{N}  \ ,
\label{complexity}
\ee
where $\cal{N}$ is the number of amorphous minima visited by the system at equilibrium.
Hence, the total entropy $S_\mathrm{LQ}$, which counts the total number of 
configurations participating to the liquid state, has been decomposed into a sum
of the vibrational entropy $S_\mathrm{vib}$, which counts how many configurations there 
are within each minimum, plus the configurational entropy $S_\mathrm c$, counting the
number of minima.
We should be more specific about this point: what minima are we considering?
All amorphous minima, or just those at a given energy level? In this last case, what energy level?
We will address all these questions later on, when we will discuss the configurational entropy
in the context of mean-field spin-glasses. For now, let us adopt an empirical view and say that
the configurational entropy is the rest of the total liquid entropy once we subtract the vibrational part, as 
in \eqref{splatto}. From this definition we can infer the number of amorphous minima actually
visited by the system at equilibrium.

We now make a crude, but rather sensible approximation: the vibrational entropy of an amorphous minimum
is not {\it too} different from the vibrational entropy of the crystal. They are not identical, of course,
since the shape of a typical amorphous minimum is in general different from that of the crystalline minimum,
and so will be the vibrational frequencies, and thus the entropy \cite{sciortino-05}. 
However, this difference is likely to be sub-dominant 
compared to the difference between vibrational and configurational entropies. 
We therefore assume,
\be
S_\mathrm{vib} \sim S_\mathrm{CR}  \ .
\label{vibraz}
\ee
This equation basically derives from an harmonic expansion of the potential 
around a minimum (local for the liquid, global for the crystal). Hence, such
approximation does not make a bit of sense in all those cases, most notably
hard sphere systems, where such an expansion cannot be done and where
excluded volume effects are dominant over energetic considerations. In fact, 
many of the considerations in this section are ill-suited for hard-spheres and we suggest
the reader to look at \cite{zamponi-08} for a review dedicated to such system.

From equation \eqref{vibraz} we can rewrite the excess entropy as follows,
\be
\Delta S (T) = S_\mathrm{LQ}(T) - S_\mathrm{CR}(T)  \sim S_\mathrm{c}(T)  \ .
\label{noemi}
\ee
This is a very important equation in the physics of supercooled liquids: 
the excess entropy is roughly equal to the configurational entropy. 
Therefore, what Kauzmann actually discovered  by means of his extrapolation was
that the configurational entropy of the deeply supercooled liquid sharply decreases 
when the temperature is lowered. This means that the number of amorphous minima 
the liquid visits in its equilibrium wandering through the phase space 
becomes smaller and smaller when the temperature is lowered. 
Such result is rather intriguing, and it supports our suspicion that the decrease of the excess entropy 
is an interesting fact by itself, irrespective of its vanishing at $T_k$. The configurational
entropy is a key concept in the physics of supercooled liquids.

It is crucial to understand that the decrease of the configurational entropy is {\it not} 
due to a lack of time to visit all available minima: we are talking about equilibrium here, 
the system is ergodic, it has enough time to visit whatever the Gibbs-Boltzmann distribution 
dictates and to jump all necessary barriers. Yet, the number of visited minima
decreases with the temperature. Why is that? Let us answer this question in two steps.

First, the equilibrium potential energy decreases with the
temperature. At low temperatures, where dynamics is dominated by vibrations around minima,
this obvious fact implies that also the average energy of the minima visited by the
liquid is lower the lower $T$ \cite{sdbs}. This is rather reasonable. Second, the number of
amorphous minima with given potential energy is smaller the smaller is the energy.  This
second fact is less obvious than the first, even though it is still
quite intuitive. Let us make a very simple example. Imagine a system made
up of $N$ non-interacting particles, each of them living into an
asymmetric double well that has an energy difference $1$
between the lowest and highest minimum. Above the ground state
(with conventional energy zero), this system has a trivial spectrum of
excited states with discrete energies $1, 2, 3, \dots$. From purely combinatorial 
considerations, the number of minima $\cal N$ with energy equal to $K$ is given by,
\be
{\cal N}(K)= \binom{N}{K}  \ .
\ee
If we consider excited states with extensive energy, $K=N u$, we get for their 
configurational entropy (eq. \eqref{complexity}),
\be
S_c(u) = -u\log(u)-(1-u)\log(1-u) \ .
\ee
Therefore, the configurational increases with increasing energy, it
is zero at $u=0$ and it reaches a maximum at $u=1/2$.
This is a trivial example, of course. In real liquids the interaction plays a
crucial role and the situation is much more complicated than this.
In particular, we will show that the drop of the
configurational entropy is intimately connected to the growth of a
correlation length. For now, this simple example helps us to accept as plausible 
the general idea that the number of minima with a certain energy
decreases with decreasing energy.

Now that the decrease of the configurational entropy with the
temperature is somewhat clearer, we may ask: what are the physical
consequences of this fact?  Can we understand what happens at the
Kauzmann temperature?  At $T_k$ the configurational entropy
\eqref{complexity} vanishes. Hence at this temperature a system at
equilibrium visits only a {\it few} amorphous minima, more precisely a
sub-exponential number of them.  If we now lower the temperature below
$T_k$, the number of visited minima cannot become negative. Therefore,
the configurational entropy $S_c$ remains stuck to zero, and the only
thing to change if we lower the temperature below $T_k$ is the
vibrational entropy $S_\mathrm{vib}$ within these minima.  We conclude
that the total entropy $S_\mathrm{LQ}$ must have a kink at $T_k$.  Below $T_k$ the
systems is confined in few lowest lying amorphous minima (we
do not call them ground state because there is always the crystal).

If we are prone to theoretical speculation, from this state of affairs
we may argue that at $T_k$ there is a  thermodynamic phase 
transition from the supercooled liquid to a new amorphous phase.
Abiding to the convention, we will call this phase the {\it
thermodynamic glass} or {\it ideal glass phase}.  In fact, this is a possibility
that Kauzmann himself mentioned in his 1948 paper \cite{kauzmann48}. He
acknowledged that the vanishing of the excess entropy at
$T_k$ could suggest the existence of a new disordered phase of
matter. However, he dismisses the hypothesis two lines after having
mentioned it, proposing (as we have seen) the kinetic spinodal as a
better resolution of the entropy crisis paradox, and dropped what seemed
to him an unphysical hypothesis. Nevertheless, the hypothesis of a
transition at $T_k$ enjoyed great success, mostly thanks to
later theories that we will study in the next chapter.

The idea of a phase transition at $T_k$ is a fascinating one,
especially within the physics of glass-formers, where all
notable temperatures are invariably defined in a fishy way.  Contrary
to $T_g$, $T_k$ would be a well-defined phase transition, independent from
any conventional time scale. As we have seen, also $T_x$ is
independent from the experimental time scale; however, at $T_x$ there
is a crossover between two physical mechanisms, rather than a sharp
transition, and for this reason the position of $T_x$ is somewhat
blurred. On the other hand, $T_k$ would be the underlying phase
transition ruling the whole phenomenology of supercooled liquids at
higher temperature. No surprise it is an appealing concept.

Yet, we must agree with Kauzmann that, given the arguments we have
provided until now, the hypothesis of a thermodynamic phase transition at
$T_k$ seems a bit of a long shot. The idea that the system remains
confined within few, deep amorphous minima seems reasonable, because
it agrees with the extrapolated behaviour of the excess entropy.
However, a couple of questions raise naturally. Is ergodicity
broken at $T_k$?  This would be the landmark of a thermodynamic phase
transition. A measurable manifestation of broken ergodicity would be
the divergence of the relaxation time $\tau_\mathrm R$ at $T_k$. Hence,
the incumbent ergodicity breaking at $T_k$ would cause a very
significant increase of the relaxation time even {\it above}
$T_k$. This is indeed what we meant when we said that the
entropy crisis at $T_k$ could be relevant even for a system that never
reached $T_k$.  Is there any hint of a divergence of $\tau$, beyond
the obvious increase leading to $T_g$?  Moreover: broken ergodicity
and an infinite relaxation time would imply infinite energy
barriers. In other words, the system should remain trapped in just {\it
one} of the lowest lying minima below $T_k$. Why the barriers between
amorphous minima would become infinite at $T_k$?  What physical,
real-space mechanism would cause this drastic phenomenon?  Let us
first focus on the relaxation time, leaving the issue of barriers for later.

\subsection{The VFT law: just a fit?}

As we know, the relaxation time $\tau_{\mathrm R}$ grows quite sharply when we get
close to the dynamic glass transition $T_g$. On the other hand, we
also know that, by definition, $\tau_{\mathrm R}$ is finite at $T_g$; in fact,
$T_g$ is defined as the point where relaxation time is of order 
$10^2-10^3$ seconds, which is finite.  Therefore, as we
already said several times, the dynamic glass transition is not a real
transition at all.  If something funny (like a divergence) happens to
$\tau_{\mathrm R}$, it must be below $T_g$. We therefore need (again!)  to extrapolate
our data. However, this time we have to be more careful than
with the extrapolation of the entropy. Unlike the
entropy (or volume, or enthalpy), the relaxation time $\tau_{\mathrm R}$ has a
{\it very} nonlinear temperature dependence (see Figs.5 and 7). This means
that we cannot simply extrapolate linearly the data of $1/\tau_\mathrm R$ 
by eye and see where it goes to zero. What we need is a solid nonlinear 
formula that interpolates well equilibrium data above $T_g$, and see what this formula tells
us about the $T<T_g$ regime. 

It would be nice to have a theoretically motivated formula. Problem is
that up to know the only theoretical scheme we have seen, able to say
something about the dynamics of glassy systems, is Mode Coupling
Theory (MCT), and we cannot use MCT to go
below $T_g$. As the reader may remember, MCT predicts a power law
divergence of the relaxation time at a temperature $T_c$ {\it larger} than $T_g$.
We need something else.

It turns out that the best fit to the relaxation data was found a long time ago,
quite independently from any theoretical framework. This is the
celebrated Vogel-Fulcher-Tamman (VFT) law \cite{vogel, fulcher, tamman}, 
which hit the physics journals almost one century ago. 
For the relaxation time, it reads,
\be
\tau_\mathrm R(T)=\tau_0 \; \exp\left(\frac{A}{T-T_0}\right) \ .
\label{vft}
\ee
If we use this formula as a three parameters ($\tau_0, A, T_0$)
fit to the data, it does a very decent job for many different
glass-formers in a wide range of temperature (see \cite{stickel} for a
test of the validity of the VFT law in many materials). In particular,
the VFT law fits the data also in the region that is
off-limits for MCT, i.e. for $T_g<T<T_c$. Note that the increase of
$\tau_\mathrm R$ in \eqref{vft} is steeper than a simple Arrhenius, 
but smoother than the MCT power law $\tau_\mathrm R\sim (T-T_c)^{-\gamma}$, 
since $T_0 \ll T_c$.

Even though we did not provide any
theoretical argument, the VFT formula looks reasonable. Its exponential
nature  agrees quite well with the the very physical
idea that activation rules the dynamics at low temperatures.  At the
same time, the non-Arrhenius argument of the exponential is justified
by what we already know to be the difference between strong and
fragile systems: by varying the parameter $T_0$ we can interpolate
between the purely Arrhenius behaviour typical of strong liquids
($T_0\sim 0$), to a sharper, and thus more fragile-like, growth of
$\tau_{\mathrm R}$, when $T_0$ increases. 

To be more precise, we recall that the
fragility of a substance is a measure of how steeply the relaxation time
(or viscosity) increases at the glass transition, and it can thus be quantified as the slope of 
$\log \; \tau_\mathrm R$ evaluated at $T_g$, 
in an Arrhenius $1/T$ plot. With such definition, eq.\eqref{vft} gives
for the fragility,
\be
\left[\frac{d \log(\tau_\mathrm R/\tau_0)}{d (T_g/T)}\right]_{T=T_g} = 
\kappa \left( 1+\kappa\frac{T_0}{A}\right)  \ ,
\label{fragiloso}
\ee
where $\kappa = \log[\tau_\mathrm R(T_g)/\tau_0]$.
The fragility of the system therefore depends on the ratio $T_0/A$.
The prefactor $\tau_0$ is a microscopic time scale. When
data are fitted, one should get a reasonable value for it, even though
far too often its actual value is completely disregarded, as long as
the fit is good.

If we decide to trust the VFT law even below $T_g$, we conclude that the relaxation
time diverges at $T_0$.  Even though the location of $T_0$ varies from
system to system, it is fair to say that $T_0$ is below but close to $T_g$
in fragile systems, while it is far below $T_g$, and quite close to $0$ in
strong systems. Thus, while at $T_g$ we
observe a sharp slowing down of the dynamics, the VFT law tells us
that at $T_0$ we eventually meet a real divergence of the relaxation
time, that is a dynamical phase transition. All this would happen at
equilibrium, thus this dynamical singularity cannot be simply due to
the system remaining trapped within some metastable state (as it could
happen in mean-field), and it must be associated to a true
thermodynamic phase transition at $T_0$, accompanied by ergodicity
breaking.

We now start to understand the link between the VFT empirical law and
Kauzmann's entropy crisis.  This first predicts (on purely empirical
basis) a divergence of the relaxation time, and thus a
phase transition, at a temperature $T_0<T_g$. The second indicates a vanishing of the
configurational entropy at a temperature $T_k< T_g$, which may be
interpreted as a sign of a thermodynamic phase
transition. It is important to
stress that both these conclusions stretch to some degree the scenario
they were developed within: the VFT law is meant to fit the data in
the region where {\it there are} data, so that, in principle, $T_0$ is
just a fitting parameter, without any particular relevance; the
entropy crisis, as we have seen, can be interpreted in several other
ways than a transition at $T_k$, including Kauzmann's proposal that there
is no $T_k$ at all. Yet, we cannot deny that the correspondence between
$T_k$ and $T_0$ seems intriguing.

\subsection{A link between dynamics and thermodynamics}

The most relevant question at this
point is: is there a {\it quantitative} link between $T_k$ and $T_0$? 
Even though this rather crucial question is still debated, on balance we 
may say that the answer is: yes. 
Indeed, in most systems one can show that \cite{angell-97, angell-98},
\be
T_0\sim T_k  \ .
\label{identita}
\ee
Before diving into the implications of this relation, a few words of
caution. In \eqref{identita} we are comparing two temperatures, none
of which is directly measured. One comes from a linear extrapolation of the data
($T_k$), the other is a parameter in a highly nonlinear fit ($T_0$), so that fixing them
sharply is quite difficult. For what concern $T_0$, we must note that
the VFT fit parameters depend  on the temperature
interval over which we fit the data, and in particular on the highest
temperature included.  As for $T_k$, the situation is perhaps even
worse. First, Kauzmann's extrapolation fixes the point $T_k$ where the
excess entropy vanishes with a certain degree of uncertainty, which
is the higher the larger the gap between $T_g$ and
$T_k$. Secondly, we must remember that the configurational entropy is
not exactly the same as the excess entropy, because of approximation
\eqref{vibraz}, so that $T_k$ is not exactly the point where a
thermodynamic transition is supposed to be. The bottom line is that
equation \eqref{identita} is accepted by those who mostly agree with
its theoretical implications, whereas it is discarded by those who
disagree. Personally, I believe the empirical evidence is in
favour of \eqref{identita}, even though one should never forget the
caveats above. 

A second important remark is in order. Some investigations claim
 that there is no empirical evidence what-so-ever for a 
divergence of $\tau_\mathrm R$, and thus conclude that the VFT law must be 
wrong \cite{dyre-novft}. This is similar to another objection we already 
met regarding $T_k$: if there is a kinetic spinodal, there cannot
be a $T_k$, and thus all the entropy crisis business is useless. We 
believe this is not the case. The way the relaxation time increases
in the VFT law, as well as the way the configurational entropy goes
to zero in the entropy crisis scenario, are more important than the
actual singularities at $T_0$ and $T_k$. In fact, equation \eqref{identita} is very
important even if none of two temperatures really existed! The 
relevance of \eqref{identita} is to point out a deep connection between
two different worlds, dynamics and relaxation time on one side,
thermodynamics and configurational entropy on the other side. This is what matters, not
the actual existence of the underlying singularities.

We therefore have this `coincidence': the dynamic divergence predicted
by the VFT fit is approximately equal to the point where the configurational
entropy vanishes. This pseudo-fact seems indeed to establish a direct
link between the dynamic and thermodynamic frameworks
and it supports the scenario we discussed above of a true phase
transition from the liquid to the thermodynamic glass at $T_k \sim
T_0$. According to this scenario, the number of thermodynamic
amorphous minima becomes sub-exponential at $T_k$, and the system
remains trapped within one of these minima. For this reason
ergodicity is broken and the relaxation time diverges, as in VFT.

As usual, we cannot avoid raising some conceptual concerns.
The states that presumably trap the system, breaking the ergodicity at $T_k$, 
have nothing to do with the crystal, of course, but are rather
the lowest lying amorphous minima of some free-energy functional. The crystal, on the other hand, 
will always be the absolute stable state of the system. Thus, below $T_k$ the thermodynamic glassy
phase has a lower free energy than the liquid, and it is thus stable with respect to it,
but it is metastable with respect to the crystal.
This is a tricky point: how to define a thermodynamic phase 
transition between different metastable phases, excluding the crystal from the game, is far from clear. 

The main problem is the following: a thermodynamic transition at $T_k$
requires the (free-energy) barriers around the lowest lying amorphous
states to diverge. If this is true, however, not only the system would
be unable to jump between the amorphous minima, but it could also
never make the transition to the crystal. However, the crystal has a
lower free-energy than the thermodynamic glass, and thus standard
nucleation theory predicts that the amorphous states {\it must}
collapse to the crystal. In which case, though, they cannot be
surrounded by infinite free energy barriers. In other words, it is
difficult (but perhaps not impossible) to imagine an ergodicity
breaking mechanism that only affects the amorphous sector of the phase
space, but that leaves open a transition channel to the crystal.  

We can only see three ways out of this contradictory loop: first, there
is no thermodynamic transition at $T_k$, relation \eqref{identita} is
just a coincidence, or perhaps it is wrong; second, elastic effects
inhibits nucleation and stabilize lowest-lying amorphous minima,
making them effectively `equilibrium' and cutting the crystal out of
the game; third, `equilibrium' amorphous minima have the same
free-energy as the crystal, or even lower than that, thus ruling out crystal 
nucleation.

Once we think about it, however, we realize that all these conceptual
problems stem from the singularity at $T_k$.  Imagine that the liquid has a kinetic spinodal at
$T_k+\epsilon$. In this case, there would be no thermodynamic glassy
phase and no conceptual problems. Yet, the relaxation time at the
spinodal would still be enormous and the configurational entropy very
small, and it would still be crucial to understand {\it why} the
relaxation time becomes so large when the
configurational entropy becomes so small. In other words, we believe
that investigating whether or not there is a relationship between
relaxation time and configurational entropy, and why it is so, is more
important than speculating about the properties of the thermodynamic
glass phase. The reason for this is that a relationship between these two
concepts may have important consequences also on the supercooled phase
that we can observe experimentally. As we already said, given that 
equilibrating the system at very low $T$ is, and always will be, 
an impossible task, the phenomenology associated to the
incumbent transition is more relevant than the transition itself.

Therefore, our most urgent objective is to understand what is the relation, if any,
between the growth of the relaxation time and the decrease of the
configurational entropy. This must be done in a direct physical way, 
we can no longer invoke the existence of a transition where these two quantities are respectively
infinite and zero, in order to link the two concepts.

%%%%%%%%%%%%%%%%%%%%%%%%%%%%%%%%%%%%%%%%%%%%%%%%%%%%%%%%%%%%%%%%%%%%%%%%%%%%%%%%
%%%%%%%%%%%%%%%%%%%%%%%%%%%%%%%%%%%%%%%%%%%%%%%%%%%%%%%%%%%%%%%%%%%%%%%%%%%%%%%%
%%%%%%%%%%%%%%%%%%%%%%%%%%%%%%%%%%%%%%%%%%%%%%%%%%%%%%%%%%%%%%%%%%%%%%%%%%%%%%%%
%%%%%%%%%%%%%%%%%%%%%%%%%%%%%%%%%%%%%%%%%%%%%%%%%%%%%%%%%%%%%%%%%%%%%%%%%%%%%%%%
%%%%%%%%%%%%%%%%%%%%%%%%%%%%%%%%%%%%%%%%%%%%%%%%%%%%%%%%%%%%%%%%%%%%%%%%%%%%%%%%
%%%%%%%%%%%%%%%%%%%%%%%%%%%%%%%%%%%%%%%%%%%%%%%%%%%%%%%%%%%%%%%%%%%%%%%%%%%%%%%%

\section{The quest for a correlation length}

Everybody accepts that the low-$T$ dynamics of supercooled liquids is
dominated by activation. Thus, understanding the behaviour of the
relaxation time amounts to understand the behaviour of potential
energy barriers.  If barriers were constant in $T$, the relaxation
time would follow a simple Arrhenius law.  However, we have seen that
fragile liquids display a much faster, super-Arrhenius increase of
$\tau_{\mathrm R}$. Indeed, they are described by the VFT formula
\eqref{vft}, with $T_0$ well above zero. This means that, at least in
fragile systems, potential energy barriers to relaxation increase with decreasing
temperature
\footnote{As already noted before, hard spheres are an altogether
different case, in that barriers are of purely entropic origin in that
case. Even though we could play general, talking always of
{\it free energy} barriers, which include both energy and entropy,
we prefer not to do so, since we would barely be able to specify what
free energy are we talking about at this point of our study. It is better to talk as much as possible 
about energy barriers, which are dominant in thermal systems, treating the
hard-sphere case apart \cite{zamponi-08}.}.
Furthermore, equation \eqref{identita} is a clear indication
that energy barriers between amorphous minima have
something to do with the number of these minima. What this
link is, however, is not clear at the present stage of our study.
What is the physical mechanism behind barriers growth?

In a finite dimensional system with short range interactions a
rearrangement process must be local in space. This means that barriers
arise because a certain (finite) number $n$ of particles have to locally
rearrange in real space in order for the system to relax.  Therefore,
it is natural to think that when barriers increase, it is
because the number $n$ of particle involved in the rearrangement
increases, which is to say that the size $\xi$ of the region being
rearranged becomes larger.

To our knowledge, the idea of a temperature
dependence of the size of the rearranging regions was first introduced by
Jenckel in 1939 \cite{jenckel}, and it is quite a natural view. Almost invariably 
in physics the fundamental cause for a sharp increase (divergence) of
the relaxation time $\tau_{\mathrm R}$, is the increase (divergence) of the
correlation length $\xi$. The precise nature of the time-length
relation depends on the system. In critical phenomena \cite{cardy},
it is a power law, 
\be 
\tau_\mathrm R\sim\xi^z \ , 
\ee 
whereas in supercooled liquids we expect some sort of exponential law, due to 
the key role played by activation \cite{montanari-sem},
\be 
\tau_\mathrm R\sim\exp\left(\frac{\xi^\psi}{k_B T}\right) \ .
\label{cocorito}
\ee 
In this equation we have assumed that the barrier $\Delta$ scales like some power of $\xi$,
\be
\Delta  \sim \xi^\psi \ .
\label{psi}
\ee
This view implies that super-Arrhenius relaxation in fragile supercooled liquids 
is due to the fact that larger and larger regions becomes correlated 
as the temperature is lowered,  so that larger ensembles of particles 
have to be rearranged to relax the system. The aim of this chapter is to 
look for such increasing correlation length.

The problem in supercooled liquids is: how to detect $\xi$?  
In a standard statistical system
(for example the Ising model) we would extract the
correlation length from the decay of the correlation function of
the order parameter \cite{parisi-book, binney}.  We already tried to do that in
liquids, by using the standard correlation functions of density fluctuations.
Unfortunately, it did not work: we did not find any clearly growing correlation length.  
As we already said, supercooled liquids are structurally unexciting and density fluctuations
seem to do nothing extraordinary close to glassiness.  

We have two paths, then.  
The first is to look for a growing lengthscale using nonstandard thermodynamic 
methods. However, what these methods may be is unclear at this level of our study.
Thus, we will, in a way, invert the problem: we will first try to understand what is the nature
of the {\it static} correlation length $\xi$ in low $T$ supercooled liquids, and then use this
(hypothetical) knowledge to develop the tools needed to unveil $\xi$. This will be a long path, 
but it will allow us to see some interesting theoretical frameworks meanwhile, as the 
Adam-Gibbs-Di Marzio theory and the mosaic theory.
The second path is to go dynamical. We have seen that dynamical
correlation functions {\it are} exciting and that dynamically heterogeneous
clusters become larger as the temperature decreases. Hence,
so we can try to define a {\it dynamic} growing lengthscale $\xi_d$.
These two paths are different, but not necessarily in contradiction with each
other. We will illustrate both of them.

\subsection{The Adam-Gibbs-Di Marzio theory (and a hand from Flory)}

The first successful attempt to connect at the theoretical level the
increase of the correlation length to the decrease of the configurational
entropy is due to a remarkable series of papers published by Adam, Gibbs
and Di Marzio between 1956 and 1965, even though one could argue that
everything started with a calculation made by Flory in 1956, which had
nothing to do with glass physics.

Before we proceed, it is essential to make a remark.  Throughout the
work of Flory's and later the ones of Adam, Gibbs and Di Marzio, the
word `configurations' is used very much in the sense we now assign to
the word `states'. In particular, these authors were not interested in
the decomposition between vibrational and configurational entropy (see
eq.\eqref{splatto}), they only focused on the configurational part,
disregarding the (somewhat more trivial) vibrational part. In the case
of the polymer on a lattice (Flory) the use of the word
`configurations' is quite approriate, whereas in the context of
glass-forming liquids (Adam-Gibbs-Di Marzio) such word is somewhat
misleading, because one expects a `state' at finite $T$ to be actually
made up of many `configurations', contributing to the vibrational
entropy of the `state'. Hence, unfaithful to the original papers, when
discussing Adam-Gibbs-Di Marzio theory I will use as much as possible
the word `states'. Unfortunately, I cannot accordingly change the wording
for `configurational entropy' (and promote it to `state entropy' or similar),
because this is far too much a well established notation in glass-forming 
liquids.

What Flory did in \cite{flory} was to calculate the number of
configurations available to a semi-flexible chain molecule, i.e. a
polymer, on a lattice. The aim of the calculation was  to
compute the free energy of the disordered phase and compare it to
that of the ordered one, thus finding the crystallization
point. Flory's analytic result demonstrated that below a certain temperature $T_m$,
\footnote{In fact, Flory considers the chain flexibility as the tuning parameter in
his calculation, but then he connects flexibility to temperature by 
making the free energy stationary. Thus, one can equally well use temperature
as the tuning parameter in his theory.},  the disordered phase has a larger free energy than the 
ordered one, and this was the main result he was after. 
However, while commenting his result, Flory noted {\it en passant} that at an
even lower, but {\it finite} temperature, $T_k<T_m$, the number of disordered configurations available
to the polymer becomes smaller than $1$, a fact that he deemed as
physically unacceptable.  He formally resolved this paradox by
recalling that there is always at least one configuration, i.e.  the
perfect crystal, which goes undetected  simply 
because the theory was designed to describe only disordered configurations, 
thus excluding the crystal. Moreover, Flory noted that it was not
necessary to bother about this funny temperature $T_k$, because
it was anyway below the crystallization point $T_m$, in a regime where the 
disordered phase is not stable, a regime he was not interested about.

We clearly see that, despite the different contexts and the different
methods, the similarity with Kauzmann's paper is striking. Kauzmann made
an extrapolation of equilibrium empirical data in supercooled liquids,
and found a temperature below which the excess entropy becomes
negative. He resolved the paradox by arguing in favour of a
kinetic spinodal: crystallization would avoid the paradox. 
Flory made an analytic calculation in the context of
polymers, and found a temperature below which the number of accessible
configurations in the disordered phase becomes smaller than $1$. He too, 
like Kauzmann, resolved the paradox arguing that crystallization would save 
the day.

The first to seriously bother about this `funny' temperature $T_k$ and to
make the crucial leap between the two approaches, was Gibbs \cite{agdm-1}.
In a very short Letter, written few months later than Flory's
work appeared, Gibbs noticed the similarity of the two paradoxes and
saw the deep implications of the connection between Flory's number of
available configurations and Kauzmann's excess entropy.  Regarding Flory's 
paradoxical temperature $T_k$ Gibbs asked the
question \cite{agdm-1}: {\it ``Could this be the glass transition?''} .

Gibbs was not satisfied by Flory's (and Kauzmann's) resolution of the
paradox, namely crystallization. He noted that in some amorphous
systems a crystal is not even present, so that it could not be the
conceptual way out of the problem. Gibbs was the first to tackle the
paradox, instead of avoiding it, by arguing that a vanishing
configurational entropy is the signal of a thermodynamic second order
phase transition \cite{agdm-1}. According to this hypothesis, the
configurational entropy has a kink at the transition $T_k$, and
it remains zero below this point, thus resolving Flory's and
Kauzmann's paradox. Gibbs finally noted that the system's dynamics had to
become very sluggish when the configurational entropy becomes so
small, even though in this first paper he did not elaborate about
this rather crucial point.

Gibb's arguments had to be developed in order to become a theory. This
was done in 1958 in a paper by Gibbs and Di Marzio (GDM) \cite{agdm-2},
where they apply Flory's technique to tackle the problem of the
glassy state. They showed that a second order phase transition has indeed to be
associated to the vanishing of the configurational entropy at $T_k$
\footnote{In GDM paper what we call here $T_k$ is indicated with $T_2$. Strictly
speaking, this is necessary, since AGDM introduce a temperature where
the analytical configurational entropy vanishes, whereas Kauzmann
introduced a temperature where the empirical excess entropy
vanishes. We know that excess entropy and configurational entropy are
only connected by an approximation, rather than being equal, so this
distinction is strictly speaking necessary.  However, I prefer not to complicate
the notation. Thus, I use
the symbol $T_k$ also for the GDM transition.}.  According to GDM, at the
transition the system remains trapped into {\it one} of the lowest-lying
amorphous minima, whose number is
sub-exponential in the size of the system, so that their configurational
entropy is zero. Hence, below $T_k$ ergodicity is broken. This is, according to GDM,
the thermodynamic glass, {\it "the fourth phase of matter"} \cite{agdm-2}.

The work of GDM is of purely thermodynamic nature, with no reference
to relaxation times or energy barriers. They argue that the dynamic
glass transition $T_g$ that we observe empirically is just a kinetic
manifestation of the underlying transition at $T_k$. However, after reading
their paper one is left with the same questions we had in the last chapter:
why the relaxation time becomes very large when the number of 
accessible states becomes very small? What is the {\it precise} mechanism
linking configurational entropy to barriers? From the theoretical
point of view it is clear that a thermodynamic transition must 
have implications also at the dynamical level. Yet, this is a very nonstandard
kind of transition, therefore having an intuition on how dynamics is
practically linked to thermodynamics is essential. 

GDM probably felt this urge as well and wrote a paragraph about this point,
\cite{agdm-2}: {\it ``Furthermore, the (free energy) barrier restricting flow of a
system from one of these states (configurations) to another is very
high in the neighbourhood of $T_k$ because, in this region ...
the few states that could conceivably occur are widely
separated in phase space, and proceeding from one to another involves
a considerable change in the topology of the molecular
entanglements. Relaxation times ...  should become very long as
$T_k$ is approached from above.''}

First, we cannot help noting the rather smart use of the brackets,
which allow GDM to skip all tricky explanations about two very
nontrivial issues, namely the difference between states and
configurations, and how to define free energy barriers. Apart from this,
we have to admit that this clarification is far from satisfying. Where this
`considerable change in the topology of the molecular entanglement'
comes from? And why should it be larger the fewer the states?  An
answer to these questions was provided seven years later by Adam and
Gibbs \cite{agdm-3}.

The key idea of Adam and Gibbs (AG) is that at low temperature
relaxation proceeds through the rearrangement of larger and larger
regions of correlated particles, which they called Cooperative
Rearranging Regions (CRR). AG define the typical CRR as the {\it smallest}
region that can be rearranged independently from its surrounding. This
means that different portions of one CRR cannot choose their own
configuration independently from each other, and thus cannot
contribute to a combinatorial proliferation of the number of
states available to the entire CRR.  As a consequence of this
fact, a typical CRR can be found in only a very small
number $\Omega$ of locally stable states.  According to AG,
this number $\Omega$ is a constant: it does not depend on $T$, nor on
the size of the CRR.  The only requirement is that it must be larger
than $1$, but even $\Omega=2$ would be a reasonable value
\cite{agdm-3}.

At this point AG ask: how many states $\cal N$ the global system can be found in?
Given that, by definition, different CRRs are weakly interacting with each other,
the answer to this question is very simple,
\be
{\cal N} = \Omega^{N/n}
\ee
where $N$ is the total number of particles of the system, $n$ the typical number 
of particles in each CRR, 
and thus $N/n$  is the total number of independent CRRs. The configurational
entropy $S_c$ is just the logarithmic density of the number of locally
stable states, and thus we obtain,
\be
S_c = \frac{1}{N}\log\;{\cal N} = \frac{\log\; \Omega}{n}  \ .
\ee
Inverting and including the temperature dependence, we get, 
\be
n(T)=\frac{\log\; \Omega}{S_c(T)} \ .
\label{agdm}
\ee
Considered how little we had to work to get this relation, we must say
that it is a remarkable result. As a consequence of the combinatorial
product of the $\Omega$ states of each different CRR, the number of states
of the entire system is exponential and the size $n$ of the CRR naturally 
increases when the configurational entropy $S_c$  decreases.  Let us
note that AG always talk about the size of CRRs in number of
particles, $n$, rather than their linear size $\xi$. Of course, the two concepts are strictly
related, since $n\sim \xi^d$.  Thus, the AG mechanism explains why the
correlation length increases when the configurational entropy
decreases. This was the result we were after.

In order to use equation \eqref{agdm} to obtain the relaxation time,
we need to work out a relation between size of the CRR and energy 
barrier $\Delta$ to be crossed in order rearrange the region. 
According to AG \cite{agdm-3}, the barrier scales with 
the number $n$ of particles that must be rearranged cooperatively,
\be
\Delta \sim n \sim \frac{1}{S_c}  \ .
\ee
Given that $n\sim \xi^d$,
AG result amounts to set $\psi=d$ in equation \eqref{psi}.
Even though at the present stage this seems a reasonable result, 
we will see later on that $\psi=d$ is not necessarily the most 
natural choice. 
Anyway, at this point we can sum-up AG results and, by using Arrhenius formula, obtain,
\be
\tau_\mathrm R = \tau_0\exp\left(\frac{B}{T\; S_c(T)}\right)
\label{agdm1}
\ee
where $B$ contains all constant factors. Equation \eqref{agdm1} not only explains
why the relaxation time increases when the configurational entropy decreases, but it also gives
quite a neat interpretation of the strong vs. fragile behaviour of supercooled liquids. Let us
see this in detail.

No one knows, of course, what is the precise behaviour of $S_c(T)$
at very low $T$. However, we can say something by using 
approximation \eqref{noemi}, together with the 
thermodynamic relation between entropy and specific heat, 
\be
\frac{d S_c}{d T} = \frac{d}{dT}(S_\mathrm{LQ}-S_\mathrm{CR})=
c_p^\mathrm{LQ}-c_p^\mathrm{CR}\equiv \Delta c_p
\ee
and thus,
\be
S_c(T)-S_c(T_k) = \int_{T_k}^T dt \; \frac{\Delta c_p}{T}
\ee
AG assume that the liquid-crystal difference of the specific heat $\Delta c_p$ is approximately 
independent of temperature. Within this approximation, and using $S_c(T_k)=0$, we have,
\be
S_c(T) = \Delta c_p \, \log( T/T_k) \ , 
\ee
and expanding the logarithm for $T\sim T_k$ we finally obtain,
\be
S_c(T) \sim \Delta c_p \; \frac{T-T_k}{T_k} \ ,
\label{cratone}
\ee
which agrees with the linear behaviour of the excess entropy observed in the empirical data
\footnote{Of course, someone arguing that the entropy crisis paradox is resolved by a flattening of
the configurational entropy at low, inaccessible temperatures, would say that this result is just 
unsupported by the real data.}. Equation \eqref{cratone} tells us something interesting, although
not surprising: the rate of change of the configurational entropy is proportional to the crystal-liquid
difference of the specific heat $\Delta c_p$. We recall that at the dynamical glass transition $T_g$, when a liquid
slips into the off-equilibrium glassy phase, its specific heat drops to a value comparable to 
that of the crystal (Fig.7). Therefore, $\Delta c_p$ is approximately equal to the jump of the specific heat
at $T_g$ and equation \eqref{cratone} is telling us that the larger is this jump (which is an
off-equilibrium feature), the sharper is the decrease of the configurational entropy (which 
is an equilibrium quantity). 

By plugging \eqref{cratone} into \eqref{agdm1} we obtain,
\be
\tau_\mathrm R = \tau_0\exp\left(\frac{T_k}{\Delta c_p\, T(T-T_k)}\right)  \ .
\label{agdm2}
\ee
Using the definition of fragility provided in equation \eqref{fragiloso}, we see that 
the AG relaxation time tells us that the fragility of a system is 
proportional to the jump of the specific heat at the glass transition, 
$\Delta c_p$. Furthermore, if we are far from $T=0$ and closer to $T_k$, we can 
approximate \eqref{agdm2} as,
\be
\tau_{\mathrm R} = \tau_0\exp\left(\frac{A}{T-T_k}\right)  \ ,
\label{agdm3}
\ee
where, as usual,  we have wrapped up in $A$ all constant factors.
Equation \eqref{agdm3} is nothing less than the VFT law \eqref{vft}. 

There is no doubt that Adam-Gibbs-DiMarzio (AGDM) theory gives several
interesting results. First, it provides a neat resolution of Kauzmann
entropy crisis in terms of a thermodynamic phase transition.  Second,
it provides a direct link between configurational entropy and size of
the correlated regions (and thus size of the barriers), thus
explaining the super-Arrhenius increase of the relaxation time in
fragile systems. Third, it relates fragility to the decay rate of the
configurational entropy, and thus to the jump of the specific heat at
the glass transition. Fourth, it provides a formula for the relaxation
time, which coincides with the most widely used fit in the empirical
analysis of equilibrium data, the VFT law. Last, but not least, all
these results were obtained at a very low price: the theory is simple
and elegant, and yet with far reaching consequences. It is therefore
no surprise that the AGDM theory has enjoyed such an enduring
favour. Indeed, this theory has shaped much of the thermodynamic
approach to glasses down to these days, and it has been a guide to
some spin-glass physics too.

As usual, though, it is not all good news, so let us see the down
sides. First of all, as we discussed in the previous chapter, the idea
of a glass phase as the fourth equilibrium phase of matter, with sharp
ergodicity breaking at $T_k$, has some interpretational problems.
Secondly, in AGDM view, the dynamic glass transition at $T_g$ is nothing
more than a mere precursor of the thermodynamic transition at $T_k$,
which is, in this framework, the only relevant physical phenomenon.
However, we have seen in the previous chapters that something quite
interesting happens already at higher temperatures, with the emergence
of two steps relaxation. Moreover, the crossover from nonactivated to
activated dynamics at $T_x >T_g$ is quite important, and it is quite
independent from the (possible) transition at $T_k$.  Thus, in
this respect, the claim of AGDM that {\it all} relevant glassy
phenomenology is regulated by the transition at $T_k$ is a bit too
extreme. 

But there is a third, most tricky point in AGDM theory, and this
regards the number of configurations $\Omega$ accessible to a typical
cooperative rearranging region. Why it is constant? It seems awkward
that the number of accessible configurations does not scale up with
the size of the region that is actually rearranged. The issue is rather 
subtle, and it mainly concerns the
difference between the number of configurations that are {\it
virtually accessible} by a cooperative rearranging region and those
that are {\it actually visited} by that region. It is not a purely
semantic issue, even though it may seem so.  In their argument in
\cite{agdm-3}, AG claim to be talking about the first number, while
they are, in fact, considering the second. This aspect of the AGDM theory 
is somewhat misleading.

In the context of the mosaic theory this ambiguity about the number of
configurations accessible to a CRR will be resolved. We will see that
each region has an exponentially large number of accessible
configurations, a number which {\it does} scale up with the region's
size. However, if the region is smaller than a certain critical size,
the number of configurations actually visited by that region is in
fact equal to $1$. We will show that this happens because of the
constraints imposed on each cooperative region by the surrounding
particles. 

The mosaic theory was developed building on some key
results obtained in the context of the $p$-spin model. It is thus
essential first to go back briefly to this mean-field spin-glass
model.

\subsection{The $p$-spin model strikes back}

One of the great advantages of the mean-field $p$-spin model is that
metastable states can be rigourously  defined. They are
local minima of the mean-field free energy, which is function of the
averaged local degrees of freedom.  In spin-glasses this is the 
Thouless-Anderson-Palmer (TAP) free-energy, function of the local magnetizations
\cite{thouless77}. It is important to understand that, unlike the
potential energy, the mean-field free energy {\it depends} on the
temperature, so that the structure of the free energy landscape depends on 
$T$. Minima can appear and disappear on changing 
the temperature. Therefore, the definition of a certain state depends on the
particular temperature at which we are working, at variance with the
definition of potential energy minimum.
In the $p$-spin, free energy barriers around these states are
infinite due to the infinite range interaction. Thus, when the system
is initially thermalized within one of these states it cannot escape \cite{burioni}, 
and the lifetime of the metastable state is thus infinite. Because of this, the separation of
the total entropy of the system in vibrational plus configurational
part (equation \eqref{splatto}), is sharp. The configurational
entropy (which is called {\it complexity} in the spin-glass context)
is simply the logarithm of the number of metastable states divided by
the size of the system $N$, and it is thus well-defined
theoretically \cite{cavagna-review}. 

As a consequence, the phase space can be uniquely and
unambiguously partitioned into basins of attraction of the various
metastable states. If we label each metastable state by an index
$\alpha$ running from $1$ to the total number of states $\cal N$, we
can write the partition function as,
\be 
Z =\sum_\sigma e^{-\beta H(\sigma)} 
= \sum_{\alpha=1}^{\cal N}
\sum_{\sigma\in\alpha}e^{-\beta H(\sigma)}   \ ,
\label{buscadero}
\ee
where $\sum_{\sigma\in \alpha}$ indicates a sum over all configurations 
$\sigma$ belonging to state $\alpha$. We can define,
\be
f_\alpha = -\frac{1}{\beta N}\log \;\sum_{\sigma\in \alpha}e^{-\beta H(\sigma)}  \ ,
\ee
as the free energy density of the metastable state $\alpha$, and write,
\be
Z= \sum_{\alpha=1}^{\cal N} e^{-\beta N f_\alpha} 
= \int df\; e^{-\beta N f} \;
\sum_{\alpha=1}^{\cal N} \delta(f-f_\alpha)  \ .
\ee
The sum under the integral is equal to the number of metastable states with
free energy equal to $f$. We can define the $f$-dependent configurational 
entropy as,
\be 
S_c(f) = \frac{1}{N} \log\;
\sum_{\alpha=1}^{\cal N} \delta(f-f_\alpha) \ ,
\ee 
and finally obtain,
\be
Z= \int df \; e^{-\beta N [f-T S_c(f)]}  \ .
\label{pinotto}
\ee 
This result is quite neat, but it must be reminded that it was only
possible because we could define sharply metastable states and thus
partition accordingly the phase space, which is hard to do out of
mean-field. In fact, as we shall see later, partitioning the phase
space into basins of the potential energy minima is not quite the same
thing.  Note also that at this point of the calculation the
configurational entropy is a natural function of $f$, rather than $T$.

In the thermodynamic limit
the integral in \eqref{pinotto} can be solved by means of the saddle-point
(or Laplace) method \cite{bender-orszag}, i.e. by finding the stationary point (in fact, the minimum)
of the expression between square brackets. The equilibrium free energy density of the system is thus,
\bea
F&&=-\frac{1}{\beta N}\log \; Z  \\
&& = \mathrm{min}_f \ [f-T S_c(f)] = f^\star-T S_c(f^\star) \ ,
\label{busu}
\eea
where $f^\star(T)$ is the solution of the saddle-point equation,
\be
S_c'(f^\star)=1/T \ .
\label{saddelina}
\ee
The meaning of this result is the following: the global equilibrium state (equivalent to the 
ergodic liquid phase)
consists in a superposition of an exponentially large number of metastable states, each one
with free energy density $f^\star(T)$, whose value is fixed by equation \eqref{saddelina}. 
The interesting point is that from \eqref{busu} we get, 
\be
F \leq f^\star \ .
\ee
The global equilibrium free energy density of the system is therefore
{\it smaller} than the free energy density of the metastable states
that dominate the partition function. This is due to the extra
entropic term $T S_c(f^\star)$ in \eqref{busu}. 
Individually taken, each metastable state has a free-energy $f^\star$ that is
too large compared to the equilibrium one. This is why each metastable
state is thermodynamically irrelevant when taken alone. It is only the
collective contribution of all the exponentially many metastable
states that gives rise to the correct free energy $F$.

The complexity $S_c(f)$ can be computed exactly in the $p$-spin model \cite{rieger92, crisanti95, monasson, recipes}: it is 
a monotonously increasing function of $f$, with a negative second derivative. 
Equation \eqref{saddelina} thus predicts that when $1/T$ increase, the solution $f^\star$
must decrease. This means that the lower the temperature, the lower the free-energy
$f^\star(T)$ of the metastable states dominating the partition function at that temperature.
We can define the equilibrium configurational entropy of the system as,
\be
S_c(T)\equiv S_c(f^\star(T)) \ ,
\ee
which is now an explicit function of the temperature. A measurement
of the configurational entropy at equilibrium at temperature $T$, gives
$S_c(T)$, which is therefore analogous to the configurational entropy we met in supercooled liquids. 
The decrease of $f^\star$ with $T$ implies that the equilibrium complexity decreases with
decreasing temperature. The reader has certainly noted the similarity with the situation already
described in supercooled liquids. In this case however, we have equations \eqref{busu} and
\eqref{saddelina} telling us exactly why the configurational entropy decreases with $T$.

If one computes the spectrum of metastable states in the $p$-spin
model, one finds a minimum free-energy density $f_0$, below which no metastable states
are found. This implies that,
\be 
S_c(f_0)=0 \ .
\label{zerollo}
\ee
Below $f_0$ the complexity is negative, i.e. the number of states is exponentially {\it small} in
the size of the system: there are no free-energy minima, neither stable nor metastable below
$f_0$, which is thus the true ground state of the system (in the $p$-spin model there is no crystal). 
The crucial point is that  the slope of $S_c$ at $f_0$ is finite, and therefore equation \eqref{saddelina} implies that the
temperature where $f^\star$ hits the lower band edge $f_0$ must be different from zero. This temperature is
called $T_s$ in the $p$-spin, but here, for obvious reasons, we will call it $T_k$. We can define $T_k$ as,
\be
f^\star(T_k)=f_0  \  ,
\ee
or, equivalently, from the equilibrium configurational entropy,
\be
S_c(T_k) = 0  \ .
\ee
At $T_k$ the configurational entropy vanishes and from \eqref{busu} we
conclude that at this temperature the two free energies, $F$ and
$f^\star$, are the same: equilibrium is (at last!) given by the states
with the lowest free-energy, they are no longer metastable and their
number is sub-exponential.  Below $T_k$, the system cannot dig
lower than $f_0$, so it remains stuck in one of the lowest-lying
states, just reducing gradually its vibrational entropy. Each one of
these states has a low enough free energy to individually
dominate the partition function. This means that below $T_k$
ergodicity is broken also at the thermodynamic level, just like it happens in
the Ising models below the critical temperature.  This is confirmed by all thermodynamic
studies of the $p$-spin model \cite{kirk-2, kirk-3, rieger92, crisanti92}:
by using the replica method \cite{parisi79, parisi80}, 
one finds a true thermodynamic transition at this same temperature $T_k$.

What we have just described is an exact realization of Kauzmann's
entropy crisis scenario at a temperature $T_k$ that is marked by a
thermodynamic phase transition. Within the $p$-spin model we do not need to extrapolate anything, all comes from
exact analytic calculations. The $p$-spin results provide a clearer 
conceptual framework for the interpretation of $T_k$ as the locus of a phase transition also in supercooled
liquids. We recall that the $p$-spin is already quite similar
to supercooled liquids at the dynamic level, i.e. close to $T_g$,
where MCT equations are the same as the $p$-spin dynamical
equations. It is therefore natural to try to extend this similarity
down to low temperatures and to export the clear
thermodynamic mechanism of the $p$-spin model to supercooled liquids.
This program, though, is not straightforward.

All exact calculations in the $p$-spin model, including that of the
configurational entropy and the replica calculation proving that there
is a phase transition at $T_k$, are made possible by the mean-field
nature of the model, which allows one to use the saddle point method
in the limit $N\to\infty$. Doing the same is not possible out of
mean-field, i.e. in finite dimensional systems as supercooled liquids
are.  Despite these difficulties, important steps have been done in
the direction of a thermodynamic theory of the ideal glass transition
\cite{mezard-96, cardenas-98, mezard-99, coluzzi-00}. By means of
nonstandard thermodynamic tools developed in the context of the
replica method \cite{monasson,recipes} and adopting an approximated
scheme to treat the liquid (typically the hypernetted chain - HNC -
approximation), one can compute analytically the configurational
entropy and the position of the thermodynamic transition $T_k$. In the
case of hard spheres, this approach has been recently reviewed in
\cite{zamponi-08}.

Even though results of such thermodynamic approach to the glassy phase
are remarkable, there is a problem, namely metastable states. The
approach of \cite{mezard-96, cardenas-98, mezard-99, coluzzi-00}
basically {\it assumes} that we can define amorphous metastable states
also in finite dimensions, and that the partition function can be
decomposed as in equation \eqref{buscadero}.  This is no trivial
assumption. In fact, a critique that could be formulated is that by
assuming the existence of metastable states one is effectively working
within the mean-field approximation, while we have no guarantee that
the mean-field results apply to finite dimensional systems.  In other
words, one may object that the nontrivial configurational entropy
calculated within this thermodynamic approach, together with the
thermodynamic transition at $T_k$, are just artefacts of an implicit
mean-field approximation.

Such uneasiness about the role of metastable states in finite
dimension is made particularly acute by one fact: the thermodynamic
approach to the glass phase describes things purely in terms of {\it
phase} space mechanisms, while it says little about what happens in
{\it real} space (see, however, the discussion of Appendix A in 
\cite{zamponi-08}).  In fact, this is an expected limit of the
$p$-spin inspired thermodynamic approach: due to its mean-field nature, the
$p$-spin model says nothing of how its phenomenology would stage in
real space, simply because there is no real space in it. However, in
real liquids we need to reinterpret the $p$-spin thermodynamic
mechanism in real space.

Summarizing, on one hand we have the AGDM picture, which is rooted in
real space, but is somewhat unclear in the way it deals with the
configurational entropy. On the other hand, we have the $p$-spin
model, and the thermodynamic approach derived from it, where the
thermodynamic role of the configurational entropy and the origin of
the transition are clear, but where their real space role is absent.
We still need to bridge this gap.

\subsection{Metastable states in finite dimension: just a delusion?}

The great difficulty we have in exporting the $p$-spin mechanism to
finite dimensional systems is how to define metastable states. We have
long postponed this problem throughout this review, and often played a
bit dirty on the ambiguity between local minima of the energy and
metastable states, and also between energy and free energy. Even
though we are definitely not alone in playing this game (see, for
example, Gibbs-Di Marzio's quotation in Section VII.A), it is worth to
try and be more careful, just for once. In this section we will first
illustrate the most naive definition of metastable states (that of local
minima of the energy), highlight its problems, and finally discuss
the need for metastable states to be defined locally both
in time and in space.  For a less pedestrian discussion of the general
definition of metastable states in finite dimension see \cite{gaveau};
for the specific case of glassy systems see \cite{kurchan-biroli} and
\cite{zamponi-08}.

First of all, even though the supercooled liquid is of course
metastable with respect to the crystal, that is not the metastability
we are interested in.  We assume that the crystal nucleation time
is much larger than the liquid relaxation time, therefore the
supercooled liquid is for all practical purposes our reference
equilibrium phase. Its free energy is equivalent to what we called $F$
in the $p$-spin. The supercooled liquid is the ergodic phase of the
system, and we are rather interested in the metastable sub-components of
this ergodic phase, whose free energy is equivalent to $f_\alpha$ in
the $p$-spin.

But what are these metastable sub-components of the supercooled liquid
phase?  As the reader may have understood by now, there is a weak,
hand-waving and often unspoken consensus that, when the temperature is
low enough, metastable states are not too different from amorphous local
minima of the potential energy. Consider a local minimum with energy
density $E$.  If the energy barriers around this minimum are large
compared to $k_BT$, the system remains confined within this minimum
for a long time. Given an energy minimum, it is  possible to
define and compute its vibrational entropy $S_\mathrm{vib}$, since
this requires simply to know the vibrational frequencies that are
derived from the matrix of the second derivative of the potential,
i.e. the Hessian \cite{modi-entropia}. In this way, we can define, 
\be
f = E - T S_\mathrm{vib} \ ,
\label{buonanotte}
\ee 
to be the free energy density of this local minimum of the energy,
i.e. of this metastable state (as usual, this makes no sense for hard
spheres).  In this same way, one can associate a metastable state to
each energy minimum and make a perfect matching between potential
energy landscape and free energy landscape. In this framework, one
can calculate the configurational entropy of metastable states just by
counting the number of minima of the energy, rather than of the free
energy, and get some interesting thermodynamic results based on this
local energy minima approach \cite{sciortino-99, sciortino-00,
sciortino-02}.

Unfortunately, this program has got two serious shortcomings.  First
problem, the role of the {\it temperature}. Any reasonable free energy
landscape depends on the temperature. In particular, not only the free
energy $f$ of a state, but also the very existence of the state must
depend on $T$.  On the contrary, the energy landscape does not depend
on the temperature, and a local (or a global) minimum is so
irrespective of how large $T$ is \cite{biroli-00}.  Second, and far
 more serious, problem, the role of the {\it size}. As we have
learnt when studying nucleation, activation time decreases when the
system's size increases, simply because there are more spots where we are
tossing our coin (to form or not to form the nucleus).  So, when we
say that at low $T$ the system remains confined within a local energy
minimum for a long time, we are very unclear, because by increasing
the size we can actually make this time as short as we want. Hence, 
the notion of a global metastable state seems rather awkward. Let us
discuss these two problems separately.

First, the temperature. The whole program of identifying energy minima
with states rests on the rather casual sentence we wrote above: {\it
``when the temperature is low enough''}, which, we must admit, is
quite vague. At a certain temperature, minima surrounded by small
barriers do not give rise to states, whereas those with large barriers
may be good candidates. A certain minimum, or basin, with large
barriers around can in turn contain a sub-structure of many minima
separated by irrelevant barriers, so that only the larger basin may be
identified with a state. Certainly, not all minima are relevant as
metastable states.  This view gave rise to the  fruitful notion
of {\it metabasins} \cite{heuer-1,heuer-2}.  When we change $T$ the
structure of states may change more or less radically: minima that
were too shallow to be identified with states at higher $T$, may
become good candidates at lower $T$, and it may be hard to follow the
structure of states in temperature.

We could say something safer if we had a precise notion of the size of
barriers.  For example, imagine there were only two scales of barriers
(or a bimodal distribution of barrier size), a large barrier
$\Delta_1$ and a small one $\Delta_2$, with $\Delta_1 \gg \Delta_2$.
Then, in the temperature regime $\Delta_2 < k_B T < \Delta_1$, it
would be somewhat safe to assume that all configurations connected by
barriers at most of size $\Delta_2$ belong to the same `state',
whereas different `states' would be separated by barriers
$\Delta_1$. Unfortunately, in general we have no idea of what is the
structure of barriers. Moreover, as we have seen, we have the strong
feeling that barriers must depend on temperature, through the
$T$-dependence of the correlation length.  The situation is thus
horribly complicated.

To make things a bit clearer, it is important to emphasize the crucial
role of the {\it time scale}: if we fix a time scale $t$, it makes
much more sense to say whether or not a certain region of the phase
space is a metastable state: we check whether or not the time needed
to get out (via activation) of that region is larger than
$t$. Therefore, it is reasonable to talk about metastable states as
regions of the phase space with finite lifetime \cite{kurchan-biroli,
gaveau}. Unless temperature is {\it extremely} low, however, any such
phase space region does not consists of one single minimum: there are
typically many relatively small rearrangements of the particles that
contribute to the same state, simply because they can all be activated
within the fixed time-scale $t$ at that particular temperature $T$.
Of course, such view of metastable state is the safer the lower $T$,
while we have to be very careful close to $T_x$, where activated
dynamics is no longer the main mechanism of diffusion. Intuition tells
us that the Goldstein's crossover temperature $T_x$ (and thus the MCT
transition $T_c$) should work as an upper stability limit of any
sensible definition of metastable states.

The fact that the lifetime of any metastable state is finite in finite
dimension, is actually quite useful when it comes to apply the
$p$-spin mechanism out of mean field. In the $p$-spin, dynamics and
thermodynamics are completely decoupled, and this is unpleasant.
For $T_k < T < T_c$, the global equilibrium state of the system
consists of a superposition of exponentially many metastable
states. But how can the system visit all these states and be ergodic,
if, due to the mean field nature of the model, barriers are infinite?
Thermodynamics just sees one ergodic component, missing the fact that
this component is in fact made up of various metastable states. On the
other hand, dynamics tells us that a system thermalized within one of
these metastable states remains stuck there forever. In the $p$-spin,
thermodynamics is ergodic, but equilibrium dynamics is not. This is a
typical artefact of mean-field.  The finite (albeit long) lifetime of
metastable states in real supercooled liquids, however, reconciles
dynamics with thermodynamics. The exponentially many metastable states
composing the liquid phase are surrounded by large, yet finite
barriers, which can be crossed as long as the system is in
equilibrium.  At a given temperature $T$, only those states that
optimize the balance between free energy and configurational entropy
(equations \eqref{busu} and \eqref{saddelina}) dominate the partition
function and the equilibrium dynamics. Their energy is $f^\star(T)$
and their configurational entropy is $S_c(T)\equiv
S_c(f^\star(T))$. The lower the temperature, the lower $f^\star(T)$,
and thus the lower the configurational entropy.  As long as $T>T_k$,
$S_c(T)>0$ and thus there are exponentially many metastable states the
system actually visits.  Their number becomes sub-exponential at
$T_k$, where $S_c$ is zero and the entropy crisis occurs. In this
picture, there is no difference between equilibrium dynamics and
thermodynamics, and the configurational entropy is a measurable
contribution to the total entropy of the system.

All is well, then? Not really. We still have to face the second, and
definitely more serious problem in defining metastable states: their size.
There is something badly missing in the notion of metastable states as
described in this section, and in previous parts of these notes, and
this is the role of {\it real space}. We have already briefly remarked
this before, but it is now time to face directly this point, because
it is really crucial.  All along these notes we always talked about
{\it global} metastable states, as if the entire system were at any
given instant in just one state. However, if we go back to Goldstein's
remarks about the local nature of particle rearrangements in real
space, we understand that this global view must be wrong. Goldstein
himself noted that different parts, far from each other, of an
infinitely large system, rearrange continuously, so that the whole
system, represented by a single phase space point, is {\it always} on
top of a barrier, rather than vibrating around the bottom of a global
minimum. This means that, globally, the lifetime of a single minimum
is {\it zero}, and all our nice picture above falls apart!

We already faced an identical problem when discussing crystallization:
the nucleation rate is defined as the number of events per unit time
per unit {\it volume}, so that the nucleation time becomes as small as we
want when we increase the system's size
\footnote{I hope the reader does not get confused by the fact that in
this crystallization example the role of metastable state is played by
the supercooled liquid phase.}.  Similarly, in a low temperature
liquid, even though activated local rearrangements are rare, we
can always increase the system size so much as to make the time
between two consecutive rearrangements as small as we want: at any
time, somewhere in the system a bunch of particles exits from its {\it
local} state, thus bringing the whole system out of its {\it global}
state.  Clearly, the life-time of the state we are interested in is
the local one, not the global one, which is trivially related to the
occurrence of different independent rearrangements. Hence, the global
landscape scenario must somehow `factorize' when we consider local
dynamics \cite{bu-bello}.

Indeed, as we remarked when discussing Goldstein's scenario, it is
only {\it locally} that the intuitive notion of activation, and thus
of metastable state, is recovered. If the observation time scale is
short `enough' and the region of the system we are observing is small
`enough', then it makes sense to say that this region is in a certain
metastable state.  Hence, metastable states must be defined locally
not only in {\it time}, as we discussed above, but also in {\it
space}.  Of course, the key issue is how we define `enough' in the
sentence above. As for time, activation tells us that the time scale
for rearrangement is approximately $\exp(\xi^\psi/T)$, where $\xi$ is
the typical size of the rearranging region. So, any time smaller than
this is fine for the existence of the metastable state. Regarding
space, the crucial scale is $\xi$ itself: if the system under observation is
much larger than $\xi$, different regions rearrange independently, and
the whole system is most of the time on top of a barrier, as noted by
Goldstein.  If, on the other hand, we observe the system over a scale
$\xi$, or smaller, then we really see a sound metastable state.

In conclusion, metastable states must go local. But how? We have
understood that locality in both time and space depend on the typical
size of the rearranging regions $\xi$. But what fixes $\xi$? In the
next few sections we shall try to answer these questions and  
reconcile metastable states with real space structure.

\subsection{The mosaic theory}

The mosaic theory (also known as Random First Order Transition -
RFOT), was formulated by Kirkpatrick, Thirumalai and Wolynes in the
late 80s. The original aim of their papers was to investigate the
potential relationships between supercooled liquids and the $p$-spin
class of mean-field spin-glasses, which was conjectured in
\cite{kirk-0}. After studying the dynamic and thermodynamic behaviour
of the $p$-spin model \cite{kirk-1, kirk-2, kirk-3}, it was introduced
in \cite{mosaic} a real space thermodynamic description of metastable
states in supercooled liquids, partly inspired by the $p$-spin
results. This was the mosaic theory.

Let us start with a rather crucial statement: if we accept that a
finite dimensional system may have many states, then different
physical portions of the system can be found in different states.
When this happens, there must be some sort of interface separating the
two states (let us call them $\alpha$ and $\gamma$) and it is
reasonable to believe that an energy cost is associated to the
creation of this interface, due to the mismatch of the two amorphous
configurations along the interface.

This is the first key ingredient of the mosaic theory: in a finite
dimensional system we can have interfaces between different states,
and thus a {\it surface tension}. Note that the price we must pay in
going from one state to another is the only sizeable consequence of
the existence of different amorphous states; thus, assuming the
existence of a surface tension is basically the same as assuming the
existence of many amorphous states. Indeed, all metastable states we
are talking about are amorphous, so that locally they all look the
same, and it is not possible to distinguish them by using a standard
local order parameter. Therefore, were it possible to crossover from
state $\alpha$ to state $\gamma$ with no energy expense, it would be
hardly reasonable to say that they are two different states.

Strictly speaking, surface tension is defined as the free energy cost
per unit area to create an interface. This very definition assumes
that the cost of an interface of linear size $L$ scales like
$L^{d-1}$, i.e. $L^2$ in three dimensions. In supercooled liquids,
however, we cannot be a priori sure about this point, and a safer
assumption is to say that the free energy cost is, \be \Delta F = Y \;
L^\theta \quad , \quad \theta \leq d-1 \ .  \ee In this case $Y$ is a
`generalized' surface tension, simply because if $\theta$ turns out to
be strictly smaller than $d-1$, we cannot talk about free energy per
unit area.  There are at least two reasons why $\theta$ could be
smaller than $d-1$: first, the free energy cost associated to an
interface depends on the order parameter, for example it is different
in the Ising model ($\theta=d-1$) and in the Heisemberg model
($\theta=d-2$). This exponent is typically smaller, the softer the
order parameter \cite{shukla}. We are not quite sure of what is the
right order parameter for amorphous states in liquids, so we better be
careful. Second, the very effect of disorder can be such as to reduce
the value of $\theta$; this may happen when the interface can decrease
its energy by passing through more favourable points and thus take a
curly shape \cite{huse-85, henley-85}. Also in this case, we have no
idea how an interface between amorphous states in liquids looks like,
so it would be a mistake being hasty about the value of $\theta$. The
only assumption, thus, is that whenever two states are in contact one
with another, there is a free energy cost ruled by a generalized
surface tension $Y$.

Why is the surface tension important? Because in finite dimensional systems the surface tension is the main ingredient of the free energy 
barriers between states. Imagine that the system is in a state $\alpha$. Thanks to thermal fluctuations,
there is a chance that a certain region (a droplet) of linear size $R$ rearranges. This is the idea of cooperatively rearranging region (CRR) in AGDM description. 
In a multi-state framework, it is reasonable to assume that the particles within the rearranging region do not rearrange to a random configuration, 
but rather make a transition to another of the many available states, say $\gamma$, 
so that the bulk free energy of the region in the new configuration will be similar as before. When this happens, there is a 
 free energy price that has to be paid due to the $\alpha - \beta$ mismatch at the interface between the rearranging region and the rest of the system, 
\be
\Delta F_\mathrm{cost}=Y \, R^\theta \ .
\label{cost}
\ee
The crucial question we must answer now is: 
what fixes the typical size of the rearranging region? The AGDM theory 
fixes this size in a purely combinatorial way, completely disregarding the surface tension between different CRRs.
The mosaic theory proceeds in a different way, getting its main inspiration from classic nucleation theory (CNT) \cite{kirk-1,mosaic}. 
In CNT the size $R_c$ of the critical nucleus is fixed by the balance between the surface tension cost $\sigma R^{d-1}$ and 
the thermodynamic gain $\delta f R^d$, where $\delta f$ is the free energy difference between the background state and the nucleated one
(see Section III.A). In liquids, we do have a surface tension term, albeit with a nonstandard exponent,  $Y R^\theta$. However, it is unclear
what the thermodynamic gain should be: it cannot be $\delta f$, since the states dominating the partition function all 
have the same free energy density $f^\star$, fixed by equation \eqref{saddelina}. 
Indeed, the free energy of the system does not decrease by subsequent rearrangements, otherwise the system would not be at equilibrium. 
According to the mosaic theory \cite{kirk-1, mosaic}, 
the thermodynamic drive to rearrange a droplet of radius $R$ is provided by the fact that such a region
has an exponentially large number of available states. There is an entropic price that must be paid by the region to 
stay in just one of these many states, and this entropic price can be released if the rearrangement takes place. Therefore, 
the thermodynamic drive is the free energy contribution from the total configurational entropy available to that region, which is equal to,
\be
\Delta F_\mathrm{gain}=-T S_c(T)\; R^d  \ ,
\label{gain}
\ee
where the minus sign emphasizes that this is a free energy gain, opposed to the free energy cost in \eqref{cost}. In \eqref{gain} and 
in the following by $S_c(T)$ we actually mean $S_c(f^\star(T))$. Given that $\theta \leq d-1$
the balance of powers between surface tension cost and configurational entropy gain works very similarly to standard nucleation: for small $R$ the cost
dominates and there is no net thermodynamic gain in the formation of a droplet. On the other hand, for large $R$ the entropic drive dominates;  
this happens for any size $R$ larger than a critical value $\xi$ fixed by the balance of \eqref{cost} and \eqref{gain},
\be
\xi = \left( \frac{Y(T)}{T S_c(T)}\right)^\frac{1}{d-\theta}  \ ,
\label{ukulele}
\ee
where we have emphasized that both surface tension and configurational entropy depend on temperature.
According to the mosaic theory, $\xi$ is the typical size of the rearranging regions. This size is inversely proportional to the configurational
entropy, although with a nontrivial exponent, and in this respect we recover the most important result of the AGDM theory, that is the fact
that the correlation length increases when the configurational entropy decreases. Note, moreover, that $\xi$ is also proportional to the
surface tension, and this is an ingredient that is completely absent in AGDM.
Expression \eqref{ukulele} for the correlation length was first derived in \cite{kirk-1} in the specific case $\theta=d-1$.

The mosaic mechanism has deeper consequences than the growth of $\xi$
with decreasing $S_c$. Imagine we prepare a zero temperature system in
a global local minimum of the potential energy.  There is no
contradiction in doing this and it can actually be done numerically. 
Now we heat up the system at our working temperature $T$, with
the (naive) purpose to promote our global energy minimum to a global
metastable state. We would soon find that this is not possible! What
happens is that, due to thermal fluctuations, the global
configurations is immediately {\it unstable} against the formation of
droplets of size $R\sim\xi$ of other locally stable arrangements, so
that after a while the system's configuration would look like a
`mosaic' of many different local states, in a continuous process of
rearrangement on a scale $\xi$ 
\footnote{In fact, to us the system would look like nothing at all:
local arrangements are all equally amorphous, so they would all look
the same. We certainly would not `see' the mosaic the way we can `see'
domains within a magnetic system.}. We conclude that the very notion
of a system globally living in a single metastable state does not make
sense at all. The best we can do is to globally prepare it in a single
energy minimum, which is a slightly different matter. This means
that metastable states can only be defined over a scale $R < \xi$. The
notion of metastable state is meaningless for portions of the system
larger than $\xi$. This solves the real-space dilemma we faced in the
last section. Note that this local definition of metastable state is
very similar in spirit to Goldstein's remark quoted in Section V.A:
the notion of a system spending a long time vibrating in a local
minimum of the energy makes sense only on a local scale.

Let us summarize the mosaic results up to now. First, compared to AGDM
scenario, the role of the number of accessible states is much clearer:
this number is of order $\exp(R^d S_c)$ and it thus scales up with the
size $R$ of the region. However, when $R<\xi$ the region is unable to
explore all its available states, because the surface tension price it
would pay in the rearrangement is larger than the entropic gain in
getting out of the original state. Thus, the number of explored states
is just $1$, when the region is not large enough.  This clears up the
AGDM ambiguity between the number of configurations that are virtually
accessible, which is always exponentially large, and the number of
configurations actually visited by a region, which is $1$ if the
region is smaller than $\xi$, and exponentially large if he region is
larger than $\xi$.  Second, the mosaic introduces and deals with the
key concept of generalized surface tension $Y$, thus confirming a
rather reasonable expectation: the typical lengthscale, and thus the
barrier, can be large not only as a result of a low configurational
entropy, but also because there may be a large energy cost $Y$ in
putting in contact two different states.

Despite the differences with AGDM theory, the mosaic still brings home
the crucial result, namely a characteristic lengthscale that increases
when the configurational entropy decreases. This is at the basis of
any relationship between the sharp raise of the relaxation time
$\tau_{\mathrm R}$ and the empirically observed drop of the
configurational entropy in supercooled liquids. However, the result is
not quantitatively the same as in AGDM: within the mosaic picture
$\xi\sim (1/S_c)^{1/d-\theta}$, whereas within AGDM $n\sim 1/S_c$, and
thus $\xi\sim n^{1/d}\sim (1/S_c)^{1/d}$ ($n$ is the number of
particles in the rearranging region).  In temperature, this implies,
\be
\xi_\mathrm{mosaic}\sim \left(\frac{1}{T-T_k}\right)^\frac{1}{d-\theta}  \ ,
\ee
whereas,
\be
\xi_\mathrm{AGDM}\sim \left(\frac{1}{T-T_k}\right)^\frac{1}{d}  \ .
\ee
Hence, given that $\theta \leq d-1$, we conclude that the increase of the correlation length predicted by the mosaic
theory is sharper than that in AGDM theory. 

The two frameworks are also different for what concerns the dependence
of the relaxation time on $S_c$. According to AGDM the barrier scales
as $\Delta\sim n\sim1/S_c$, and thus the relaxation time is
$\tau_{\mathrm R}\sim\exp(A/TS_c)$, which naturally leads to VFT law
(see eq. \eqref{agdm3}).  The way the mosaic fixes the barrier is
different \cite{mosaic}.  As we have seen, one of the main sources of
inspiration for the mosaic theory is standard nucleation. In this
context, the barrier is fixed by the maximum of the free energy cost
for droplet formation, \be \Delta F= Y R^\theta - T S_c R^d \ .  \ee
The position of the maximum is (up to a numerical factor) $R=\xi$,
whereas the value of $\Delta F$ at the maximum, i.e. the barrier
$\Delta$ is, 
\be 
\Delta =\frac{Y(T)^\frac{d}{d-\theta}}{[T
S_c(T)]^\frac{\theta}{d-\theta}} \ ,
\label{baronazzo}
\ee
and thus for the relaxation time we get,
\be
\tau_{\mathrm R} = \tau_0 \; \exp\left(\frac{Y(T)^\frac{d}{d-\theta}}{T\; [T S_c(T)]^\frac{\theta}{d-\theta}}\right)  \ ,
\label{bollo}
\ee where we recall that, \be S_c(T) \sim \Delta c_p \;
\frac{T-T_k}{T_k} \ .  \ee We therefore see that the mosaic theory
does not provide directly the VFT law \eqref{vft}, because the precise
dependence on $S_c$, and thus on $T-T_k$, depends on the exponent
$\theta$.  In \cite{mosaic}, however, it was claimed, using
renormalization group arguments, that $\theta=d/2$. In this case,
formula \eqref{bollo} boils down to $\tau_{\mathrm
R}\sim\exp(A/TS_c)$, and despite their many differences, the mosaic
and AGDM theories would give the same final result. It is hard to say
whether the claim $\theta=d/2$ is correct or not.  Certainly, such claim is
useful to get VFT law. However, one can argue that a fit where the
exponent of $S_c$ (and thus that of $T-T_k$) is a free parameter as in
\eqref{bollo}, would make an even better job than VFT. Thus, it is
probably not worthwhile to make uncertain claims on the value of
$\theta$, just for the sake of getting the exact VFT law.

It is more interesting to note that the barrier $\Delta$ in \eqref{baronazzo} can be rewritten as,
\be
\Delta_\mathrm{mosaic} \sim \xi^\theta\; Y   \ .
\label{zumpazumpa}
\ee
This result is at variance with AGDM, according to which, 
\be
\Delta_\mathrm{AGDM} \sim n \sim \xi^d  \ .
\ee
Given that $\theta\leq d-1$,
the mosaic barriers grow slower than than the number of particles contained in the
rearranging region. Conceptually, this makes a lot of sense: the barrier, according to the mosaic, 
is dominated by the surface tension term, because at the surface is where we actually pay energy. But the surface tension term is ruled 
by the exponent $\theta$, and so is the barrier. As anticipated before,  the exponent $\psi$ linking the barrier to the linear size of
the rearranging region (equation \ref{psi}) is not necessarily equal in all frameworks: for AGDM $\psi=d$, while for the mosaic $\psi=\theta$.
As we have seen, if $\theta=d/2$, as advocated in \cite{mosaic}, the barrier scales with temperature
 in the same way in the mosaic and in the AGDM theory, $\Delta \sim (T-T_k)^{-1}$.

Equation \eqref{baronazzo} shows that the
surface tension critically enters into the barrier for
rearrangement. Thus, barriers depend on the temperature also through
the $T$ dependence of $Y$. Up to now we have not discussed how the
surface tension can change in temperature. In fact, at very low $T$ it
is reasonable to believe that $Y$ does not vary strongly with $T$,
and that it levels to some finite value. Thus, the increase of the
barrier $\Delta$ at low temperatures is predominantly triggered by the
drop of the configurational entropy. At higher temperatures, on the
other hand, the situation is different. From the discussion in the
last section, we do not expect the multi-state scenario we are
adopting here to hold also at high temperatures and we already
suggested that states (whatever they are) become ill-defined above
Goldstein's crossover $T_x$.  This requires the surface tension $Y$ to
go to zero close to $T_x$.  In the context of the mosaic theory,
therefore, the Goldstein's temperature $T_x$, and thus also the 
MCT transition $T_c$, acts like a thermodynamic
spinodal point, where the surface tension vanishes, setting the limit
of validity of the theory itself. We recall that $T_x$ can also be
interpreted as the locus of a topological transition between stable
minima and unstable saddles \cite{grigera-02}. Saddles cannot sustain
a finite surface tension, because the system can always rearrange by
exploiting the negative-unstable modes. Therefore, the topological
view of $T_x$ fits well with the idea of a spinodal fixed by the
equation, \be Y(T_x)\sim 0 \ , \ee in the mosaic context. 

It is 
interesting to note that the interpretation of $T_x$ as a spinodal
point is supported also by the $p$-spin model. As we have seen, $T_x$
must be identified with the MCT dynamical transition $T_c$, and this
in turn is exactly the same as the dynamical transition $T_c$ of the
$p$-spin model. In \cite{recipes} it has been introduced an effective
potential, given by the free energy cost one must pay to thermalize a
system $B$ at fixed distance (in phase space) from an equilibrium
system $A$. This effective potential (expressed as a function of the
distance between the two systems), works similarly to the standard
mean-field free energy in first-order phase transitions: above $T_c$
it has just one minimum, while at $T_c$ it develops a secondary,
metastable minimum, due to the existence of metastable states in the
system. Thus, in the first-order transition framework provided by this
effective potential, the dynamical transition $T_c$ acts like a
thermodynamic spinodal, i.e. the temperature above which the
metastable minimum of the free energy disappears and the free energy
barrier between states vanishes \cite{recipes}.

Despite its many interesting aspects, there are some unclear points about the mosaic mechanism. First, in what precise sense the configurational entropy is the drive to rearrangement?
How can the droplet know about the many states available to it? We would not ask the same in standard nucleation, because energy (and free energy) is a 
much clearer driving force.
The same is not true for entropy: the new configuration would just be {\it one} of many others, exactly like the {\it one} configuration the droplet started from. Where is the gain?
We need a pedestrian definition of entropic drive to rearrangement. Second, why rearranging droplets do not get {\it larger} than $\xi$? If standard nucleation is the inspiration of mosaic, then one
would naively conclude that once $R>\xi$, the droplet is encouraged to grow indefinitely, just like a crystal nucleus does. Is it so? Third, at first sight the mosaic dynamics does not seem 
steady-state: the system starts in a homogeneous configuration, and then it breaks up in a multi-state pattern, just like a mosaic. What is this pattern? 
Is it a typical configuration of the system or not?
Apparently, it is not, because there are many inter-states interfaces where surface tension energy is concentrated, which seem to be absent in the original homogeneous configuration. 
How do we solve this paradox? Let us see this three points one by one.

The expression `entropic drive to rearrangement' is catchy, but in
fact quite a misleading one. What actually drives the rearrangement,
in the mosaic theory as in any other scenario, is simply thermal
fluctuations. However, once the region has been pushed out from its
original state by thermal fluctuations, it makes sense to ask whether
it is more likely for the droplet to go back to the original state or
to stay in the new, or any other, state. This is where entropic
considerations about the number of available states come into play. If
there are very few states, then the memory of the original state
provided by the surface tension at the interface is sufficient to pull
the droplet back. If, on the other hand, the droplet is large, there
are exponentially many states available; once the transition has been
done it is harder for the droplet to go back. Therefore, we see that
it is somewhat easier to understand the mosaic mechanism if we think
about an already rearranged region and talk about `surface tension
drive to go back', rather than the `entropic drive to get out'.

Standard nucleation theory seems a natural paradigm for the mosaic
picture, but we must be in fact very careful about this comparison.
In standard nucleation the growth of a nucleus beyond its critical
size is guaranteed by the fact that this process {\it decreases} the
total free energy of the system. In the mosaic scenario, nothing of
this sort happens, because the newly formed droplet of a different
state typically has the same free energy density as the rest of the
system, so there is no gain in making it bigger. In fact, if we
remember that the droplet formation is ruled by thermal activation and
that the barrier will scale as a power of the size, it is reasonable
to believe that the regions that rearrange have a typical size equal
to the {\it smallest} thermodynamically stable droplet, and this is
$\xi$. Smaller droplets are quick to form (small barrier), but are
also thermodynamically unstable, because surface tension pulls them
back; droplet larger than $\xi$, on the other hand, are
thermodynamically stable, but need to cross an exponentially larger
barrier and thus take a very long time to rearrange.  We conclude that
$\xi$ is indeed the typical size of the rearranging regions in the
mosaic picture.

Finally, is the mosaic mechanism steady-state?  Yes.  Whenever a
rearrangement takes place, the global mosaic pattern that is produced
is in turn a typical configuration of the system, with the same
physical and statistical properties as any other equilibrium
configuration. Therefore, the misleading concept that must be rejected
is the one of a homogeneous configuration, as opposed to a
inhomogeneous mosaic configuration crossed by energy-leaden
interfaces. To fix ideas, let us forget about states and consider a
certain potential energy minimum $\alpha$. Now, imagine that a region
of the system of size $\xi$ is rearranged, making a transition to a
different configuration that locally would be part of a different
minimum $\beta$.  If we now re-minimize the energy, and relax the
large stress that is produced at the interface, the system will be
found again in a new energy minimum, which is very similar to $\alpha$
far from the rearranged droplet, and to $\beta$ within the droplet.
Such new minimum $\gamma$ is statistically identical to $\alpha$ and
$\beta$. In particular, the residual surface energy of $\gamma$
located close to the droplet interface is just a local etherogeneity
of the system, as the typical stress and strains that are always found
in disorder systems.  The initial minimum $\alpha$ has exactly the
same spatial etherogeneities in energy even before the formation of
the new droplet $\beta$. These are the result of the continuous
rearrangements taking place in the system. If we now switch from
minima to states, we conclude, as already done before, that the very
notion of global metastable state is in fact misleading. Only on scale
$\xi$ it makes sense to talk about states, whereas for $R\gg\xi$ the
system is unstable against the fragmentation in sub-regions of
correlated particles. The concept of a homogeneous global state,
crossed by no interfaces, does not correspond to anything real.

\subsection{A Gedankenexperiment}

In \cite{gedanken} it was proposed a reformulation of the mosaic theory that clarifies some of the points raised above and that puts on a firmer basis the theory.
Let us make the hypothesis that a liquid, at low temperature, is trapped into a global metastable state $\alpha$. We want to prove that this hypothesis, i.e. the fact
that the state is global, is thermodynamically unstable and show that only below a certain length scale we can talk about metastable states. 

Within the system, let us focus on a sphere of
radius $R$.  We want to calculate the thermodynamics of such sphere, and in particular its probability to rearrange into a different state. To do this we assume
that all particles outside the sphere are frozen (this is the `gedanken' part of the experiment). 
This assumption is not that crazy: the particles far from the region do not interact with it, so that we can forget about them (if they rearrange, they do that independently
from the considered region), whereas the non-rearranging 
particles in the vicinity of the rearranging region are quite immobile at low $T$, with only small vibrations around their local equilibrium positions.
The main effect of these external particles is to produce a pinning field at the interface that tend to keep the sphere in the original state $\alpha$. 

To compute the partition function
we make a decomposition similar to that 
of equation \eqref{buscadero}, where from a sum over all configurations of the sphere we go to a sum over all states available to the sphere. 
In this case, however, we have to remember that one of these states, namely $\alpha$, is privileged compared to all the others, because
it does not have to pay any  surface tension cost.
We can therefore split the partition function of the sphere in two parts, $Z_\mathrm{in}$ given by the sum over all configurations belonging 
to state $\alpha$, plus a term $Z_\mathrm{out}$ given by the sum over all other configurations. We have,
\be
 Z_\mathrm{sphere} = Z_\mathrm{in} + Z_\mathrm{out} ,
\ee
with,
\bea
&& Z_\mathrm{in}  = \exp(-\beta f_\alpha R^d) \nonumber \\
&& Z_\mathrm{out} = \sum_{\gamma\neq\alpha} \exp(-\beta f_\gamma R^d -\beta Y R^\theta) = \nonumber \\
&=& \int df \, \exp[S_c(f) R^d -\beta f R^d -\beta Y R^\theta] \ ,
\label{zuppo}
\eea
where we have used the definition of configurational entropy of the sphere,
\be
\sum_{\gamma} \delta(f-f_\gamma) = \exp[S_c(f) R^d]   \ .
\ee
The integral in \eqref{zuppo} can be approximately calculated by using the saddle point method, provided that $R^d$ is large enough. 
Similarly to what we did for equation \eqref{pinotto}, this amounts to find the solution $f^\star$ of the saddle point equation,
\be
S_c'(f^\star) = 1/T   \ .
\ee
State $\alpha$ is one of the many metastable states dominating the total partition function
of the system, and thus also its free energy is equal to $f^\star$: as expected, 
the rearrangement of the sphere does not bring the system to a lower free-energy level (on average).
Thus we have,
\bea
&&Z_\mathrm{sphere} =  Z_\mathrm{in} + Z_\mathrm{out} \nonumber \\
&&=\exp(-\beta f^\star R^d)\nonumber \\
&& + \exp[S_c(T) R^d -\beta f^\star R^d -\beta Y R^\theta] 
\eea
where, as usual, $S_c(T)=S_c(f^\star(T))$.
From this partition function we can work out the probability $p_\mathrm{in}$ for a sphere of radius $R$ to be found in its original state $\alpha$,
and the probability $p_\mathrm{out}$ for it to switch to one of the exponentially many other states,
\bea
p_\mathrm{in}(R)&=& \frac{Z_\mathrm{in}}{Z_\mathrm{sphere}} = \frac{\exp[\beta Y R^\theta]}{\exp[\beta Y R^\theta]+  \exp[S_c R^d]}
\label{pin}
\\
p_\mathrm{out}(R) &=& \frac{Z_\mathrm{out}}{Z_\mathrm{sphere}} = \frac{\exp[ S_c R^d]}{\exp[\beta Y R^\theta] + \exp[S_c R^d]}  \ ,
\label{pout}
\eea
where we remember that,
\be
\theta \leq d-1 \ .
\ee
Relations \eqref{pin} and \eqref{pout} are the core of the mosaic theory. The role of the surface tension $Y$ and of the 
configurational entropy $S_c$ can be clearly read from these equations: $Y$ is the force trying to keep the sphere {\it in} the original state, and thus it enters in 
the exponential of $p_\mathrm{in}$, whereas $S_c$ is the reason to get {\it out} of the original state (now we now how to interpret this statement),
so that it enters in the exponential of $p_\mathrm{out}$. The two probabilities balance for $R=\xi$ given by,
\be
\xi = \left( \frac{Y(T)}{T S_c(T)}\right)^\frac{1}{d-\theta}  \ ,
\nonumber
\ee
which is the mosaic correlation length \eqref{ukulele}.
For small values of $R$, that is for $R< \xi$, the surface tension term dominates, and
thanks to the sharp exponential form of the probabilities we have,
\bea
p_\mathrm{in}(R<\xi) &\sim& 1 \nonumber \\
p_\mathrm{out}(R<\xi) &\sim & 0  \ ,
\label{minore}
\eea
so that the sphere has a very small probability to change state due to the overwhelming effect of the pinning field at the interface.
On the other hand, for $R > \xi$ we have the opposite,
\bea
p_\mathrm{in}(R>\xi) &\sim& 0 \nonumber  \\
p_\mathrm{out}(R>\xi) &\sim& 1   \ ,
\label{maggiore}
\eea
and the sphere is found in a different state with probability one: the pinning field and the surface energy are thermodynamically overwhelmed by the
configurational entropy. 

We have therefore proved that only on scales smaller than $\xi$ we can talk about a region of the system 
being trapped by a metastable states. On larger length scales, the notion of global homogeneous state does not make sense, or, better,  it is 
thermodynamically unstable against fragmentation into a mosaic-like pattern of correlated regions of smaller size $\xi$.

Note that, if the control parameter regulating the balance between $p_\mathrm{in}$ and $p_\mathrm{out}$ were the temperature or
the magnetic field, rather the radius of a funny sphere, 
what we have just described would look very much like the standard energy vs. entropy interplay at the basis of 
first order transitions, like the crystal-liquid one: at low $T$ the energy dominates, so that equilibrium is given by the broken-ergodicity phase,
i.e. the crystal, whereas at large $T$ the entropy dominates, and equilibrium is given by the ergodic phase, i.e. the liquid. 
In the mosaic theory
something quite similar happens, due to the different powers of $R$ in the prefactors of $Y$ (the energy) and $S_c$ (the entropy): at low $R$ energy
dominates and ergodicity is broken (the sphere remains trapped in a single state), at large $R$ entropy dominates and ergodicity is restored (the sphere is
free to visit all available states). Such `first-order transition' interpretation of the mosaic mechanism is conceptually very useful. 
Whenever we will talk about a `transition' in the following, it will be understood as a transition in $R$.

Unlike the original mosaic formulation of \cite{mosaic}, the thermodynamic framework of \cite{gedanken}  
does not establish a link with nucleation theory. 
As a consequence,  the exponent $\psi$ connecting the free energy barrier to the size of the rearranging region,
$\Delta F \sim \xi^\psi$, remains as a free parameter of the theory, rather than being equal to $\theta$
as in the original mosaic formulation \cite{mosaic}. Another consequence of avoiding the comparison with nucleation is that the question
of why the region does not grow beyond $\xi$ does not seem an issue anymore. The rearrangement of regions larger than $\xi$ 
is thermodynamically favoured, but the time needed to actually make this rearrangement scales as $\exp(\xi^\psi/T)$, so that
typically only regions of the minimal stable size, that is $\xi$, will be rearranged.

The thermodynamic rephrasing of the mosaic mechanism tells us once again that metastable states
can only be defined up to a scale $\xi$. Beyond that size the liquid ergodic phase (containing {\it all} the other states) dominates
the partition function. In this sense, we may say  that the mean-field scenario inspired by the $p$-spin model
is valid only on lengthscale smaller than $\xi$. Given that $\xi$ increases (diverges) when $S_c$ decreases (vanishes), the 
validity of the mean-field scenario is the more robust the closer we are to Kauzmann's temperature $T_k$. 
However, we stress once more that whether the singularity at $T_k$ exists or it is avoided, is 
irrelevant for the validity of the physical scenario we have described. The mosaic relaxation mechanism 
is valid even in presence of  
a kinetic spinodal, or a change in slope of $S_c$ (just to make two examples of avoided Kauzmann transition). It is the 
{\it incumbent} transition, rather than the transition itself, that defines the physics of the system.

At the end of this long road, from Goldstein to the mosaic, passing through Kauzmann's entropy crisis and Adam-Gibbs-Di Marzio
theory, we have finally answered most of the questions we asked  about the low $T$ phase of supercooled liquids.
We understood that barriers increase when lowering the temperature because larger and larger regions of the systems must be
rearranged in order to restore ergodicity, and because the size of the barrier scales as some power of the size of the rearranging region.
Growing barriers explain the super-Arrhenius increase of the relaxation time in fragile liquids. Moreover, and
perhaps more importantly, we understood that the empirical correspondence between the raise of the relaxation time
and the drop of the configurational entropy noted by Kauzmann is not just a coincidence. We found two different mechanisms relating the
size of the rearranging regions to configurational entropy, AGDM and the mosaic. Both mechanisms make sense, and at the end they provide
similar results for the relaxation time, even though the mosaic scenario is somewhat more convincing and can also be better formalized.
All boils down to a competition between volume and (generalized) surface effects, i.e. to a competition between configurational 
entropy and surface tension. At lower temperatures the system spends most part of its time in low energy portions of the phase space, 
where the number of different amorphous minima is smaller. As a consequence,  it takes larger and larger regions to build up 
enough states to overcome the surface tension cost.

The problem now is: are these answers correct? Do we have any empirical evidence, beyond the validity of VFT law, of 
the mosaic scenario? Within the mosaic theory, and by using various approximations,  one can predict how different empirical 
quantities correlate to each other, and also make some quantitative statements \cite{xia-00,xia-01}. However, one may argue that 
other theories fare equally well when it comes to this kind of indirect tests. It would be important to make a {\it direct} check of the key mosaic prediction,
namely the existence of the sharp first-order-like transition encoded in equations \eqref{pin} and \eqref{pout}. In fact,
the thermodynamic set-up proposed in \cite{gedanken}, and that we have described in this section (a free sphere in a frozen environment),
seems to be ideal for a numerical implementation. Moreover,  the transition predicted by the mosaic, that is the way $p_\mathrm{in}(R)$ 
decays from $1$ to $0$ at $R=\xi$, is very sharp indeed, so that it should possible to clearly observe whether or not this transition takes place
in a numerical experiment.  This kind of test is what we will see next. In so doing, we will discover 
a new static correlation function able to uncover the growth of amorphous order.

\subsection{The growth of amorphous order}

When we think about it, we realize that there is something more
fundamental than the mosaic scenario that needs to be empirically
tested: all thermodynamic frameworks we have analyzed in this chapter
rely on the existence of a {\it static} correlation length $\xi$ that
grows when the temperature is lowered. Are we sure that such a thing
exists? After all, we have seen a couple of pages ago that all
standard structural correlation functions in supercooled liquids are
quite boring, and do not show any evidence of a growing lengthscale
connected to the spectacular raise of the relaxation time. So, why
should it be easier now to find a static correlation length? The
answer is that the theoretical frameworks we have analyzed in this
chapter, and in particular the mosaic theory, have taught us something
new about the supposed nature of $\xi$, and we can exploit this
knowledge to build a new kind of correlation function.

The problem about detecting amorphous order is that it is
amorphous. Irrespective of how stupid this sentence may seem, it is
true: a region of correlated particles all belonging to state $\alpha$
look to us exactly the same as a region of particles in state
$\gamma$. In a configuration full of patches of different states (a
situation typical of the mosaic scenario), we are unable with
standard technical tools to detect all these patches, nor the
interfaces between them. Yet, if states all look the same to us, they
do not look the same to each other: a key ingredient of the mosaic
picture is that when two different states are in contact a surface
tension price must be paid at the interface.  If a region wants to
rearrange, it has to fight against the stabilizing pinning field at
the interface between that region and the rest of the system, which
tries to keep the region in its original state $\alpha$, and
eventually to pull it back. The mosaic tells us that in order to
overcome the surface tension, the region needs to be larger at lower
temperature. This can be rephrased by saying that at lower $T$ the
stabilizing effect of the pinning field at the interface penetrates
deeper within the bulk of the region. This fact is clearer if we
understand that the pinning field at the border of the rearranging
region acts as an amorphous boundary condition imposed by the external
state $\alpha$: the growth of the rearranging region at low $T$ is
thus a consequence of the deepening effect of such amorphous boundary
condition.

Using boundary conditions to check whether in a system there is
growing correlation is a classic of statistical physics
\cite{parisi-book}.  In nondisordered, ordinary systems, this is
mostly a mathematical tool needed to sharply define concepts as
ergodicity breaking and the existence of different thermodynamic
states, whereas standard correlation functions are used to measure a
growing lengthscale. In disordered systems, however, the use of {\it
amorphous} boundary conditions seems the only viable tool to measure a
correlation length. If we want to detect the growth of amorphous order
in the system, we need to measure how long-ranged is the effect of
amorphous boundary conditions on the system. Such a study has been
done at a dynamical context in \cite{scheidler-02}. Here, however, we
want to focus on the thermodynamic approach.

The set-up we have used to give a thermodynamic formulation of the
mosaic theory suggests a very practical way to carry out this program
in numerical experiments on supercooled liquids \cite{cavagna-07, biroli-08}
(the same method can also be successfully used in spin systems 
\cite{garr-jack, cammarota}). 
Thermalize a supercooled liquid at low temperature $T$, take a sphere of radius $R$
within this system, and freeze all particles outside this
sphere.  The frozen border of the sphere acts as an amorphous
boundary condition, trying to keep the sphere in its original state $\alpha$. We
now let the particles within the sphere thermalize under the effect of
this boundary condition, and ask what is the asymptotic equilibrium
state of the particles within the sphere.  If the radius is small, the
$\alpha$ boundary condition will rule over the entire sphere making it
hard to change state, whereas if $R$ is large it will be easier for
the particles to rearrange in a state different form $\alpha$ and thus
decorrelate from their initial configuration.
 
Hence, the prediction we want to test is the following: by lowering
the temperature, we should see that it takes larger $R$ to decorrelate
the central particles from the boundary.  Note that this is a weaker
prediction compared to the existence of a sharp transition at $R=\xi$
prescribed by the mosaic; what we want to confirm here is just that
the effect of amorphous boundary conditions is the more long-ranged
the lower the temperature, without much thought about the precise
functional form of this effect.

We have just one technical problem: how do we check whether or not particles have changed state?
Remember, all amorphous states look the same to us, we do not have a single-state order parameter able to say: this is $\alpha$, this
is $\gamma$. In fact, this is a common problem to all disordered systems with many states, including spin-glasses \cite{virasoro}. 
The standard way adopted to deal with this problem is to introduce an 
order parameter that {\it compares} different states, rather than measuring something about one given state. 
The idea is that, if it is impossible to say in what state a certain region is, it may be possible to say whether it has 
changed state or not. The order parameter that compares two states, $\alpha$ and $\gamma$, is normally called overlap 
$q$, and in spin-systems is defined as \cite{cavagna-review},
\be
q_{\alpha\gamma} = \frac{1}{v} \sum_{i\in v} \langle \sigma_i\rangle_\alpha \langle \sigma_i\rangle_\gamma
\ee
where $v$ is the volume of the region over which the overlap is computed
(we have already met something similar to the overlap in the dynamical context, see equation \eqref{corrspin}).
In this formula $\langle \cdot \rangle_\alpha$ indicates
a thermal average performed by summing over all configurations belonging to state $\alpha$. 
In spin systems $\sigma_i=\pm 1$, so that the overlap measures the similarity between state 
$\alpha$ and $\gamma$, by comparing the average value of the local degrees of freedom 
in the two states. 

The overlap is not restricted to spin systems, as $\sigma_i$ can be
any degree of freedom able to differentiate the local arrangement of
particles in different states.  For a liquid, we can proceeds as
follows: we divide space into small cells (i.e. we introduce a fine
grid) and we assign occupation numbers $\sigma_i=1,0$ to each cell
$i$, according to whether or not a particle is present in the cell
\cite{grigera-04}. In this way it is possible to define the overlap
between two states of a certain region. In particular, let us consider
a small volume $v$ at the centre of the unfrozen sphere of radius $R$
and measure the overlap $q(R)$ between the sphere configuration in the
initial state $\alpha$ and the configuration in the equilibrium state
reached by the sphere under the constraint of the $\alpha$ boundary
conditions, \be q(R)=\frac{1}{v} \sum_{i\in v} \langle
\sigma_i\rangle_\alpha \langle \sigma_i\rangle_{R_\alpha} \ , \ee
where by $\langle \cdot \rangle_{R_\alpha}$ we mean an average
performed under the effect of the $\alpha$ boundary conditions on the
interface of a sphere of radius $R$. In this way, a large value of
$q(R)$ means that $\langle \cdot \rangle_{R_\alpha}\sim \langle \cdot
\rangle_\alpha$, the volume $v$ at the centre of the sphere has
remained in a configuration belonging to the initial state $\alpha$;
on the other hand, a value of $q(R)$ close to $0$ means that $\langle
\cdot \rangle_{R_\alpha}\sim \langle \cdot \rangle_\gamma$, where
$\gamma\neq\alpha$ is one of the exponentially many other states
available to the sphere, meaning that the region $v$ has significantly
decorrelated.  The good thing about $q(R)$ is that it can be used
irrespective of the validity or not of the mosaic theory: it is just a
tool to detect the growth of amorphous order in supercooled liquids.

It can be shown that $q(R)$ is a correlation function whose mathematical definition can be formalized more precisely 
\cite{montanari-sem}. Standard correlation functions
measure the correlation between two different points of the system at mutual distance $R$, and for this reason 
they may be called {\it point-to-point} correlation functions.  As we have said many times, this kind of correlation 
functions do not give any exciting result in low $T$ liquids. What $q(R)$ does, on the other hand, is to measure
the correlation between one point (the centre of the sphere) and a set of particles, that is the external frozen
system at distance $R$ from the centre of the sphere. For this reason the function $q(R)$ is also called {\it point-to-set} 
correlation function, and hopefully it carries more information than standard point-to-point functions.

Note, finally, that the order parameter we are using in $q(R)$ is essentially the density fluctuations, because the grid and its
occupation numbers $\sigma_i$ are just a way to translate a continuous density field in a set of discrete spin 
variables. Therefore, what distinguish $q(R)$ from previous correlation functions is not really the choice of
the order parameter (which is always density fluctuations), but rather the fact that it compares two states
rather than focusing on a single configuration, and that it is a point-to-set correlation rather than a point-to-point one.

When $q(R)$ is measured in numerical experiments \cite{cavagna-07,
biroli-08}, the results are rather interesting (see Fig.17).  The
function decays with increasing $R$.  This is quite obvious, it simply
says that the effect of the boundary on the equilibrium inner
configuration of the sphere is weaker the larger the sphere. 
From this decay it is possible to define a
lengthscale as the point where $q$ is smaller than a conventional
small number $\epsilon$ (say, $\epsilon =0.05$),
\be q(R>\xi) < \epsilon \
\label{dizione}.
\ee
The crucial point is that the decay of $q(R)$ gets significantly
slower the lower the temperature, and thus $\xi$ as defined in
\eqref{dizione} increases when the temperature is decreased. This
effect is rather strong and unambiguous. Nothing similar is found in any
structural correlation function, whose change with $T$ is
disappointing, to say the least. The growth of $\xi$ confirms at last
the crucial hypothesis of this whole chapter: there {\it is} a growing
thermodynamic lengthscale in supercooled liquids. The slower decay of the
overlap $q(R)$ is a direct indication of the growth of amorphous
order: when $T$ is low and $R\ll\xi$ all the particles contained in
the sphere are correlated and their state is ruled by the amorphous
boundary condition $\alpha$.

\begin{figure}
\includegraphics[clip,width=3.4in]{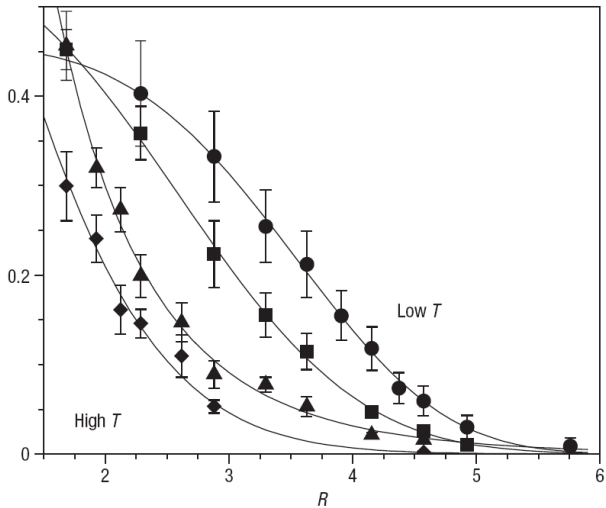}
\caption{ {\bf The growth of amorphous order} - 
The overlap $q$ as function of the radius $R$ of the sphere at different temperatures, from 
above $T_c$ (diamonds, $T=2.1 T_c$) to below $T_c$ (circles, $T=0.9 T_c$). The
decay of $q(R)$ is slower the lower the temperature $T$, indicating that there is a
clearly growing static correlation length $\xi$. The radius $R$ is expressed in units of
particle diameters. If we define the correlation length by the point where (for example)
$q(R=\xi)=0.05$, we obtain that $\xi$ grows from $2.7$ for $T=2.1 T_c$, to $4.9$ for $T=0.9 T_c$. 
Full lines are fits using equation \eqref{pulloz}.
 (Reprinted  with permission from \cite{biroli-08}).}
\end{figure}

Before we proceed, we must note that a different correlation length,
$\hat\xi$, can be extracted from the correlation function of the
energy fluctuations \cite{iran1, fernandez,iran2}. Similarly to the
lengthscale we are discussing here, $\hat\xi$ is of thermodynamic
nature, and yet not derived from any standard structural correlation
function. However, $\hat\xi$ it is associated to a different order
parameter compared to $\xi$, namely the energy, rather than the
density fluctuations, and for this reason its physical interpretation
is somewhat different. This notwithstanding, it is reasonable to
believe that all thermodynamic lengthscales are similar to each other
and thus it is encouraging that $\xi$ and $\hat\xi$ increase of a
comparable factor ($\sim 2$) at the lowest temperatures currently
obtained in numerical simulations \cite{iran1,fernandez,iran2}.

There is another result provided by $q(R)$ that is also quite
relevant: independently of the value of $\xi$, at high temperature
$q(R)$ decays exponentially, as an ordinary correlation function,
while at low $T$ the decay deviates significantly from an exponential
(Fig.18). The anomalous (i.e. nonexponential) decay of $q(R)$ gets
stronger the lower the temperature.  This means that the change in the
shape of the correlation function is not simply encoded in the growth
of $\xi(T)$. In other words, there is no length-temperature scaling of
$q(R)$, 
\be 
q(R;T)\neq q(R/\xi(T)) \ .  
\ee 
This result is important: the anomalous decay of $q(R)$ is the first
{\it qualitative} landmark of the deeply supercooled phase that we
find at a purely thermodynamic level.  As the reader may remember,
when we introduced the glass transition we were concerned that $T_g$
could be a purely conventional point, marking no qualitative change in
the fundamental physics of the liquid phase.  Disappointed by the
standard static observables, we finally found a crucial equilibrium
signature of glassiness in the behaviour of the dynamic correlation
function, and in particular in the two steps relaxation displayed at
low temperatures.  Nothing comparable, however, was found at the
thermodynamic level: it really looked like a glass-former was
statically the same at low as at high $T$, and this made all attempts
to give a thermodynamic interpretation of glassy phenomenology
somewhat flimsy.  The anomalous, nonexponential decay displayed by
$q(R)$ at low $T$ indicates that the equilibrium viscous phase can be
qualitatively distinguished from the high $T$ fluid phase also at the
thermodynamic level, not only a the dynamical one. This is a support
for all the thermodynamic scenarios for the glassy phase described in
this chapter.

\begin{figure}
\includegraphics[clip,width=3.4in]{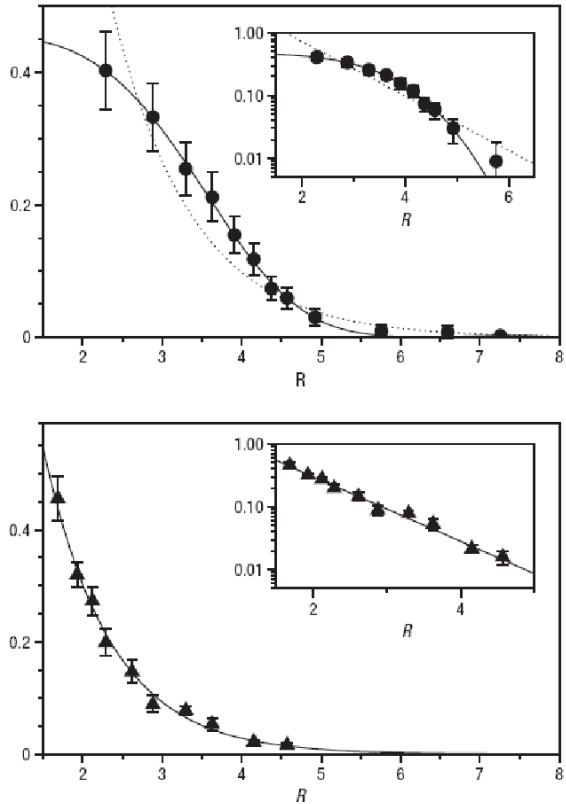}
\caption{ {\bf From exponential to nonexponential static relaxation} - 
The overlap $q$ as a function of the radius $R$ of the sphere at two different temperatures.
Lower panel: at high temperatures, $T=1.5 T_c$, the decay of the overlap is exponential, compatible 
with a one-state scenario. Upper panel: below the MCT transition, at $T=0.9 T_c$, the decay becomes strongly nonexponential.
The radius $R$ is expressed in units of particle diameters. The insets show the same data in semi-log representation,
where a pure exponential decay is a straight line. Note that the power law corrections typical of 
critical phenomena would give an opposite curvature in this representation. Dotted lines are the best exponential fit to 
the data; full lines are fits using equation \eqref{pulloz}.
(Reprinted with permission from \cite{biroli-08}).
 }
\end{figure}

\subsection{Mosaic reloaded}

Even though the empirical behaviour of $q(R)$ is quite illuminating, it does not seem to be good news for the mosaic theory.
According to equations \eqref{minore} and \eqref{maggiore}, one would expect that for $R<\xi$ the equilibrium state of the sphere under the influence of the amorphous
boundary condition is the same as the original state $\alpha$ and thus $q(R)$ to be large; on the 
contrary,  for $R>\xi$ the sphere thermalizes into one of the exponentially many available states $\gamma$, giving rise to small
value of $q(R)$. So, $q(R)$ should have a sharp decay at $R\sim \xi$, where
\be
\xi=\left(\frac{Y}{T S_c}\right)^\frac{1}{d-\theta}  \ ,
\label{musallo}
\ee 
is the usual mosaic length scale.
Let us call $q_1$ the overlap of state $\alpha$ with itself. This is the self-overlap of a state, which at
nonzero temperature is smaller than $1$, even though it is typically quite large \cite{cavagna-review}. 
If we assume that the overlap between two different states is zero, $q_{\alpha\gamma}\sim 0$ \cite{cavagna-07}, we can write,
\be
q(R) = q_1\; p_\mathrm{in}(R)  \ .
\label{cetriolo}
\ee
Therefore, according to the mosaic scenario, the sharp jump of $p_\mathrm{in}$ described by equation \eqref{pin} gives rise
to an equally sharp jump of the overlap,
\bea
q(R<\xi)&&\sim q_1  
\\
q(R>\xi)&&\sim 0  \ .
\eea
Contrary to this prediction, numerical data show that the decay of $q(R)$ is rather smooth at the analyzed temperatures 
(see Figs.16 and 17).

As we have seen the decay of $q(R)$ is nonexponential. A possible fit of the data (but by no means the only one), is,
\be
q(R)=\Omega\; \exp\left[-\left(R/\xi\right)^\zeta\right]
\label{pulloz}
\ee
with,
\be
\zeta \geq 1 \ .
\ee
The anomalous exponent $\zeta$ grows larger than $1$ when $T$ decreases below $T_c$ \cite{biroli-08} and this fact
means that the decay of $q(R)$ becomes moderately sharper at low $T$ (see Figs.17 and 18). Despite this fact, however,
the shape of $q(R)$ even at the lowest observed temperature, remains very different from the exponential jump of
\eqref{pin}.
Therefore,  there is no reason to believe that the growing lengthscale defined through \eqref{dizione} has anything to do with the mosaic 
correlation length \eqref{musallo}.  Let us go through the hypothesis of the mosaic theory, to see where things could have
gone wrong.

First of all, there was the very nontrivial assumption of the
existence of many `metastable states'. We put quotes here, just to
remember how problematic this definition was. States are not simply
potential energy minima, but not even too different from them. Their
existence should be guaranteed by energy barriers much larger than
$k_BT$, but we have no idea of the real size of these barriers; in
fact, we started looking for a correlation length in order to have a
better understanding of barriers! Given all this, and given the smooth
form of $q(R)$, one would be very tempted to say that there is only
{\it one} state, that is the liquid, and that the growth of $\xi$
displayed by $q(R)$ has nothing to do with the existence of many
states and the configurational entropy, but it is simply the ordinary
increase of a standard correlation length. For example, in second
order phase transitions, above the critical temperature the system is
ergodic, there is just one state, but the correlation length increases
nevertheless when lowering the temperature.  The situation in liquids
could be the same, with perhaps a critical temperature equal to
zero. We would still need to understand what is the physical mechanism
behind the increase of $\xi$, but we could argue that it has nothing
to do with the multi-state picture described by the mosaic, and
inherited from the $p$-spin model.

Yet, there is one problem in accepting this one-state scenario: the
anomalous decay of $q(R)$ at low $T$ does not seem to be compatible
with any standard picture of critical phenomena. Working within a
one-state framework, which is qualitatively the same at high and low
temperature, it is hard to get a non-exponential form of a correlation
function as the one displayed by $q(R)$. In particular, in a semi-log
representation as the one in the inset of Fig.18, the power law
corrections that show up in critical phenomena for $T\sim T_c$ and
$R<\xi$ \cite{cardy, binney}, give a curvature of the correlation
function opposite to the one of $q(R)$.  Therefore, if the original
mosaic formulation, giving a sharp drop of $q(R)$ does not seem to be
correct, also the opposite one-state framework does not match the
empirical results.

Let us see what else could be wrong. Due to the increase of $\zeta$,
the decay of $q(R)$ is somewhat sharper at lower $T$ (see equation
\eqref{pulloz}), even though not nearly as sharp as equations
\eqref{pin} and \eqref{pout} would require. What is the origin of the
step-like behaviour in the mosaic theory? The original argument is
based on a competition between the terms $Y R^\theta$ and $S_c R^d$:
if $R<\xi$, then the surface tension term is larger than the entropic
term, the probability to get out of the state is exponentially
suppressed, and vice-versa. This gives rise to the exponentially sharp
step of $p_\mathrm{out}(R)$, which can be roughly approximated by a
$\Theta$-function, \be p_\mathrm{out}(R)= \frac{\exp[ S_c
R^d]}{\exp[\beta Y R^\theta] + \exp[S_c R^d]} \sim \Theta(R-\xi) \ ,
\label{wasa}
\ee
with $\xi$ given by \eqref{musallo}.
This argument, however, assumes that the surface tension $Y$ is the {\it same} for all pairs of states and for all values of $R$. 
This assumption seems quite strong, once we consider that we are dealing with a 
disordered system where all states are supposed to be different. It is natural to imagine that the surface tension
depends on the particular pair of states and also on the spatial position of the interface within the system. 
These considerations suggest that it may be more appropriate to assume the existence of a {\it distribution} of surface tensions values, 
rather than a single value $Y$ equal for all \cite{biroli-08}. What are the consequences of this generalization of the mosaic scenario?

Consider a region of size $R$ originally in the same state $\alpha$ as the surrounding particles. The probability to make a transition to 
a new state $\gamma$ depends now on the particular surface tension of the pair, $Y_{\alpha\gamma}$. In particular, even when $R$ is very small,
it may be possible for the region to rearrange into a new state $\gamma$ with a surface tension $Y_{\alpha\gamma}$ that is low enough to 
make the transition convenient. In the original mosaic argument we asked: given a fixed value of $Y$, what is the {\it minimum} size $R$ 
necessary to perform the transition? The answer was $R=\xi$, with $\xi$ given by \eqref{musallo}.
Now we invert this question and ask: given a certain value of $R$, what is the {\it maximum} surface tension $Y$ that allows the transition?
To answer this question we have to invert relation \eqref{musallo} and get, 
\be
Y_\mathrm{max}= T S_c R^{d-\theta} \ .
\ee
All states having a surface tension with the external (frozen) state $\alpha$ lower than $Y_\mathrm{max}$ are potentially a target 
for rearrangement, while states with surface tension larger than $Y_\mathrm{max}$ are inaccessible (note that we are adopting the 
simplifying $\Theta$ approximation of \eqref{wasa}). Therefore the probability to make the transition, and thus get out of the original
state, $p_\mathrm{out}$, gets a contribution from all possible states with surface tension smaller than $Y_\mathrm{max}$.
If we introduce a surface tension probability distribution $P(y)$, we can formalize this argument by writing, 
\be
p_\mathrm{out}(R) = 
\int_0^{T S_c R^{d-\theta}} dy\; P(y)  \ .
\label{ciulla}
\ee
Given that $p_\mathrm{in}=1-p_\mathrm{out}$, from \eqref{cetriolo} we get for the overlap $q(R)$,
\bea
q(R) &=& q_1\left(1-\int_0^{T S_c R^{d-\theta}} dy\; P(y)\right)\nonumber \\ 
&=& q_1\; \int_{T S_c R^{d-\theta}}^\infty dy\; P(y)  \ .
\label{reloaded}
\eea
The argument is slightly more complicated than this (see ref. \cite{biroli-08}), but the explanation above captures the essence of the idea.
What is the the typical size $\xi$ of the rearranging regions in this new version of the mosaic? Indeed, equation \eqref{musallo} is no longer defined, 
once we give up a single value of $Y$. In fact, equation \eqref{reloaded} provides a very intuitive answer to this question. Whatever is the 
structure of $P(y)$, it is reasonable to expect that it has a scale $\bar y$, be it the average or any other typical scale. Under this very general assumption,
on dimensional grounds we conclude from \eqref{reloaded} that,
\be
q(R) = g\left(\frac{TS_cR^{d-\theta}}{\bar y}\right) = \hat g\left((R/\xi)^\frac{1}{d-\theta}\right)   \ ,
\ee 
where,
\be
\xi=\left(\frac{\bar y}{T S_c}\right)^\frac{1}{d-\theta}  \ ,
\ee
to be compared with \eqref{musallo}.
We therefore see that having introduced a surface tension distribution $P(y)$ leaves the fundamental structure of the mosaic mechanism exactly the same:
the correlation length is fixed by the competition between the typical scale of surface tension $\bar y$ and the configurational entropy $S_c$.
 The original mosaic hypothesis of a single value of surface tension is equivalent to assume,
\be
P(y)=\delta(y-Y) \ ,
\ee
which, once plugged into \eqref{ciulla} consistently gives back the original mosaic step-like form \eqref{wasa}.

On the other hand, if the surface tension distribution is {\it not} delta-peaked on a single value $Y$, we get for 
$p_\mathrm{out}$, and thus for $q(R)$, a softer transition: the broader the distribution $P(y)$, the softer the decay 
of $q(R)$. Hence, the nonexponential, but quite soft decay empirically observed for $q(R)$ may be
reinterpreted in the context of a generalized mosaic theory, where the surface tension is distributed according
to some broad probability density $P(y)$, rather than having a sharp constant value. Under this interpretation,
the change in the anomalous decay of $q(R)$ with changing temperature, is the consequence of the change of 
$P(y)$. In particular, we have seen that $q(R)$ becomes sharper at lower temperature,
i.e. the anomaly $\zeta$ in \eqref{pulloz} increases when $T$ decreases. This implies that $P(y)$ narrows and becomes 
more peaked at lower temperature. Let us see whether this prediction makes sense or not.

There are physical reasons to believe not only that the surface tension fluctuates, but also that these fluctuations 
may be quite large when the temperature is not very low. The fundamental origin of the surface tension fluctuations is 
that, due to the disorder, the degree of mismatch of the two states could vary significantly along the interface, with different portions paying a different
price. The contributions coming from the various portions along the interface add up to give the total surface tension
of that region. Therefore, a central limit theorem argument suggests that when the size $\xi$ of the rearranging region is large the 
different contributions along the interface add up to a sharply defined average surface tension, i.e. to a narrow $P(y)$. On the other hand,
when $\xi$ is not very large, fluctuations around the average will be large and thus $P(y)$ quite broad.

Beside this central limit theorem argument, there is a more intrinsic role of the disorder in making the surface tension fluctuate \cite{random-interface, martin-03, garel-07}, 
but one nevertheless concludes that $P(y)$ should get narrower at larger sizes. 
Given that at high $T$ the size $\xi$ of the rearranging region is smaller, and vice-versa, the conclusion of this argument
is that the surface tension distribution is expected to get sharper when lowering $T$.  According to equation \eqref{reloaded}, this means that
also the decay of $q(R)$ should become sharper at lower temperatures, which is indeed what is observed empirically.

Let us summarize the reloaded version of the mosaic presented in this
section. The soft, yet nonexponential decay of the overlap (or
point-to-set correlation function) $q(R)$ is in contrast both with the
original mosaic theory and with a standard one-state approach. If we
give up the hypothesis of a sharp single value for the surface tension
and introduce a probability distribution $P(y)$ for this quantity, we
obtain relation \eqref{reloaded} between $q(R)$ and the integral of
$P(y)$: the decay of $q(R)$ is still ruled by a correlation length
$\xi$ fixed by the competition between surface tension and
configurational entropy, but the transition at $R\sim\xi$ is now
softer, depending on how broad $P(y)$ is.  This is more in line with
the empirical behaviour of $q(R)$. Moreover, general considerations
suggest that $P(y)$ should be broader at higher $T$ and sharper at
lower $T$, and this fact too agrees with the observed sharpening of
the anomalous decay of $q(R)$ at lower temperatures.

On balance, we may say that the reloaded mosaic theory of deeply
supercooled liquids fares decently well. The mosaic recovers many of
the classic ideas about the link between configurational entropy and
lengthscale, it smartly exploits the exact mean-field scenario of the
$p$-spin model and it mixes all that in a fairly convincing physical
mechanism based on the competition between surface tension and
configurational entropy. The (very scarce) empirical data available
show that a static correlation length indeed exists and grows when
lowering $T$. Moreover, the behaviour of the associated novel
correlation function is at least not in plain contrast with the mosaic
mechanism. If we are optimistic, we may conclude that many of the
ideas presented in this chapter are not only fascinating, but in fact
also true.

Yet, we must be honest about one point. A direct evidence of the fact
that the drive to a region's rearrangement is the configurational
entropy is still lacking. In other words, the prefactor of $R^d$ in
all mosaic relations, a prefactor that we call $S_c$ and interpret as
a configurational entropy, could in fact be anything. The fact that we
cannot think about anything better than the configurational entropy as
an interpretation of this prefactor is no proof. In fact, by using the
method explained above, one can find a static length scale even in
kinetically constrained models \cite{garr-jack}. In such systems
thermodynamics is trivial and the static length scale is not given by
the competition between configurational entropy and energy, at variance with
the mosaic scenario. Moreover, the relation between length and relaxation
time is not the same as the mosaic approach in these systems.

As we have seen, the role of configurational entropy as a
thermodynamic drive to rearrangement is tricky.  Moreover, the very
concept of configurational entropy is deeply rooted in the physics of
Kauzmann's transition and in its $p$-spin mean-field counterpart, so
that it stirs uncomfortable feelings in those parts of the glass
community sceptical about exporting mean-field results to finite
dimensional systems. The configurational entropy is the keystone not
only of the mosaic theory, but of all the thermodynamic approach to
the deeply supercooled phase, hence this is not a scarcely relevant
the question.  A direct empirical test of the role of the
configurational entropy in determining the growth of $\xi$ and of the
relation between $\xi$ and the relaxation time $\tau_\mathrm R$, would
put our understanding of the thermodynamics of supercooled liquids on
a firmer basis.

\subsection{The dynamical correlation length}

The correlation length $\xi$ we have discussed in the previous
sections is {\it static}, meaning that it comes from the decay in
space of the (nonstandard) static correlation function $q(R)$.  The
increase of $\xi$ indicates the growth of amorphous structural order
in the liquid. Indeed, both the AGDM and
the mosaic theory, from which $\xi$ is conceptually derived, are
purely thermodynamic frameworks.  As the reader may remember, though,
when we discussed the nonexponential decay in time of the dynamical
correlation function $C(t)$ and the role of dynamical heterogeneities,
we learned that close to the glass transition mobility fluctuations
becomes large, and that particles moving significantly different from
the average tend to form clusters. This phenomenon immediately
suggests that it may be possible to define a {\it dynamical}
correlation length $\xi_d$. This is what we will do now.

First of all, let us remember briefly the phenomenological
evidence. If we take two snapshots of the system separated by a time
interval $t$ in the plateau regime of $C(t)$ and we measure the
displacement of each particle (i.e. the mobility), we observe 
that the spatial distribution of displacements is very heterogeneous.
Particles mobility can vary of orders of magnitude. 
Such mobility fluctuations increase when the temperature is lowered 
(see \cite{sillescu-99, ediger-00, glotzer-00} for reviews). 
Furthermore, particles with similar mobility cluster together and this is
suggestive of some sort of cooperative dynamics. In order to extract a
correlation length, though, we need to go from this clear, but rather 
qualitative picture of the two snapshots, to the robust definition of a suitable correlation
function. Moreover, we need to find a well defined way to select the time 
interval $t$ at which dynamical heterogeneities are strongest. In fact, we
did not provide until now any  convincing argument explaining why
$t$ should be of the same order as the plateau regime. We just stated it
as a fact. We shall see that the system itself selects the right time interval
over which we can best observe dynamical cooperative behaviour.

The existence of clusters of highly mobile (or immobile) particles of size $r$
suggests that the movement of particle $i$ over the time interval $t$ is
correlated to the movement of particle $j$ at distance $r$ from $i$,
over the same time interval.  For example, it
may be that the only way for particle $i$ to get out of its cage over
a time $t$ is to participate to a synchronous collective movement of a
certain number of particles; if particle $j$, at distance $r$ from $i$
 participates to the same collective movement, there will be a
correlation between the displacements of $i$ and $j$.  
Hence, we need
to measure the correlation between the displacements over a time interval $t$
of particles at mutual distance $r$.
The displacement $u_i(t)$ of particle $i$ is simply, 
\be
u_i(t) =  x_i(t)- x_i(0)  \ ,
\ee
where to avoid to burden the notation with vectors and norms we are assuming to be in one dimension. 
As usual, we need to subtract from this quantity its average, in order to compute a connected correlation 
function. Thus, we define the mobility fluctuation as,
\be
\delta u_i(t) =  u_i(t) - \langle u(t) \rangle \ .
\ee
Note an important point: even though the system is heterogeneous, 
once we average over the Gibbs distribution $\langle\cdot\rangle$, i.e. over many
thermal realizations, we restore homogeneity. Of course, the system is
statistically homogeneous. This is the reason why dynamical heterogeneities
disappear if the time interval $t$ between the two snapshots is much larger than
the relaxation time $\tau_\mathrm R$: in this case the time average is equal to 
the ensemble average, and thus statistical homogeneity holds. 

We can now introduce the mobility-mobility correlation function,
\be
G_u(r,t) = \frac{
\langle  \sum_{i,j} \delta u_i(t) \, \delta u_j(t) \; \delta(r-r_{ij}(t)) \;  \rangle}
{\langle  \sum_{i,j} \delta(r-r_{ij}(t)) \rangle}
\label{gu}
\ee
where $r_{ij}(t)=|x_i(t)-x_j(t)|$. Note the formal similarity of this correlation function 
with the radial correlation function $g(r)$ defined in \eqref{gidierre}: 
in $G_u(r,t)$ the contribution of each of the two particles at distance $r$ is weighted 
with its displacement over the time interval $[0,t]$. 
The normalizing denominator in \eqref{gu} is basically $g(r)$ and it grants 
that when the mobility fluctuations are
uncorrelated, $G_u(r,t) = \langle \delta u \rangle^2 =0$.
The correlation function $G_u(r,t)$ depends both on space
and time; in order to assess how long-ranged is this
function at fixed time $t$ we introduce the space integral of $G_u(r,t)$.
We obtain in this way the dynamic susceptibility of the mobility,
\be
\chi_u(t)=\int dr\; G_u(r,t)  \ .
\label{petrella}
\ee
A very large value of $\chi_u(t)$ simply means that $G_u(r,t)$ is very long-ranged
\footnote{In general, it would be more correct to normalize the space integral of $G_u(r,t)$ 
by $G_u(0,t)$. In fact, there are cases (as some kineticaly constrained models in one dimension)
where the mobility correlation function can grow for very trivial reasons, 
which have nothing to do with cooperative dynamics; therefore if the normalization is disregarded 
the susceptibility may give misleading results \cite{toninelli-05}. However, this normalization is not 
uniformely used in the literature, so I do not use it here.}.

The mobility-mobility correlation function \eqref{gu} was first
introduced in \cite{lancaster-97}, as a tool to discover cooperative
regions in numerical simulations of glass-forming liquids. However,
the analysis of $G_u$ done in \cite{lancaster-97} was very limited and
thus no increasing correlation length was detected.  A similar
mobility-mobility correlation function was later studied in
\cite{donati-99} (simulated supercooled liquids) and in
\cite{bennemann-99} (simulated polymer melts). In both cases clear
evidence was found that $G_u(r,t)$ becomes increasingly long-ranged
when the temperature is decreased, thus pointing out the existence of
a growing correlation length. This can be understood from the behaviour of $\chi_u(t)$ (Fig.19) 
\cite{donati-99, bennemann-99}: it first grows, when $t$ is 
in the ballistic and early $\beta$ regime of the dynamic correlation function, and it then reaches
a maximum for $t=t^*$, decreasing to a smaller value for later times (see \cite{toninelli-05}
for a careful study of all dynamical phases of $\chi_u(t)$). 

The peak of $\chi_u(t)$ confirms what we anticipated in the discussion
of dynamical heterogeneities: the clusters of dynamically correlated
particles are {\it transient} in time. For times too short, there is
simply no correlation because interaction among the particles is
absent, or trivial; for times too long, the time average equals the
ensemble average and thus statistical homogeneity is restored. It is
only at intermediate times that dynamical correlations are relevant.
The behaviour of $\chi_u(t)$ is also very important because it clearly
selects the value $t^*$ as the time at which heterogeneous dynamics is
the most intense. As expected from our previous discussion of
dynamical heterogeneities, one finds that $t^*\sim \tau_\mathrm R$,
i.e. mobility fluctuations are largest when the displacement is
computed over a time window of the order of the $\alpha$ relaxation
time.

\begin{figure}
\includegraphics[clip,width=3.7in]{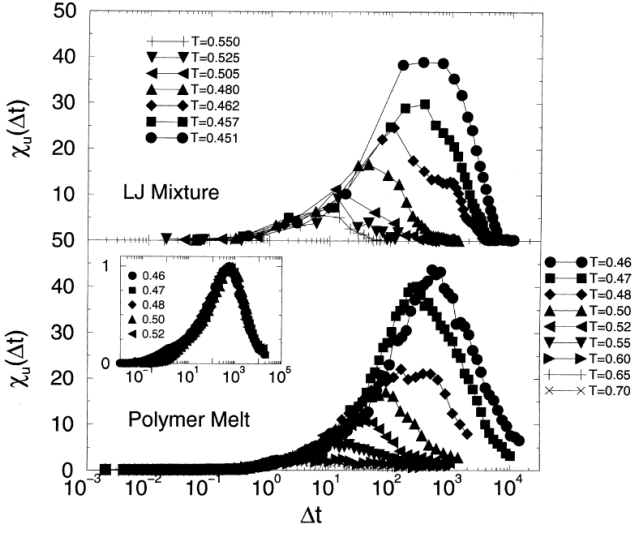}
\caption{ {\bf The dynamical susceptibility $\chi_u(t)$} - The mobility-mobility
susceptibility $\chi_u(t)$ as a function of time at different
temperatures, in glass-forming liquids (upper panel) and polymer melts
(lower panel). The curves reach a peak at $t=t^*$, which corresponds
to the time regime over which dynamics is most heterogeneous. If we
rescale susceptibility and time by their peak values and plot 
$\chi_u(t/t^*)/\chi_u(t^*)$ we obtain an excellent collapse
of the curves at various temperatures (inset).
(Upper panel: reprinted with permission from \cite{donati-99}; copyright 
of American Physical Society. Lower panel: reprinted with permission from \cite{bennemann-99}.)
}
\end{figure}

The susceptibility $\chi_u(t)$ is the space integral of the
mobility-mobility correlation function $G_u(r,t)$ and thus a large
susceptibility implies a slower spatial decay of $G_u(r,t)$. This
means that $G^*_u(r)= G_u(r, t=t^*)$ is the slowest decaying
correlation function and we can therefore extract from $G^*_u(r)$ the
largest lengthscale at that temperature. This is the definition of dynamical
correlation length $\xi_d$. The crucial result is that when the temperature is lowered, both the
position $t^*$ and the height $\chi_u(t^*)$ of the peak of the dynamical
susceptibility grow significantly \cite{donati-99, bennemann-99}. In turn, this fact implies that
the mobility correlation function $G^*_u(r)$ is more long-ranged and that the
dynamical correlation length $\xi_d$ increases.

The mobility-mobility correlation function $G_u(r,t)$ connects
directly to the physical picture of dynamical heterogeneities and it is thus quite
illuminating. However, the essential ingredient to find the dynamical correlations is not 
really the mobility, but rather the fact that we are calculating a 
{\it four}-point correlation function, in contrast with standard {\it two}-point functions, as $g(r)$.
This is clear from \eqref{gu}: each one of the two displacements $\delta u$
contains in turn the position of the particle at two instant of time.
The four-point nature simply derives from the fact that to unveil cooperative dynamics we
must check what happens in two spatial locations, separated by $r$, at two different 
instants of time, separated by $t$.

The correlation function can thus be defined using any reasonable observable, 
provided that we keep the four-point structure. We call this general correlation function
$G_4(r,t)$, dropping the displacement label $u$, and emphasizing its four-point nature.
In the particular, but important, case of the density fluctuations, we have, 
\bea
&&G_4(r,t) =\langle \delta\rho(0,0)\delta\rho(0,t) \; \delta\rho(r,0)\delta\rho(r,t) \rangle  \nonumber \\
&&- \langle \delta\rho(0,0)\delta\rho(0,t) \rangle \; \langle \delta\rho(r,0)\delta\rho(r,t)\rangle  \ .
\label{dasgupta}
\eea
The structure is conceptually the same as before: $\delta\rho(0,0)\delta\rho(0,t)$
measures what is going on at a certain position over the time interval $t$, and we compute
the spatial correlation of this product with its counterpart at distance $r$.
This definition of $G_4$ allows us to give a deeper
interpretation the correlation functions we are discussing. 
As we have repeated many times, the most remarkable dynamical signature 
of the glass transition is the two steps relaxation 
of the dynamical correlation function, $C(t)$. 
The lower the temperature, the longer $C(t)$ 
remains at the plateau. Ergodicity is thus `imperfectly' broken: the dynamical
correlation function is stuck to a nonzero value for times that are very long, 
although shorter than the correlation time. We may express this fact by writing,
\be
\lim_{t\to\infty} C(t, T\sim T_g) = C^\star \neq 0 \ ,
\ee
where the limit is imperfect, in the sense that $t$ must remain smaller than $\tau_\mathrm R$, which is
however very large close to $T_g$. As we have seen, this imperfect transition becomes perfect in MCT and
in the mean-field $p$-spin model. On the other hand, at higher temperatures,
\be
\lim_{t\to\infty} C(t, T\gg T_g) =  0 \ .
\ee
This means that the long time limit of the dynamical correlation function can be interpreted as the 
order parameter of the (imperfect) discontinuous glass transition 
\footnote{As we have seen previously in these notes, the onset of glassiness in 
supercooled liquids, as well as in MCT and the $p$-spin model, has a distinct discontinuous 
nature, in the sense that the order parameter $C^\star$ is already finite at the transition}. 
This fact is ripe of consequences on the interpretation of $G_4$.
By using the density fluctuations, we can write the dynamical correlation function as,
\be
C(t)=  \langle \delta\rho(x,0)\delta\rho(x,t) \rangle \ .
\label{spompo}
\ee
Let us now compare \eqref{dasgupta} to \eqref{spompo}. 
If we define the time-dependent field, 
\be
\varphi(x,t)= \delta\rho(x,0)\delta\rho(x,t) \ ,
\ee
we have that $C(t)$ is indeed the order parameter, i.e. the average of the field, 
\be
C(t)=\langle \varphi(x,t) \rangle \ ,
\ee
which, thanks to the average, does not depend on $x$.
On the other hand, $G_4(r,t)$ is the spatial correlation function of the same field, 
i.e. the connected average of the two-point product at distance $r$,
\be
G_4(r,t) = \langle \varphi(0,t)\varphi(r,t)\rangle - \langle\varphi(0,t)\rangle\langle\varphi(r,t)\rangle \ .
\ee
Therefore, $G_4(r,t)$ is nothing else than the (spatial) correlation function associated to the order 
parameter most relevant to the glass transition, namely the dynamical correlation function $C(t)$. 
This conclusion makes the definition of $G_4(r,t)$ certainly more natural. The order parameter
$C(t)$ is in turn a (dynamical) correlation function and this explains the four-point nature of $G_4$.

Even though we have conceptually defined the dynamical susceptibility as
the space integral of $G_4(r,t)$, one can see that, as any other susceptibility, 
$\chi_4(t)$ measures also the fluctuations of the bulk order parameter, $\Phi(t)$, defined
as the space integral of the field,
\be
\Phi(t)= \frac{1}{V} \int dx  \, \varphi(x,t) = \frac{1}{V} \int dx  \, \delta\rho(x,0)\delta\rho(x,t) \ .
\label{villaro}
\ee
Indeed, we can write,
\bea
\chi_4(t)&=& V\left[ \langle \Phi(t)^2 \rangle- \langle \Phi(t) \rangle^2 \right]=  \nonumber \\
&&\frac{1}{V} \int dx\; dy\; \langle \varphi(x,t)\varphi(y,t)\rangle - \langle\varphi(x,t)\rangle\langle\varphi(y,t)\rangle= \nonumber \\
&&\int dr \; \langle \varphi(0,t)\varphi(r,t)\rangle - \langle\varphi(0,t)\rangle\langle\varphi(r,t)\rangle = \nonumber \\
&=&\int dr\; G_4(r,t)
 \ .
\label{castro}
\eea
When the field carries a discrete space variables, $\varphi_i(t)$, 
we have to simply change normalization,
\be
\Phi(t)= \frac{1}{N} \sum_i \varphi_i(t) \,
\label{villaro2}
\ee
\be
\chi_4(t)= N\left[ \langle \Phi(t)^2 \rangle- \langle \Phi(t) \rangle^2 \right]  \ .
\label{castro2}
\ee
Whatever is the specific choice of the field $\varphi(x,t)$ (or $\varphi_i(t)$),
the average of its bulk counterpart, $\Phi(t)$, is equal to the dynamical correlation function,
\be
\langle \Phi(t) \rangle=C(t)  \ .
\label{mirella}
\ee
From \eqref{castro}  and \eqref{mirella} we therefore learn something interesting: the
dynamical susceptibility $\chi_4(t)$ measures the fluctuations of the
dynamical correlation function $C(t)$. The peak of $\chi_4$ shows that 
these fluctuations reach a maximum at
$t=t^*\sim\tau_\mathrm R$. Following this same
interpretation, we have that the mobility dynamical susceptibility,
$\chi_u(t)$, related to the particle displacements $u_i(t)$, measures
the fluctuation of the mean-square displacement (MSD). As we have seen
in Section IV-G, the MSD has a plateau structure completely analogous
to that of the dynamic correlation function.  We therefore expect
that $\chi_4(t)$ (i.e. the fluctuation of the dynamic
correlation function) and $\chi_u(t)$ (i.e. the fluctuation of the
MSD), give qualitatively the same results.

The need to define a four-point object in order to have a nontrivial
susceptibility has been long known in the context of spin-glasses.  A
very early and particularly illuminating discussion of this point was
given in \cite{kirk-chi4}, where it was also hinted that a similar
tool was probably essential also in glass-forming liquids.  For what
concerns numerical  simulations in liquids, the four-point correlation function
$G_4$ defined in \eqref{dasgupta} was first introduced in
\cite{dasgupta}, well before mobility-mobility correlation functions
were analyzed. However, the 
attempt of \cite{dasgupta} to find a growing
lengthscale through $G_4$ gave negative results. 
Later, in \cite{parisi-97},
the dynamical susceptibility $\chi_4(t)$ was defined as the fluctuation of the 
overlap between two configurations at different times, 
\be
q(t) = \frac{1}{N} \sum_i \sigma_i(0)
\sigma_i(t) \ ,
\label{bulkq}
\ee
where the definition of the `spins' $\sigma_i$ has been given in Section VII-G.  
The associated dynamical susceptibility is,
\bea
\chi_4(t) &=& N\left[ \langle q(t)^2 \rangle- \langle q(t) \rangle^2 \right] \nonumber \\
&=&\frac{1}{N}\sum_{i,j} \langle \sigma_i(0) \sigma_i(t) \sigma_j(0) \sigma_j(t) \rangle \nonumber \\
&-& \langle \sigma_i(0) \sigma_i(t)\rangle\langle \sigma_j(0) \sigma_j(t) \rangle 
\label{maminkia}
\ ,
\eea
which is clearly the space integral of a four-point correlation function.
The numerical analysis of $\chi_4(t)$ done in \cite{parisi-97}, though, 
focused on off-equilibrium, rather than equilibrium dynamics, and it
found a dynamical correlation length that increased with the waiting
time. A successful equilibrium study of the dynamical susceptibility in numerical 
experiments was
finally performed in \cite{donati-franz-99}. The behaviour of
$\chi_4(t)$, was found to be qualitatively the same as the susceptibility
associated to the mobility fluctuations, $\chi_u(t)$: $\chi_4(t)$ has a peak at
$t=t^*\sim \tau_\mathrm R$, and the height of this peak $\chi_4(t^*)$ 
increases when $T$ decreases (Fig.20). From the decay of $G^*_4(r)=G_4(r,t=t^*)$ one can
extract a growing dynamical correlation length, $\xi_d(T)$. 
This behaviour of $\chi_4(t)$ and the associated growth of $\xi_d$ has been also 
confirmed experimentally \cite{cipelletti-06}.

\begin{figure}
\includegraphics[clip,width=3.7in]{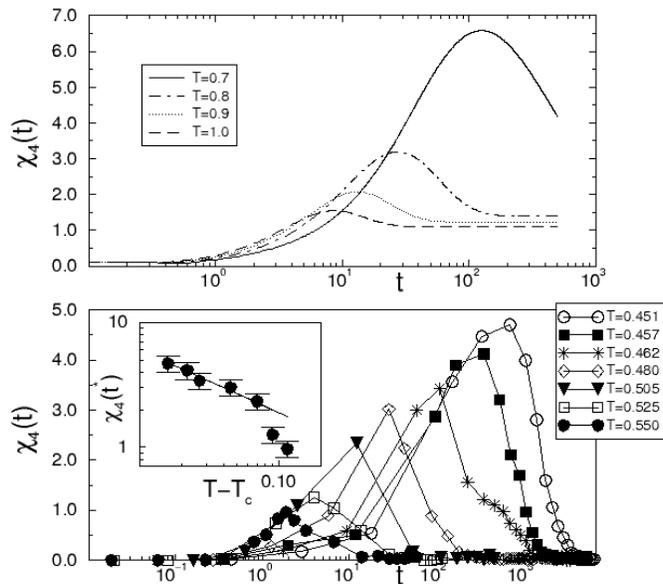}
\caption{ {\bf The dynamical susceptibility $\chi_4(t)$} - The dynamical
susceptibility $\chi_4(t)$ as a function of time at different
temperatures, in the mean-field $p$-spin model (analytic - upper panel) 
and in glass-forming liquids (numerical - lower panel). In this 
case $\chi_4(t)$ is obtained from the fluctuation of the overlap,
equation \eqref{maminkia}. Inset: the peak value of the 
susceptibility $\chi_4(t^*)$ scales reasonably well as a power law
with (avoided) divergence at the MCT $T_c$.
(Reprinted with permission from \cite{donati-franz-99})
}
\end{figure}

Up to now we have seen that $\chi_4(t^*)$, and thus $\xi_d$, increases
with decreasing temperature. What is the form of this growth? The data
seems to be compatible with a power law divergence of $\chi_4(t^*)$ at
the MCT temperature $T_c$ \cite{donati-99, bennemann-99,
donati-franz-99, lacevic-03}.  Of course, we must apply to this kind
of power-law fit all the caveats we learned when discussing MCT. In
particular, this divergence cannot be really present and $\xi_d$
remains finite at $T_c$. Still, the fact that above $T_c$ data are
reasonably well fitted by a power law behaviour typical of MCT agrees
with the interpretation of $T_c$ as a dynamical mean-field divergence,
smoothed out in finite dimension by activated barrier crossing. This
interpretation receives further support from the following two
facts. First, in the context of MCT it has been proved the existence
of a dynamical correlation length diverging (as a power law) at $T_c$
\cite{biroli-04}. Second, $\chi_4(t)$ can be exactly computed in the
$p$-spin mean-field model and its qualitative form is the same as the
one found in supercooled liquids (Fig.20). In particular,
$\chi_4(t^*)$ diverges as a power law at the dynamical transition
$T_c$ \cite{donati-franz-99}. Furthermore, in the $p$-spin below the
dynamical transition it can be defined a four point static
susceptibility, similar in spirit to the dynamic $\chi_4$, which
diverges for $T\to T_c^-$ \cite{donati-franz-99, franz-static-xi}.
These MCT and mean-field results strengthen the idea that the increase
of the relaxation time on approaching $T_c$, and thus the `imperfect'
ergodicity breaking of the dynamic correlation function $C(t)$, is due
to the `imperfect' divergence of the dynamical correlation length at
$T_c$.

In the context of mean-field spin-glasses the divergence of
$\xi_d$ has a neat interpretation \cite{kirk-chi4, recipes, recipes-2,
franz-comment}.  In these systems, the dynamical transition 
corresponds to the formation of a secondary minimum of an effective
potential $V(q)$, which is roughly the free energy price we must pay 
to keep a system at fixed overlap $q$ from an equilibrium configuration
\cite{recipes, recipes-2}.
The lowest minimum of this potential at $q=0$ corresponds to the ergodic phase,
whereas a secondary (metastable) minimum at $q\neq 0$ appears below $T_c$
and it corresponds to the broken ergodicity phase. In this context, the dynamical
transition $T_c$ acts like a spinodal instability \cite{kirk-chi4,
recipes-2}: the curvature of the potential at the point where the
secondary minimum disappears (appears) is zero, and thus the associated
susceptibility and correlation length diverge. It was this physical
mechanism that suggested to calculate $\chi_4(t)$ also in supercooled liquids 
\cite{franz-comment}.

What happens in deeply supercooled liquids below $T_c$? Clearly, the MCT fit only holds
above $T_c$; the {\it empirical}
values of $\chi_4(t^*)$ and of $\xi_d$ are finite at $T_c$, and they go
on growing below this temperature, just as the plateau of the
dynamic correlation function $C(t)$ does not really diverge at $T_c$, and it
get longer at lower temperatures. However, the {\it theoretical}
interpretation of the growth of $\chi_4(t^*)$, in terms of an
approaching spinodal at $T_c$, does not make sense below
$T_c$. Moreover, the static susceptibility defined for the $p$-spin model 
in \cite{donati-franz-99, franz-static-xi} diverges going to
$T_c^-$, and thus it {\it decreases} when lowering the
temperature; hence, it cannot help us in giving an interpretation of
the empirical growth of $\chi_4(t^*)$ below $T_c$. What then?

It is not easy to give an answer to these questions.  Perhaps, a
partial understanding can be reached if we note that the role of
$T_c$ as a spinodal temperature was also emphasized in a rather
different context, that of the mosaic theory: $T_c$ (or the
Goldstein's temperature $T_x$, which is the same) is the point where
the generalized surface tension between amorphous metastable states
goes to zero, basically because minima below $T_c$ turn into saddles above $T_c$. 
In the context of the (reloaded) mosaic theory we
measured a static correlation length $\xi$, which is not defined 
above the spinodal $T_c$. Therefore, we see a certain specular
symmetry of the two frameworks where the dynamic and static correlation
lengths have been defined. The pivot of this symmetry is $T_c$.  In
the dynamical mean-field context, $\xi_d$ diverges at $T_c^+$ and it
is undefined below $T_c$. In the static context, $Y$ vanishes 
at $T_c^-$, so that $\xi$ is undefined 
above $T_c$, but it increases below $T_c$. This considerations suggest
that around $T_c$ there may be a crossover between dynamic and static
correlation length. 

Under this interpretation, the growth of $\xi_d$ 
above $T_c$ would be driven by the approaching mean-field 
spinodal at $T_c$, whereas below $T_c$ the empirically measured $\xi_d$ would in fact be slave to the 
static correlation length $\xi$, whose growth is driven by a thermodynamic mosaic-like mechanism.
In fact, we have already seen that the growth of the relaxation time $\tau_\mathrm R$
has different origins above $T_c$ (vanishing number of unstable directions) and below $T_c$
(activated dynamics). It would therefore be not surprising if something similar held
for the correlation length. After all, rather than having different correlation 
lengths (dynamical and static), it is perhaps more satisfying to have different
physical mechanisms leading to the growth of the {\it same}
correlation length. On this highly speculative, utterly unsupported, and rather boring
considerations, we conclude this long chapter.

%%%%%%%%%%%%%%%%%%%%%%%%%%%%%%%%%%%%%%%%%%%%%%%%%%%%%%%%%%%%%
%%%%%%%%%%%%%%%%%%%%%%%%%%%%%%%%%%%%%%%%%%%%%%%%%%%%%%%%%%%%%
%%%%%%%%%%%%%%%%%%%%%%%%%%%%%%%%%%%%%%%%%%%%%%%%%%%%%%%%%%%%%
%%%%%%%%%%%%%%%%%%%%%%%%%%%%%%%%%%%%%%%%%%%%%%%%%%%%%%%%%%%%%
%%%%%%%%%%%%%%%%%%%%%%%%%%%%%%%%%%%%%%%%%%%%%%%%%%%%%%%%%%%%%
%%%%%%%%%%%%%%%%%%%%%%%%%%%%%%%%%%%%%%%%%%%%%%%%%%%%%%%%%%%%%
%%%%%%%%%%%%%%%%%%%%%%%%%%%%%%%%%%%%%%%%%%%%%%%%%%%%%%%%%%%%%
%%%%%%%%%%%%%%%%%%%%%%%%%%%%%%%%%%%%%%%%%%%%%%%%%%%%%%%%%%%%%
%%%%%%%%%%%%%%%%%%%%%%%%%%%%%%%%%%%%%%%%%%%%%%%%%%%%%%%%%%%%%

\section{I couldn't disagree more!}

When I lived in the UK, I enjoyed to watch a TV show called {\it
``If I Ruled the World''}. My favourite part of the program was when 
the dialectic ability of the guests was challenged by
asking them to disagree with some amazingly obvious statements, like
`War is a very bad thing', or `Jesus was a nice guy', and so on. Each
reply had to start with the ritual sentence: `I couldn't disagree
more!'.

I wish the content of these notes were so obvious as to make it
equally hard to disagree with it. Yet, this is not the case. As I said
in the Introduction, what I have presented here is a partial view, and
the reader should keep in mind that there are several other
theoretical frameworks.  In this final chapter, I will give a very
superficial overview of few of these alternative theories, not
really with the aim of explaining them in depth, but rather to provide
the necessary references for the reader to build her/his own view.

\subsection{Frustration-limited domains}

The frustration-limited-domains (FLD) theory \cite{kivelson-95,
tarjus-95, tarjus-96, tarjus-98, kivelson-98, viot-98, tarjus-00,
grousson-01,tarjus-02,tarjus-08} agrees on the fact that the super-Arrhenius
increase of the relaxation time in fragile glass-formers must be due to
the growth of some sort of cooperative order in the system. Similar to
the AGDM and mosaic theories, FLD also recognizes that such ordering
must have a thermodynamic origin. What FLD disagrees about is the
physical origin of the cooperative regions.

FLD is based on two basic concepts: locally preferred structure and
geometric frustration (sometimes called structural, or topological
frustration). The theory assumes that in the liquid there is a locally
preferred structure (LPS) that differs from the local structure of the
thermodynamically stable crystal. A LPS can be defined as an arrangement
of molecules that minimizes some local free energy. In absence of frustration, the
liquid would freeze into the LPS at a second order critical point
$T^*$. However, the system is prevented from doing this by the fact
that the LPS is incompatible with long-range order, and thus it cannot
tile the whole space. This is geometric frustration. One of the key
points of the theory is that the presence of geometric frustration,
however small this is, kills the phase transition at $T^*$, and yet
this avoided critical point controls much of the physics of the system,
and in particular the sharp slowing down typical of glass-forming liquids.

An interesting point emphasized by the FLD theory is that the strongly
first order nature of ordinary liquid-crystal phase transitions is in
fact evidence of an avoided critical point. The growth of LPS order
causes a superextensive strain that grows when the temperature is
lowered. Beyond a certain point, this strain makes the system
thermodynamically unstable and the strain is released by an abrupt
restructuring of the liquid, which therefore undergoes a (sharp) first
order transition to the crystal at $T_m$ \cite{tarjus-02}. For this
reason $T^* > T_m$ according to FLD theory \cite{kivelson-95}. In this
context, the first order character of the liquid-crystal transition is a
consequence of frustration.

The prominent effect of avoided criticality, is that the long range
order that would be established in the unfrustrated system, cannot be
sustained when frustration is present. For this reason the systems
breaks up into domains of {\it frustration-limited} size $\xi$. In
other words, {\it ``molecules in a liquid ... tend to arrange
themselves into a locally preferred structure corresponding to the
minimization of an appropriate local free energy; but the spatial
extension of this local arrangement is thwarted by ubiquitous
structural frustration that prevents a periodic tiling of space''}
\cite{tarjus-00}.
This static lengthscale must not be confused with the
correlation length $\xi_0$ associated to the (avoided) critical point.
In absence of frustration $\xi_0\to\infty$ at the critical point and
it {\it decreases} when the temperature is decreases below $T^*$; on
the other hand, $\xi$ is infinite at any temperature below the
critical point if the frustration is zero. The theory assumes that 
frustration is in general rather small, so that $\xi \gg \xi_0$.

Different domains will have different `orientations' of the LPS, so that
there is a surface tension energy on the domain walls. In these conditions, 
the rearrangement (or restructuring) of the domains becomes activated, and
at low $T$ this causes a stupendous slow down of the dynamics. 
This observation suggests that it may be possible to incorporate the
main ingredients of FLD within relatively simple coarse-grained
models. This is what FLD has done \cite{kivelson-95, tarjus-96,
viot-98, tarjus-00}. In FLD models, a first part of the Hamiltonian
reproduces the unfrustrated critical point; this can be any
short-range, ferromagnetic (Ising-like) interaction term of strength
$J$. The important point is that the order parameter (the spin) must
have many `orientations' or colours, in order to reproduce the mismatch
between different domains. To the unperturbed part one adds a weaker,
long-range, anti-ferromagnetic term of strength $K$, which represents
the geometric frustration in the system. The transition prescribed by
the ferromagnetic $J$ term is suppressed by the presence of the (weak)
long-range frustration $K$.

From the statistico-mechanical analysis of such (or similar) model,
one finds that $\xi\sim \sqrt{J/K}\,\xi_0^{-1}$. Given that the
critical length $\xi_0$ decreases getting further from $T^*$, the
frustration-limited domains size $\xi$ grows on lowering $T$.  Apart
from this rather crucial result explaining super-Arrhenius relaxation,
the theory obtains a number of other interesting results. In
particular, FLD models link the amount of geometric frustration in the
system to the fragility of the liquid, so that many different kinds of
relaxation behaviour can be reproduced within the theory. Moreover, if
the long time $\alpha$ relaxation is linked to the rearrangement of 
frustrated-limited domains on scale $\xi$, according to FLD $\beta$
relaxation corresponds to the unfrustrated, and therefore faster,
dynamics over the critical scale $\xi_0 \ll \xi$.

The idea of linking glassiness to geometric frustration dates back to
the early 80s \cite{nelson-83, nelson-83-bis, sethna-83, sethna-85},
when geometric frustration was studied in relation to the curvature of
space (see also \cite{tarjus-08}).  Moreover, the fact that the competition between
nearest-neighbour ferromagnetic and next-nearest-neighbour
anti-ferromagnetic interaction leads to a logarithmic slowing down,
similar to that observed in glassy dynamics, was first explored in
\cite{shore-91,shore-92}. Yet, the FLD theory was the first to link in a 
coherent way geometric frustration to cooperative behaviour.
Indeed, one of the great virtues of FLD is that it explains in clear physical
terms {\it why} the system breaks up into domains, whereas other schemes sometimes
just assumes that domains exist, and focus on the reason why they grow.
On the other hand, it must be said that, even though the scaling arguments
and mathematical details leading to the result $\xi\sim 1/\xi_0$ are perfectly
fine, the growth of $\xi$ in the context of FLD is perhaps not as physically 
transparent as in alternative theories.

It is interesting to note that the FLD theory provides quite a natural
explanation for the phenomenon of polyamorphism, i.e. the coexistence
of inequivalent amorphous phases \cite{tarjus-02}. In particular, the
presence of a structure factor somewhat intermediate between
completely amorphous glass and ordered crystal in the glacial phase of
some fragile liquids (as triphenylphosphite), suggests that
polyamorphism could be a poorly crystallized phase, i.e. a powder
of polydisperse crystallites, resulting from the competition between 
LPS and geometric frustration. This would fit perfectly within FLD 
theory. On the other hand, one may wonder why in more ordinary 
(non-polyamorphic, non-glacial) systems, standard tools, as the static
structure factor, do not detect the existence of frustration-limited
domains. This could be due either to the excessive polydispersivity 
of the LPS polycrystal, or to its unusual (and still unknown) symmetry
properties. Of course, the FLD theory would benefit a lot from a direct 
observation of frustration-limited domains.

\subsection{Dynamical facilitation}

The dynamical facilitation theory (DFT)
\cite{kcm-1,kcm-2,kcm-3,kcm-4, kcm-5} is utterly critical 
of the theoretical ideas presented in these
notes. According to DFT, every important aspect in the
physics of glass-forming liquids can be qualitatively and
quantitatively understood purely at the dynamical level, with no need
of any thermodynamic or landscape description. 
This is a partial list of what is at best useless, and at worst wrong,
according to DFT: any phase space or topological approach
to the glass transition; minima and saddles of the potential energy;
mode coupling theory and the MCT transition $T_c$; the idea of a
crossover from nonactivated to activated dynamics at $T_x$; 
the landscape approach; any
mean-field description, and in particular the $p$-spin model; the
Adam-Gibbs theory; the mosaic theory. Not to mention Kauzmann's
transition $T_k$. Clearly, DFT couldn't disagree more
with the content of the present review.

The main starting observation of DFT is that dynamical heterogeneities
are the most distinctive trait of glassiness.  In this way, DFT sets
its two pillars: dynamics and real space.  Dynamical heterogeneities
are, indeed, a dynamical phenomenon, associated, as we have see to a
dynamical correlation length; moreover, dynamical heterogeneities
clearly show that there are large mobility fluctuations in real
space. Clearly, a purely static, phase space approach is in great pain
in describing dynamical heterogeneities.

The idea of DFT is simple, elegant and actually quite old
\cite{glarum-60, phillips-72, fred, fred2}.  In a low $T$ liquid
everything is stuck, and only few particles will be mobile. We can
interpret these particles as mobility defects. The key point is that
these mobility defects will not remain isolated: the mobility of one
particle prompts the mobility of other {\it nearby} particles. In the
words of Glarum, {\it ``the relaxation of a molecule is more probable
immediately after one of its neighbors has relaxed than at an
arbitrary time''} \cite{glarum-60}. This is dynamic facilitation. Because of this
effect, one expects clustering of mobile regions and thus a {\it
``mesoscopic demixing of mobile and static regions''} \cite{kcm-1}.
Note that this simple mechanism encodes by itself the origin of
dynamical heterogeneities. Moreover, in this context glassiness is the
consequence of the effective dynamical constraints in the interaction
among particles when the temperature is lowered, irrespective of the
thermodynamics of the system.

The DFT approach owes a considerable part of its credibility to the
existence of kinetically constrained models where the scenario
described above is exactly realized. These models display a
phenomenology that is, in some respects, quite similar to the one of
real supercooled liquids.  Important examples are the
Fredrickson-Andersen (FA) model \cite{fred} and the East model
\cite{east}. In both models the Hamiltonian (and thus all
thermodynamics) is completely trivial: a chain of one dimensional
noninteracting spins. So much for the thermodynamic approach!  On
the other hand, the dynamics is nontrivial due to the kinetical
constraints (see \cite{evans-02, ritort-03} for reviews on kinetically 
constrained models).  For example,
in the FA model a spin can flip if either of its nearest neighbours is
in the up state, where up represent a more mobile state.  This dynamical 
rule embodies dynamical facilitation: if a spin is isolated, with no mobile (up)
spins around, it can do nothing; on the other hand, clusters of mobile
spins are free to move. Such simple rules are enough to cause a sharp
slowing down of the dynamics.  In particular, there are kinetically
constrained models with a strong-like behaviour and others with a
fragile, super-Arrhenius behaviour.
In the context of kinetically constrained models, then, mobile spins  play the role
of mobility defects in real liquids, and they are the ones that prompts the
dynamics. We clearly see that DFT is essentially a theory of defect diffusion \cite{glarum-60}. 

The mobility feedback triggered by facilitated dynamics is sharply suppressed at low
$T$, when mobility defects become more and more dilute. The typical
distance between defects is naturally interpreted as a correlation
length. In particular, there are three phases: at high $T$ the system
is very reach in defects, which form a cluster percolating throughout
the entire system; in this phase the dynamics is very fluid. At a
lower, but intermediate temperature, cluster of defects of high
mobility coexist with isolated, and thus immobile, defects. Finally, at
very low $T$ defects are typically isolated and the dynamics is stuck 
\cite{kcm-4}.

The DFT approach would remain rather limited without a way to map a
normal liquid system into one of these lattice models with facilitated
dynamics. DFT  achieves this result by making a coarse
graining of the dynamics over a spatial scale of the order of the
structural correlation length given by the pair correlation function
\cite{kcm-3,kcm-4}. This procedure is necessary in order to have
statically uncorrelated cells in the system. This is a tricky point,
however, because we have seen that there is another static correlation length
which grows when lowering $T$, in contrast with the standard one used
in \cite{kcm-3,kcm-4}. Hence, it is not {\it a priori} clear what is the correct
coarse graining lengthscale.

The DFT description has a particularly vivid realization if we
consider as elementary object of our study the space-time trajectories
of the defects in one dimension, rather than simple static
configurations.  This is a particularly useful representation, and its
is valid in general, irrespective of its particular role within
DFT. In the trajectory space, one can clearly see that when the
temperature is lowered, bubbles are formed in the space-time plane
with a typical size $l(T)$, which grows when the temperature is
lowered.  This scale is naturally connected to the dynamical
correlation length \cite{kcm-4}. These kind of space-time domains are
conceptually distinct from the cooperative rearranging regions we
discussed in the thermodynamic context, from AGDM theory down to the
mosaic. This is no surprise, because the aim of DFT is to explain
dynamical heterogeneities, denying any thermodynamic origin of
cooperative behaviour. The situation is not that simple, though,
as we have seen that nontrivial thermodynamic correlations 
{\it do} exist and it is probably necessary to reconcile dynamics
and thermodynamics, rather suppressing one of the two. Still, the
picture of dynamical heterogeneities as real domains in a larger 
$(x,t)$ space is very fruitful.

The DFT approach challenges virtually all fits and extrapolations that
are normally done in the glassy context, from the MCT fit of
viscosity, down to the VFT fit and to Kauzmann's extrapolation,
claiming that other theoretical forms are equally satisfying in
fitting the data and that extrapolations are, well, just
extrapolations.  In so doing, DFT questions very vividly (see in
particular \cite{kcm-4}) all theoretical scenarios that relies more or
less firmly on such things.  A partial reason for the radically
critical attitude of DFT, comes from the fact that for a long time
much of the theoretical description of glass-forming liquids, inspired
by MCT and mean-field spin-glasses, disregarded real space structure
(as we have seen, the non-mean-field approach of \cite{kivelson-95,
tarjus-95, tarjus-96, tarjus-98, kivelson-98, viot-98, tarjus-00,
grousson-01,tarjus-02} was a notable exception). Even though the need
to go local in space was very clearly stated in some of the old papers
(including Goldstein's, as we have seen), how to do that remained
rather elusive. The problem of real space was only marginally on the
map. At some point, in this scientific landscape dynamical
heterogeneities appeared, bringing the urge of a new, real space
approach. As we have seen, a purely static, phase space description is
at pain in describing dynamical heterogeneities. It was therefore not
surprising that in this context a different theory as DFT appeared,
whose starting point was the very ingredient somewhat muted in the
physics of glassy systems, namely real space structure.  Hence, apart
from its scientific value and the validity of its prediction, which I
do not evaluate here, DFT had the merit to force everyone in the field
to put at the centre of the investigation real space. In particular,
any mean-field inspired thermodynamic approach could no longer be
decoupled from the problem of how to define and measure a correlation
length.

As we have seen in these notes, the situation has changed in the last
ten years or so.  Even the purely thermodynamic approach to the deeply
supercooled phase makes some efforts to understand what is actually
going on in real space.  Some of the efforts are rather successful and
we finally have a static correlation length, and a reasonably
clear interpretation of what are local activated events. Moreover, as
we have seen, also the investigation of dynamical heterogeneities
received much help from some tools that make perfectly sense also in
the mean-field context, as the dynamic susceptibility. The main problem now
seems how to bring together the dynamic and static descriptions of
cooperativity; but all this in real space.

Of course, in those cases where quantitative predictions done by DFT
are in radical disagreement with other theories (as the mosaic or
MCT), experiments and simulations must clear up which is the
fundamentally correct view.  However, part of the DFT framework is in
fact only an alternative {\it description} of glassy phenomena, not
necessarily in contradiction with other descriptions, once real space
traits are included into them. For example, the concepts of dynamical
facilitation is certainly valid {\it per se}, and the interpretation
of dynamical heterogeneities as space-time domains is illuminating.
The fact that a model with trivial thermodynamics displays glassy
dynamics (as the FA model), is not a proof that thermodynamics is
irrelevant in all glassy systems. Similarly, the fact that a model
with no real space structure displays glassy dynamics (as the $p$-spin
model), is not a proof that real space is irrelevant in all glassy
systems.  Different descriptions, pruned of their incorrect aspects,
can ultimately coexist and complete each other. Such coexistence may
even contribute to a deeper understanding.

\subsection{From particles to quasi-species}

Another approach linking glassy phenomenology to the concentration of
defects, albeit of topological, rather than dynamical nature, was
recently introduced in \cite{procaccia-07, procaccia-07-bis,
procaccia-07-tris, procaccia-08, procaccia-09}. In two dimensions this
approach shares some similarities with the theory of volume defects in
kinetically constrained models formulated in \cite{lawlor-02,
lawlor-05}.

The main idea of the approach is to perform a coarse-graining of the
system that allows a {\it discrete} statistical mechanics formulation,
which is hopefully easier to study than the original continuous
one. The up-scaled degrees of freedom of such theory are a finite
number of quasi-species (also called `defects' in the early
formulations of the theory), each one with well-defined energy and
entropy. How to define quasi-species?

In two dimensions this was done by performing the Voronoi tessellation
of the liquid configurations of a simulated glass-forming system
\cite{procaccia-07, procaccia-07-bis, procaccia-07-tris,
procaccia-08}, building on the results of \cite{deng-1, deng-2,
deng-3, deng-4, perera-99}.  Due to the Euler topological constraint,
the average coordination number of the Voronoi cells must be equal to
$6$, so that local coordination numbers other than $6$ can be
identified with topological defects. The total concentration $c(T)$ of
defects as a function of temperature was first measured in
\cite{perera-99}, where it was found a rather mild temperature
dependence of it. However, in \cite{procaccia-07, procaccia-07-bis} a
finer classification of defects was given according to the number of
sides of the cells, thus defining a small number of different
quasi-species to whom each particle belongs.  In particular,
liquid-like quasi-species were identified and it was found that their
concentration $c_l$, in contrast with the total concentration of
defects, has a very sharp temperature dependence.  At high $T$,
$c_l(T)$ follows an Arrhenius decay, whereas at low temperature,
$c_l(T)$ seems to decrease as a power law, $c_l\sim (T-T_c)^2$, where
the fitted divergence $T_c$ is below the lowest simulated temperature,
so that (as usual) no real singularity is actually observed.  In three
dimensions Voronoi tessellation is less satisfactory, but
quasi-species can still be defined by counting the number of nearest
neighbours (within a certain threshold distance) of each particles
\cite{procaccia-09}.  In this way each particle belongs to one out of
a certain (small) number of quasi-species. As in the two dimensional
case, the concentration of liquid-like quasi-species decreases quite
sharply on lowering the temperature (although apparently no longer as
a power law).

The liquid-like quasi-species concentration is associated within this
framework to a (static) correlation length, $\xi\sim c_l^{-1/d}$,
representing the typical distance between liquid-like
quasi-species. In finite size simulations, $\xi$ becomes rapidly
larger than the size of the system and this implies that no
liquid-like quasi-species are found below a certain (size-dependent)
temperature, so that their concentration is effectively zero.  The
interesting point is that an independent determination of the
relaxation time shows that, $\tau_\mathrm R \sim \exp(\xi/T)$, both in
two and three dimensions \cite{procaccia-07-tris, procaccia-08,
procaccia-09}.  This result implies that the barrier to rearrangement
of cooperative regions scales as $\xi$, in contrast with AGDM theory,
according to which the barrier scales as $\xi^d$ (note, though, that
an early study \cite{procaccia-07} claimed that in $d=2$ the barrier
scaled as $\xi^2$; this was later corrected to $\sim \xi$ in
\cite{procaccia-07-tris, procaccia-08, procaccia-09}).

The free-energy $f$ of the  $n$-th quasi-species can be defined by inverting the formula
for its average concentration,
\be
\langle c(n) \rangle =\frac{e^{-\beta f(n)}}{\sum_n e^{-\beta f(n)}}  \ ,
\ee
where $\langle c(n)\rangle$ is computed numerically.  What one
observes is that $f(n,T)$ is a linear function of $T$, thus indicating
that the energy $e(n)$ and entropy $s(n)$ of each quasi-species do not
depend on temperature \cite{procaccia-07, procaccia-09}. This is
interpreted as a validity criterion of the coarse-graining from
particles to quasi-species.  From the free-energy, $f(n,T)=e(n)- T
s(n)$, one can obtain the energy and entropy of each quasi-species,
use them as input parameters of a model of noninteracting degrees of
freedom, and perform a semi-quantitative thermodynamic study. The results
are encouraging \cite{procaccia-07-bis,procaccia-09}, in that the
calculated liquid-like concentrations follow rather closely the
functional form obtained by the fit of the numerical data.

%%%%%%%%%%%%%%%%%%%%%%%%%%%%%%%%%%%%%%%%%%%%%%%%%%%%%%%%%%%%%%%%%%%%%%%%%%%%%%%
%%%%%%%%%%%%%%%%%%%%%%%%%%%%%%%%%%%%%%%%%%%%%%%%%%%%%%%%%%%%%%%%%%%%%%%%%%%%%%%

\subsection{The other spin-glass}

The $p$-spin model is definitely {\it not} the mother of all
spin-glasses.  Originally, spin-glass models were introduced to
describe real disordered magnets and the archetype of such models is
the Edward-Anderson (EA) spin-glass, where Ising spins interact
through some random $2$-body couplings \cite{virasoro}. In
contrast to the EA pairwise interaction, in the $p$-spin model the
interacting units are plaquettes of $p$ spins, with $p\geq 3$
\cite{cavagna-review}.

The differences in the phenomenology of these two classes of models
are deep and numerous. One of such differences, that we have already
remarked before, is that in the EA spin-glass static and dynamic
transitions coincide and such (unique) transition is distinctively continuous,
whereas in the $p$-spin model the dynamical transition has a discontinuous
nature and it is well separated from the lower static transition.  These
$p$-spin traits suit very well two distinctive features of real
supercooled liquids, namely the sudden appearance of two-steps
dynamical relaxation close to $T_g$ and the fact that all the static drama
seems to take place at lower temperature (close to $T_k$).  These
analogies, together with the fact that the dynamical
equations of the $p$-spin coincide with the MCT ones, 
were the first hint that the $p$-spin model, unlike models
for realistic spin-glasses as EA, could be the right paradigm for
glass-forming liquids \cite{kirk-0}.

In the light of this, it is perhaps surprising the recent claim of
\cite{moore-06, moore-07} that the right paradigm for describing the
thermodynamic behaviour of supercooled liquids is the `other'
spin-glass, i.e. that with continuous, rather than discontinuous
transition. By using the effective potential approach of
\cite{recipes}, the free energy of a liquid was mapped to that of the
EA model with an external magnetic field $h$ \cite{moore-06}.  This
mapping was then used to import to supercooled liquids physics all the
results, but alas! also all the open issues, of spin-glass
physics. The foremost of these issues is whether or not in $d=3$ the
EA model in a field has a thermodynamic phase transition at finite
$T$.  Even though this question is still quite disputed (see
\cite{petit-02} vs. \cite{jonsson-05} for experiments, and
\cite{leuzzi-08} vs. \cite{young-08} for numerical simulations), the
authors of \cite{moore-06} trust there is no transition and therefore
exploit the mapping to claim that also in glass-forming liquids there
is no thermodynamic (i.e. Kauzmann's) transition at finite $T$.  This
is the first result of the effective potential theory of
\cite{moore-06,moore-07}.

As I have stressed more than once during this review, whether or not a
real thermodynamic phase transition exist at finite $T$ is perhaps
{\it not} the most relevant question in supercooled liquids,
considering that we will never be able to equilibrate down to this
temperature anyway. What really matters is to uncover the correct
physical mechanism leading to the sharp (super-Arrhenius) increase of
the relaxation time  typical of fragile
systems. In this respect the effective potential theory of
\cite{moore-06, moore-07} agrees with most of the approaches we have
studied, in claiming that there is an increasing cooperative
lengthscale $\xi$ of static nature
\footnote{This lengthscale is called $R^*$ in \cite{moore-06}, to
distinguish it from the standard spin-glass correlation length $\xi$. However,
here I prefer to keep the same notation as I have used in the rest
of the notes, and call $\xi$ the cooperative length.}, which is
responsible for the increase of the barrier to relaxation and thus
ultimately of the relaxation time. Moreover, the mechanism of
formation of the cooperative regions according to
\cite{moore-06,moore-07} shares some similarities with the mosaic
theory, in that $\xi$ is fixed by the competition between a surface energy
cost and a thermodynamic advantage, where the former scales with a
power of the region's size $R$ smaller than the latter.

However, there is one {\it big} difference between the effective
potential theory and the mosaic theory and this is the physical origin
of the thermodynamic advantage. As we have seen, within the mosaic
this is provided by the {\it entropy} gain due to the existence of
many states available to the region, and this gain scales like $T S_c
R^d$. In contrast to this, according to \cite{moore-06,moore-07} the
drive to rearrangement is provided (under the spin mapping) by the
{\it energy} gained by flipping a certain number of spins, due to the
presence of the magnetic field $h$. In complete similarity with the
Imry-Ma argument for the random field Ising model \cite{shukla}, this
gain is proportional to $h R^{d/2}$: the energy of each flipped spin
is a random variable so that according to the central limit theorem
the net gain scales like the square root of the volume. We clearly see
that, within this theory, the surface tension exponent $\theta$ has to
be {\it smaller} than $d/2$ otherwise the cost is asymptotically
always larger than the advantage. According to
\cite{moore-06,moore-07} this condition is granted by the fact that in EA-like
spin-glasses one has $\theta\sim 0.2$
\cite{huse-86}. Furthermore, the usual fiddling with the largely
unknown exponent $\psi$ (connecting $\xi$ to the barrier), yields a
relationship for the relaxation time that is compatible with VFT
\cite{moore-06}.

Intriguing as it may be, the energy gain hypothesis of
\cite{moore-06,moore-07} has one rather serious problem: what is $h$
in supercooled liquids?  In doing the mapping from liquids to
spin-glass, it is very hard to keep under control the relationship
between the liquid parameters, as density and temperature, and the
magnetic field $h$. Hence, not only the theory provides no precise
functional form for $h=h(T,\rho)$, but also there is no clear (nor
unclear) physical interpretation of $h$ in the liquid context. This is
unfortunate, since according to the effective potential theory the
field is the very core of the thermodynamic drive to cooperative
rearrangement. Hence, what the theory really lacks is a physical 
justification of the energy gain to rearrangement: it is unclear 
what would be the equivalent of the Imry-Ma argument in real liquids.

As we have seen in the last chapter, the current numerics indicates
that $\xi$ is fixed by two competing terms, scaling with different
powers of $R$, without actually telling much about the physical nature
of these two terms.  Hence, one cannot exclude {\it a priori} that the
thermodynamic drive is of energetic, rather than entropic,
nature. Still, it would be important to have a physical interpretation
of this energy drive in order to compare the effective potential
theory to other approaches. In my own opinion, one possibility is that
standard local energy fluctuations, which indeed scale as the square
root of the volume, could provide the energetic gain described by the
effective potential theory: particles in a certain region may
rearrange simply because the new local configuration has slightly
(i.e. of order $R^{d/2}$) lower energy. The fact that some interface
energy cost must always be paid would assure that anyway the
rearrangement process leaves the system at the same avarage energy
level. Under this interpretation, to distinguish the mosaic from the
effective potential theory one should check numerically what is the
right balance of powers between surface cost and thermodynamic drive.
Note that if an entropic gain, scaling like $R^d$, is present and if
the value of $\theta$ is not too small (say, larger than $d/2$), then
the energetic drive would be anyway irrelevant compared to the
entropic drive. We are probably not far from being able to solve this
issue at least numerically.

%%%%%%%%%%%%%%%%%%%%%%%%%%%%%%%%%%%%%%%%%%%%%%%%%%%%%%
%%%%%%%%%%%%%%%%%%%%%%%%%%%%%%%%%%%%%%%%%%%%%%%%%%%%%%
%%%%%%%%%%%%%%%%%%%%%%%%%%%%%%%%%%%%%%%%%%%%%%%%%%%%%%
%%%%%%%%%%%%%%%%%%%%%%%%%%%%%%%%%%%%%%%%%%%%%%%%%%%%%%
%%%%%%%%%%%%%%%%%%%%%%%%%%%%%%%%%%%%%%%%%%%%%%%%%%%%%%

\section{Outlook}

As we have seen in the last two chapters, much of the current research
in glass-forming liquids is dedicated to develop new theoretical and
empirical tools able to define and detect a growing correlation
length. For some, this lengthscale finds its natural role within the
theoretical frameworks inspired by mean-field theories, and actually
contributes to their real space update; for others, the correlation
 length is the cornerstone over which to build  new
theories.
Whatever is the right way, it is out of doubt that the next frontier
in the physics of supercooled liquids is to reach a full understanding
of real space cooperativity. From this understanding, it will be
probably possible to select the most convincing theoretical
framework. What are the main open issues on this front?

We have seen that at low temperatures the growth of the correlation
length is responsible for the super-Arrhenius increase of the
relaxation time. We have also seen that this happens because the
activation barrier scales like some power of the correlation length,
$\Delta \sim \xi^\psi$.  We have shown that the Adam-Gibbs and the
mosaic scenario differ on the theoretical determination of
$\psi$. Numerically, a precise empirical determination of $\psi$ in
three-dimensional systems is still lacking. Knowledge of $\psi$ would
prove very useful to compare different theoretical scenarios. Of
course, this requires to measure the precise value of $\xi$ and
$\tau_\mathrm R$ over a large enough temperature range. If this is not
problematic for the relaxation time, we are still at the beginning for
what concerns the correlation length.  Of course, one must be sure
that the time and length that are measured are indeed correlated to
each other; in other words, they must measure the dynamical and
spatial fluctuations of the {\it same} order parameter. Yet, at low
temperatures we may expect that only one fundamental scale ruling slow
fluctuations survives, so that probably any reasonable measurement of
$\tau_\mathrm R$ and $\xi$ would do the job.

A determination of the surface tension between amorphous states, and in
particular of the exponent $\theta$, seems also quite important,
because it may shed light on the most fundamental ingredient of the
mosaic theory, namely the competition between configurational entropy
and surface energy. In particular, we have seen that the surface
tension should go to zero at the effective spinodal temperature
$T_c\sim T_x$, but we provided no empirical evidence for this. In
fact, as we have seen, one of the sharpest critique put forward by the
facilitated dynamics approach is that the infamous crossover from
high-$T$ nonactivated, to low-$T$ activated dynamics at $T_c\sim T_x$
is just a delusion of the community, and that in reality there is no
$T_c$, no $T_x$ and activation is the only relevant mechanism
\cite{kcm-4}. Thus, checking whether or not the surface tension goes
to zero at the spinodal $T_c$, seems relevant.  Furthermore, an
empirical determination of the exponent $\theta$ may also have
something to say about the real space geometry of the cooperative
rearranging regions.

A third point that seems still quite tricky is the interplay between
the dynamic and the static correlation lengths. If it seems reasonable
that at low $T$ just one scale survives, it is certainly essential to
understand how we pass from two distinct correlation lengths to one,
given that such crossover occurs in a temperature region (around
$T_c$) that is crucial both from an experimental and theoretical point
of view. As we have seen, the dynamical correlation length $\xi_d$
grows quite sharply even above $T_c$.  In fact, according to
mean-field inspired pictures, there should be a divergence of $\xi_d$
at $T_c$, which is however avoided due to the intervention of
activated phenomena. These are in turn cooperative, and ruled by the
static correlation length $\xi$, which is undefined above $T_c$. From
this picture, one would conclude that $\xi_d$ is {\it not} associated
to activated events, or at least that its growth get the main
contribution from a nonactivated mechanism.  Therefore, it remains to
be understood: 1. Why and how a nonactivated mechanism (saddles?)  is
associated to the increase of $\xi_d$? 2. How does it work precisely
the relay between $\xi_d$ and $\xi$ at $T_c$?  Regarding these two
points, some interesting theoretical progresses have recently been
done in Ref.\cite{franz-montanari}.  It would be important to support
them with an adequate empirical analysis.

This said, the reader should not be misled to think that we are close
to reach a final understanding of supercooled liquids and the glass
transition. Not only the three `little' points I have raised above
will in fact take ages to clear up and select a unified picture, but
also we must not forget that there are dozens of other open issues
that, due to the pedestrian nature of these notes, I did not discuss
at all. Hence, the student be not afraid about the lack of unsolved
problems to work on.  Supercooled liquids will remain a beautifully
open subject for quite a long time.

\vglue  1 truecm

\centerline{\bf Acknowledgments} 

\vglue  0.5 truecm

I am deeply indebted to S. Franz, I. Giardina, T.S. Grigera, G. Tarjus
and F. Zamponi for their critical reading of the manuscript and for
several important remarks and corrections.  I thank R.
Benzi and I. Procaccia for inviting me at the Weizmann Institute and
for giving me the opportunity to write this review.  I finally wish to
warmly thank J.-P. Bouchaud, G. Biroli, C. Cammarota, S. Franz,
I. Giardina, T. S. Grigera, V. Marinari, M.A. Moore, G. Parisi,
F. Ricci-Tersenghi, M. Tarzia, P. Verrocchio and F. Zamponi for many valuable
discussions on the subject of supercooled liquids while writing this
review.

\end{document}